\documentclass[]{aastex63}

\newcommand{\hi}{{\rm H}{\textsc i}}

\received{XXX}
\revised{YYY}
\accepted{ZZZ}

\submitjournal{ApJ}

\shorttitle{WALLABY: Tidal interaction in Eridanus supergroup}
\shortauthors{Wang et al.}

\graphicspath{{./}}

\begin{document}

\title{WALLABY Pre-Pilot Survey: The effects of tidal interaction on radial distribution of color in galaxies of the Eridanus supergroup}

\correspondingauthor{Jing Wang}
\email{jwang\_astro@pku.edu.cn}

\author[0000-0002-9663-3384]{Shun Wang}
\affiliation{Kavli Institute for Astronomy and Astrophysics, Peking University, Beijing 100871, China}
\affiliation{Department of Astronomy, School of Physics, Peking University, Beijing 100871, China}

\author{Jing Wang}
\affiliation{Kavli Institute for Astronomy and Astrophysics, Peking University, Beijing 100871, China}

\author{Bi-Qing For}
\affiliation{ICRAR, The University of Western Australia, 35 Stirling Highway, Crawley WA 6009, Australia}

\author{Lee, Bumhyun} 
\affiliation{Kavli Institute for Astronomy and Astrophysics, Peking University, Beijing 100871, China}

\author{Reynolds, T.N.} 
\affiliation{ICRAR, The University of Western Australia, 35 Stirling Highway, Crawley WA 6009, Australia}

\affiliation{ARC Centre of Excellence for All Sky Astrophysics in 3 Dimensions (ASTRO 3D), Australia}

\author{Lin, X.} 
\affiliation{School of Physics, Peking University, Beijing 100871, China}

\author{Staveley-Smith, L.} 
\affiliation{ICRAR, The University of Western Australia, 35 Stirling Highway, Crawley WA 6009, Australia}

\author{Li Shao} 
\affiliation{National Astronomical Observatories, Chinese Academy of Sciences, 20A Datun Road, Chaoyang District, Beijing, 100012, China}

\author{Wong, O.I.} 
\affiliation{CSIRO Space and Astronomy, PO Box 1130, Bentley WA 6102, Australia}
\affiliation{ICRAR, The University of Western Australia, 35 Stirling Highway, Crawley WA 6009, Australia}
\affiliation{ARC Centre of Excellence for All Sky Astrophysics in 3 Dimensions (ASTRO 3D), Australia}

\author{Catinella, B.} 
\affiliation{ICRAR, The University of Western Australia, 35 Stirling Highway, Crawley WA 6009, Australia}
\affiliation{ARC Centre of Excellence for All Sky Astrophysics in 3 Dimensions (ASTRO 3D), Australia}

\author{Serra, P.} 
\affiliation{INAF – Osservatorio Astronomico di Cagliari, Via della Scienza 5, 09047 Selargius, CA, Italy}

\author{Verdes-Montenegro, L.} 
\affiliation{Instituto de Astrofísica de Andalucía (CSIC)}

\author{Westmeier, T.} 
\affiliation{ICRAR, The University of Western Australia, 35 Stirling Highway, Crawley WA 6009, Australia}

\author{Lee-Waddell, K.} 
\affiliation{ICRAR, The University of Western Australia, 35 Stirling Highway, Crawley WA 6009, Australia}
\affiliation{CSIRO Space and Astronomy, PO Box 1130, Bentley WA 6102, Australia}


\author{Koribalski, B.S.} 
\affiliation{CSIRO Astronomy and Space Science, Australia Telescope National Facility, PO Box 76, NSW 1710, Australia}
\affiliation{School of Science, Western Sydney University, Locked Bag 1797, Penrith, NSW 2751, Australia}

\author{Murugeshan, C.} 
\affiliation{CSIRO Space and Astronomy, PO Box 1130, Bentley WA 6102, Australia}
\affiliation{ARC Centre of Excellence for All Sky Astrophysics in 3 Dimensions (ASTRO 3D), Australia}

\author{Elagali, A.} 
\affiliation{Telethon Kids Institute, Perth Children’s Hospital, Perth, Australia}

\author{Kleiner, D.} 
\affiliation{INAF – Osservatorio Astronomico di Cagliari, Via della Scienza 5, 09047 Selargius, CA, Italy}

\author{Rhee, J.} 
\affiliation{ICRAR, The University of Western Australia, 35 Stirling Highway, Crawley WA 6009, Australia}
\affiliation{ARC Centre of Excellence for All Sky Astrophysics in 3 Dimensions (ASTRO 3D), Australia}


\author{Bigiel, F.} 
\affiliation{Argelander-Institut für Astronomie, Universität Bonn, Auf dem Hügel 71, D-53121 Bonn, Germany}

\author{Bosma, A.} 
\affiliation{Aix Marseille Univ, CNRS, CNES, LAM, Marseille, France}

\author{Holwerda, B.} 
\affiliation{University of Louisville, Department of Physics and Astronomy, 102 Natural Science Building, 40292 KY Louisville, USA}

\author{Oh, S.-H.} 
\affiliation{Department of Physics and Astronomy, Sejong University, 209 Neungdong-ro, Gwangjin-gu, Seoul 05006, Republic of Korea}

\author{Spekkens, K.} 
\affiliation{Royal Military College of Canada, PO Box 17000, Station Forces, Kingston, Ontario, Canada, K7K 7B4}


\begin{abstract}

We study the tidal interaction of galaxies in the Eridanus supergroup, using $\hi$ data from the pre-pilot survey of WALLABY (Widefield ASKAP L-band Legacy All-sky Blind surveY). We obtain optical photometric measurements and quantify the strength of tidal perturbation using a tidal parameter $S_{sum}$. 
For low-mass galaxies of $M_* \lesssim 10^9 M_\odot$, 
we find a dependence of decreasing $\hi$-to-optical disk size ratio with increasing $S_{sum}$, but no dependence of $\hi$ spectral line asymmetry with $S_{sum}$. This is consistent with the behavior expected under tidal stripping.
We confirm that the color profile shape and color gradient depend on the stellar mass, but there is additional correlation of low-mass galaxies having their color gradients within $2R_{50}$ increasing with higher $S_{sum}$. For these low-mass galaxies, the dependence of color gradients on $S_{sum}$ is driven by color becoming progressively redder in the inner disk when tidal perturbations are stronger. For high-mass galaxies, there is no dependence of color gradients on $S_{sum}$, and we find a marginal reddening throughout the disks with increasing $S_{sum}$.
Our result highlights tidal interaction as an important environmental effect in producing the faint end of the star formation suppressed sequence in galaxy groups.

\end{abstract}


\section{Introduction} \label{sec:intro}

Lambda CDM simulations predict that groups and clusters of galaxies started to assemble in large quantities from a redshift of $\sim$2 \citep[e.g.,][]{2004MNRAS.355..819G}. By the present time, more than 20\% of the galaxies with a stellar mass above $10^{10}$ $M_\odot$ are likely satellites of groups \citep[e.g.,][]{2005ApJ...630....1Z, 2006MNRAS.365...11C}, and the ratio increases for galaxies with lower stellar masses, up to 30\% for galaxies with $M_* > 10^9 M_\odot$ \citep[e.g.,][]{2007MNRAS.376..841V, 2008ApJ...676..248Y}. Thus the environment plays an increasingly more important role in the galaxy evolution at later epochs \citep[e.g.,][]{2008MNRAS.387...79V, Wetzel2013, Haines2015}.
Theories and observations have converged on the point that galaxies grow primarily through forming stars, and the neutral hydrogen provides the raw material for forming stars \citep[see e.g.,][]{2008AJ....136.2846B}. While molecular gas may be the more direct star forming material, the $\hi$, which can be further replenished by gas cooling and accretion from the CGM, provides the reservoir to sustain the star formation \citep[e.g.,][]{Saintonge2016, Catinella2018, Wang2020, 2021ApJ...918...53G}. Because of its low density and often extended nature, the $\hi$ is an excellent probe of environmental effects. Thus the variation of star formation rate (SFR) and $\hi$ richness in different environments with respect to those in the field is an effective measure of the effect of environment on galaxy growth \citep{Boselli2006, boselli2014a, 2021PASA...38...35C}.

Several observational trends have been established which serve as benchmarks of environmental effects in galaxy evolution models. The SFR and $\hi$ richness of galaxies tend to be lower in more massive groups \citep[e.g.,][]{2009MNRAS.400.1962K, Hess2013}, and lower in satellites than in central galaxies \citep[e.g.,][]{2012MNRAS.427.2841F, 2017MNRAS.466.1275B}. For satellites, the SFR and $\hi$ richness tend to be lower in the vicinity of group centers \citep[e.g.,][]{Gavazzi2005, Gavazzi2006}, and at higher local densities \citep[e.g.,][]{2013A&A...553A..90G, 2020MNRAS.499.3233R}. These trends tend to be more prominent for low-mass galaxies than for high-mass galaxies \citep[e.g.,][]{Boselli2014}.
These environmental trends are based on general parameters of the environment, which effectively capture but mix different physical mechanisms. For example, a high level of small-scale density can be related to tidal interaction events in a compact group, or harassments \citep{Moore1996} in a cluster; a high level of large-scale density can be related to fly-by interactions \citep[e.g.,][]{1992ApJ...400..153M, 2012ApJ...751...17S} in a loose group, or ram-pressure stripping \citep{Gunn1972} from the dense hot gas of a massive cluster. Even in a given cluster or group, weak ram-pressure and tidal effects co-exist near the virial radius \citep[e.g.,][]{Balogh2000, 2004ApJ...613..866K}, while ram-pressure stripping, viscous stripping \citep{Nulsen1982}, evaporation \citep[e.g.,][]{Nipoti2007}, harassment and galaxy-cluster tidal effects co-exist at intermediate cluster-centric radii.
Moreover, those environmental parameters often ignored the in-fall trajectory of galaxies into the dense environment which plays a crucial role in determining the relative importance of different environmental processes \citep[e.g.,][]{Vollmer2001, 2015MNRAS.454.2502S, 2021MNRAS.502.1785J}.

Studies based on various galaxy samples have been designed to specifically investigate different gravitational or hydrodynamic processes. The galaxy-galaxy tidal interaction or merger, because of its prevalence in low- to intermediate-mass groups \citep[e.g.,][]{Chung2009}, and strong link to starbursts \citep{Larson1978}, has received extensive attention. These are typically based on samples of phase-space selected galaxy pairs \citep[e.g.,][]{2003MNRAS.346.1189L, Ellison2010} or morphologically selected post-mergers \citep[e.g.,][]{Toomre1972}. Whether and when the global SFR, the central SFR, and the SFR in the outer disks are boosted or quenched are key signatures that are frequently searched for. It was found that mergers or galactic interactions on average moderately and significantly elevate the total and central SFR of star-forming galaxies (e.g., \citealt{1994ApJ...431L...9M, 2007A&A...462..507L, Morales-Vargas2020}; but see \citealt{2012A&A...540A..96M} for the case in compact groups).
The central and total SFR of the primary galaxy (the more massive one) is more strongly elevated if the interacting pair has higher SFR, higher stellar mass, smaller separation, or lower relative velocity \citep[e.g.,][]{1994ApJ...430..179Z, 2004MNRAS.352.1081A, 2018MNRAS.479.3381B}. The central SFR of the target galaxy may become suppressed if the interacting neighbor is passive, possibly due to a lack of ISM (interstellar medium)-ISM hydrodynamic interaction, and/or a ram-pressure removal of the circumgalactic medium (CGM) by the dense CGM of the neighbor \citep[e.g.,][]{2005ApJ...635L..29P, 2016ApJS..222...16C}.
The SFR in the outer disk may temporarily be suppressed alongside an elevated central SFR, in the beginning of the interaction \citep[e.g.,][]{2019ApJ...881..119P}. There is also evidence that galactic interactions may fuel AGN \citep[e.g.,][]{2008A&A...486...73S}, destroy, induce or strengthen bars \citep[e.g.,][]{1991ApJ...370L..65B, 1996ApJ...471..115B}, dilute metallicities \citep[e.g.,][]{2010ApJ...721L..48K, 2020MNRAS.494.3469B}, heat the CGM \citep[e.g.,][]{2004ApJ...607L..87C} and modify molecular gas mass \citep{2011A&A...534A.102L, 2017A&A...607A.110L}. The SFR (distribution) and $\hi$ richness may be altered by these additional effects.

Theoretical studies suggest that tidal interactions affect SFR level and distribution through modifying the cold gas distribution and kinematics, and the rate of converting cold gas to stars \citep[e.g.,][]{Toomre1972, 2008MNRAS.384..386C, 2009ApJ...691.1168H}.
However, statistical studies on the response of the $\hi$ gas to the tidal interactions has mostly been conducted with single-dish total $\hi$ fluxes \citep[e.g.,][]{2015MNRAS.448..221E, Ellison2018}. The total $\hi$ mass in post-mergers has been found not to be depleted but possibly slightly increased \citep[e.g.,][]{Ellison2018}, raising the question how galaxies managed to maintain little net change in $\hi$ richness, while there can be violent inflow and conversion of $\hi$ to fuel the AGN and star formation, and accompanying energetic feedback \citep[e.g.,][]{2011MNRAS.418.2043E, 2013MNRAS.430.1901H}. Studying the $\hi$ content of galaxy pairs is harder with single-dish observations, as their resolutions are usually not high enough to resolve the individual galaxies. Thus, some studies only investigated the total $\hi$ content of the system at different merger stages \citep[e.g.,][]{2015ApJ...805....2S, 2018ApJS..237....2Z}, while some others try to deblend the total $\hi$ into the individual galaxies based on their positions within the telescope beam and assumptions about their properties \citep[e.g.,][]{2011AJ....142..170H, 2020MNRAS.499.3193B}. These studies consistently find that the $\hi$ richness of merging pairs remains normal with respect to isolated galaxies. But we caution that diffuse and extended $\hi$ tails produced in interactions can go beyond the single-dish beam or be resolved out in a targeted observation \citep[e.g.,][]{2019MNRAS.489.5723F, 2019MNRAS.487.5248L}. Such tails are prevalent as demonstrated by interferometric $\hi$ images of a number of local compact groups \citep[e.g.,][]{2001A&A...377..812V, 2013MNRAS.428..370S, 2019MNRAS.482.3591R}.

Although our knowledge about how galactic properties depend on local densities and how different types of tidal interactions modify galactic properties accumulates, statistical quantification of how tidal interactions, amongst many other environmental mechanisms in the context of fully mapped groups/clusters, affect the SFR distribution and $\hi$ richness has been limited \citep[e.g.,][]{2021PASA...38...35C}. 
Considering the possible hydrodynamical effects additionally provided by the dense intracluster medium, it is important to separate the effects of tidal interactions from hydrodynamic effects in order to better understand the initial conditions of galaxies before they infall into and evolve in the groups/clusters \citep[e.g.,][]{Wetzel2013}.
Previous results suggest that tidal enhancement of central SFR is much weaker in high-density regions than in low-density regions \citep{2006MNRAS.367.1029S, 2013ApJ...762...43K}. However, as passive galaxies are more abundant in dense environment, it is unclear whether this trend is more driven by the (lack of) ISM-ISM hydrodynamic effects \citep{Ellison2010}, or by other effects more related to the large-scale cluster properties \citep{2009MNRAS.399.1157P}.

The Eridanus supergroup observed in the pre-pilot stage of WALLABY ( Widefield ASKAP L-band Legacy All-sky Blind surveY, \citealt{Koribalski2020}), provides us with an opportunity to study the galaxy-galaxy tidal effects in a complete cosmic structure. 
It consists of three groups undergoing a major merger into one cluster \citep{Willmer1989, 2005JApA...26....1O, Brough2006}, which should increase the frequency of galactic tidal interactions with respect to isolated groups of similar masses \citep{Fujita1998, Gnedin2003}. Such systems should be more common at high redshift, but Eridanus is unique below a redshift of 0.08 \citep{2004MNRAS.352..605B}.
Its proximity enables us to explore the behavior of the extremely low-mass ($M_* < 10^9 M_\odot$) dwarf galaxies, which are building blocks of more massive galaxies, and more vulnerable to environmental effects than the massive galaxies. Simulations suggest that perturbation from even a minor neighbor with a mass ratio less than 1/10 could induce gas inflows \citep{2020MNRAS.494.4969P, 2020MNRAS.493.3716H}.
However, despite their significant tidal contribution, such low-mass dwarfs are often not included \citep[e.g.,][]{Kauffmann2004}
or suffer from incomplete sampling \citep[e.g.,][]{Tortora2010} in SDSS-based studies due to the limited depth of SDSS images.
With the availability of deeper and wider optical and radio data (e.g., WALLABY and Legacy Survey, see Section \ref{sec:LS}), we can build a more comprehensive picture of galaxy evolution across the spectrum of galaxy mass.

The moderate resolution of WALLABY is not sufficient to resolve the $\hi$ in all the galaxies, but greatly helps deblend the $\hi$ emission that could have been confused with single-dish observations, as well as capture extended $\hi$ tails that could extend beyond a single-dish beam. A sample selected by $\hi$ flux can be biased against relatively gas-poor and passive galaxies, but is advantageous for focusing on the early stage of environmental processing. Such a selection is particularly useful to break the nurture or nature degeneracy encountered in environment studies, where gas-poor galaxies in dense environments can either reflect on-going environmental processing or the cluster assembly history.

The paper is organized as follows. We introduce the samples that we construct and the data we use for our study in Section \ref{sec:data}. We illustrate how we treat the data and derive physical parameters, including color gradients and tidal parameters, in Section \ref{sec:analysis}. The results related to $\hi$ asymmetry, $\hi$-to-optical disk size ratio, color profile, color gradients and tidal parameters are presented in Section \ref{sec:results}. Then in Section \ref{sec:Dis} we link our results to previous theoretical and observational studies and discuss the implications on group galaxy evolution. At last in Section \ref{sec:con}, we summarize the key results and the conclusions on our findings. 

\section{Sample} \label{sec:data}
\subsection{The HI sample of WALLABY detected galaxies in the Eridanus supergroup} \label{sec:WALLABY}
The Eridanus supergroup was observed by Australia Square Kilometre Array Pathfinder (ASKAP) as part of WALLABY \citep{Koribalski2020} in its pre-pilot phase. With 36 antennas functioning, the data reaches a detection limit of 2.4$-$4.4 mJy beam$^{-1}$ across the field (central beams have lower rms than the beams at the edge of the field) with a spatial resolution of 30 arcsec and a channel width of 18.5 kHz, or 4 km s$^{-1}$. 
The raw data was reduced with ASKAPsoft \citep{Whiting2017}. The $\hi$ data cube was searched for detections with SoFiA \citep{Serra2015, 2021MNRAS.506.3962W}. The products including velocities, coordinates, $\hi$ total fluxes, and moment maps, were released internally to the WALLABY team. More details about the data can be found in previous WALLABY publications studying the Eridanus fields \citep{2021MNRAS.tmp.2066F, 2021MNRAS.tmp.2089M, 2021MNRAS.tmp.2077W}.

We exclude two $\hi$ clouds ($\hi$ dectections with no optical counterparts): WALLABY J033911-222322 and WALLABY J033723-235753, which are studied in detail by \citet{2021MNRAS.tmp.2077W}.
The main purpose of this paper is to study how galaxies evolve in the Eridanus supergroup, so we select from the remaining 53 $\hi$ detections using the following criteria: 1) classified as members of any of the three sub-groups by \citet{Brough2006} (Section \ref{sec:optical}), or 2) within the escape velocity curve and two virial radii in the phase-space diagram of any of the three sub-groups. The mass, virial radius, and velocity dispersion of each group are taken from \citet{Brough2006}. The selection results in 36 galaxies, which we refer to as the $\hi$ sample and serves as the main analysis sample. The radial velocities of galaxies in the $\hi$ sample range from $\sim$1200 to $\sim$2000 km s$^{-1}$.

We adopt a distance of 20.9 Mpc \citep{Forbes2006} for the Eridanus supergroup.
We take the star formation rate (SFR) estimated in \citet{2021MNRAS.tmp.2066F} following the procedure described in \citet{Wang2017}. In brief, each SFR is considered to be the sum of dust attenuated and unattenuated SFRs. The dust unattenuated SFR is estimated based on the GALEX \citep{Martin2005} FUV (NUV when FUV is not available) luminosities, and the dust attenuated part on the WISE \citep{Wright2010} W4 luminosities. When neither FUV nor NUV data is available, the W4 SFR is taken as the lower limit of the total SFR. When there is no detection in W4 band, zero dust attenuation is assumed. 

\subsection{The optical sample combining the Brough et al. (2006) catalog and Cosmicflows-3 catalog} \label{sec:optical}

\citet{Brough2006} compiled a catalog of galaxies in the Eridanus supergroup combining data from 6dFGS, NED and HyperLEDA. The flux limit is roughly 13.1 mag in the $K$ band. They assigned in total 60 members to the three sub-groups (NGC1407 group, NGC1332 group and Eridanus group) of the supergroup using the friends-of-friends (FoF) method. We exclude two galaxies (APMUKS(BJ) B033830.70-222643.7 and NGC 1331) from the supergroup members due to problematic optical images, and use the remaining 58 galaxies as the first part of our optical catalog.

Cosmicflows-3 \citep{2016AJ....152...50T} is the best existing description of the large-scale environment surrounding the Eridanus supergroup. 
We use the Cosmicflows-3 catalog to include non-member galaxies near the edge of the supergroup which are not close enough to ensure a group membership, but close enough to exert significant tidal force on the relatively outlying group members. We do not use the member identification for the Eridanus supergroup itself from the Cosmicflows-3 group catalog \citep{2015AJ....149..171T, 2016AJ....152...50T}, because its galaxies were selected with a flux limit (11.75 mag in the $K$ band) brighter than that of \citet{Brough2006}, so missed a large fraction of the WALLABY-detected, $\hi$-rich dwarf galaxies. 

For each member galaxy of the Eridanus supergroup, we search for galaxies in the Cosmicflows-3 catalog which are within a projected distance of 0.8 Mpc around the member galaxy, and differ in luminosity distance from the Eridanus supergroup center by less than 4.11 Mpc (three times the virial radius of the supergroup), but were not identified as members of the supergroup in \citet{Brough2006}. By doing so, we select in total three galaxies, which is the second part of the optical catalog. 
Although the number of three looks small, this procedure confirms that we do not evidently miss an outlying population possibly contributing to the summed tidal strength which is not included by \citet{Brough2006}.
If we increase the threshold projected distance to 1.37 Mpc (the virial radius of the supergroup), only five more galaxies are further included and these additional galaxies contribute less than 5\% of the total tidal strength to any of the relevant galaxies in the $\hi$ sample. We thus stick to a threshold of 0.8 Mpc. Low-mass galaxies are largely missed by the Cosmicflow-3 catalog, but they contribute little to the total tidal forces as we will show later.

Finally, we have 61 galaxies in total in our optical sample. We retrieve radial velocities for optical sample galaxies from NED, by selecting the measurement in the optical with the minimum uncertainty.
When no measurement is available, we take those labelled as ``preferred'' by NED. There are 22 galaxies from the $\hi$ sample which are overlapping with the optical sample. For these galaxies, we adopt the $\hi$ systematic velocities from WALLABY as their radial velocity.

\section{Analysis} \label{sec:analysis}

\subsection{Photometry}
\subsubsection{Optical total fluxes and surface brightness profiles} \label{sec:LS}
We use optical $g$, $r$ and $z$ images from DESI Legacy Survey \citep{Dey2019}, to derive photometric measurements for the two samples.
The typical full-width half-maximum of the point spread function is 1.2 arcsec. The typical depths are 23.7, 23.3, and 22.2 mags in the $g$, $r$ and $z$ bands, respectively.

We largely follow the photometry procedure of \citet{Wang2017}. The main steps are described below.

\begin{enumerate}
\item \textbf{Deblending and masks.}
We use SExtractor \citep{Bertin1996} to produce a mask image for each galaxy, through the so-called cold$+$hot source finding mode \citep[e.g.,][]{2004ApJS..152..163R}, based on the $r$-band image. In the cold mode, all clumps possibly belonging to the galaxy are merged into a master segmentation, by setting the SExtractor deblending parameter to 0.3. Then, in the hot mode, small clumps in the master segmentation of the galaxy are picked out by setting the deblending parameter to 0.001. With the aid of the SExtractor output $CLASS\_STAR$, we inspect the clumps in all three bands, so that foreground stars and background galaxies are masked. We then dilate the masks with a width of 11 pixels, to cover the scattered light of bright stars and galaxies. We inspect each masked image again, and adjust the mask when necessary.

\item \textbf{Galaxy shape and background subtraction.}
We use SExtractor to derive the center, position angle, axis ratio, and a rough estimate of the background and background rms for each galaxy. We use the python package $photutils$ \citep{Bradley2016} to derive surface brightness profile in each of the three bands. The surface brightness is derived as the $\sigma$-clipped median value of pixels in elliptical rings, which have the center, position angle and axis ratio fixed to the estimates of SExtractor.
We identify the radius where the profile flattens within the noise level (the flattening radius hereafter), and use this part of profile to estimate a local residual of the background. We remove this residual background from the radial profile. The uncertainty $\sigma$ in the surface brightness values are calculated combining Poisson error and the uncertainties introduced in the two steps of background subtraction. We cut the radial profiles at the surface brightness level of 1-$\sigma$.

\item \textbf{Clean image and total flux.}
The pixels masked within the flattening radius are replaced by the surface brightness profile value at the corresponding radius. The pixels outside the flattening radius are assigned random values following a Gaussian distribution with $\sigma$ equivalent to the measured rms of the background. By doing so, we produce a clean image in each band for each galaxy. The clean images are inspected for quality. We finally run SExtractor again on the clean image to derive the half-light radius $R_{50}$ and Petrosian magnitude measurements. 

\end{enumerate} 

Galactic extinction is corrected based on the Planck 2013 dust model \citep{Abergel2014} and the dust extinction curve of \citet{Cardelli1989}.

\subsubsection{Stellar mass, HI mass and disk sizes} \label{sec:RHI}
We estimate the stellar mass based on the Petrosian fluxes in the $r$ and $g$ bands. The $r$ band stellar mass-to-light ratio based on the $g - r$ color is calculated according to the equation of \citet{Zibetti2009}. The stellar masses are then estimated according to the $r$ band luminosity and stellar mass-to-light ratio. Optical disk sizes are estimated as $R_{25, g}$ in the $g$ band. We perform linear interpolation on $g$-band surface brightness profiles to derive the radius where the surface brightness reaches 25 mag arcsec$^{-2}$. 

$\hi$ masses of galaxies in the $\hi$ sample are from \citet{2021MNRAS.tmp.2066F}. For the optical-only galaxies (i.e. those in the optical sample but not in the $\hi$ sample), we still need their $\hi$ masses (as part of the total baryonic mass) in the calculation of tidal forces later. We approximate the $\hi$ masses to be zero for the optical-only galaxies which are covered but not detected in the WALLABY observations. The upper limits of the $\hi$ mass are no larger than 10\% the stellar mass of these galaxies, thus the $\hi$ mass does not contribute much to the total baryonic mass. We also confirm that the results are not changed if we assume the upper limits instead of zero for the $\hi$ mass.

For the optical-only galaxies beyond the WALLABY observing footprint (16 galaxies), we test two sets of approximations for the $\hi$ mass. In the first set, we approximate the $\hi$ mass to be zero which leads to a lower limit of the baryonic mass of the galaxy. In the second set, we estimate the $\hi$ mass based on the $g - r$ color and effective stellar surface density, following the equation of \citet{Zhang2009}. The second set of approximations can be viewed as an upper limit of $\hi$ mass, because the equation of \citet{Zhang2009} is derived using a sample that is strongly biased towards $\hi$-rich field galaxies (i.e. cross-match of the SDSS DR4 and HyperLeda $\hi$ data), while the galaxies in the optical sample are in a denser environment and tend be more $\hi$-deficient \citep{2021MNRAS.tmp.2066F}.
We examine the final results based on the two types of approximations and do not find a significant difference. We thus use the $\hi$ masses estimated from color and surface density for the optical-only galaxies that are not in the WALLABY footprint.

We estimate the $\hi$ disk size $R_{\rm HI}$, the radius where the $\hi$ surface density reaches 1 $M_\odot$ pc$^{-2}$ \citep{Broeils1997}, based on the size-mass relation of \citet{Wang2016}. We do not directly derive $R_{\rm HI}$ from the WALLABY images for $\hi$ sample galaxies, because most $\hi$ disks in the $\hi$ sample are barely resolved. We confirm that the resolved disks are well consistent with the $\hi$ size-mass relation with a scatter of 0.07 dex for the deviation.

\subsubsection{Color gradients for the HI sample}
Based on the surface brightness profiles, we derive color profiles and color gradients. We derive $g - r$, $g - z$ and $r - z$, but will focus on the $g - r$ color after finding that the trends are similar. 
The color uncertainty at each radius is calculated as $\sigma_{g - r} = \sqrt{\sigma_g^2 + \sigma_r^2}$, where $\sigma_g$ and $\sigma_r$ are the errors of surface brightness in the $g$ and $r$ band, respectively. We limit color profiles to the radial range where $\sigma_{g - r} < 0.1$ mag.

\citet{MacArthur2004} found that different fitting ranges for color gradients change the result, with flips in the sign of the color gradient within or beyond the effective radius \citep[see also][]{2008ApJ...683L.103B}. Using integral field spectrograph (IFS) facilities such as CALIFA and MaNGA, \citet{Marino2016} and \citet{Zheng2017} also found evidence of bending or breaks in color profiles of late-type galaxies. Thus, in this paper, color gradients are derived in two different radial ranges: $0 < R < R_{50, z}$ and $R_{50, z} < R < 2 R_{50, z}$, which are denoted as $CG_{01}$ and $CG_{12}$, respectively. We divide galaxies into pieces by $R_{50, z}$ instead of the break radius in color profiles, because it better traces the fractional growth of the stellar disk. At a radius of $2 R_{50, z}$, 80\% (90\%) galaxies in our sample have the surface brightness in the $r$ band brighter than 23.8 (24.5) mag arcsec$^{-2}$, thus restricting the analysis within this radius minimizes contamination from scattered light of neighboring galaxies or bright stars.

The color gradient ($CG$) is derived as the slope of the linear fit to the color profile $g - r$ as a function of $R/R_{50, z}$ in a given radial range, where $R$ is the radius, and $R_{50, z}$ is the half-light radius in the $z$ band. Specifically, the $CG$ is given as $CG = \delta (g - r) / \delta (R / R_{50, z})$. We use $R_{50, z}$ instead of other bands for it traces the stellar mass more closely, and 
has better signal-to-noise ratio than the half-$M_*$ radius. We experiment with conducting the analysis throughout this paper based on the half-light radius in other bands and also based on half-$M_*$ radius, and the results do not significantly change and the correlations do not become stronger.

We point out that, when deriving the $CG$ in the way described above, we have treated the galaxy as a whole. But the bulges should response less sensitively than disks to environmental effects, as the stars are older and hotter \citep[e.g.,][]{1994cag..book.....S}. It will perhaps be useful in future studies to conduct bulge-disk decomposition and derive $CG$ for disks only. However, as the $R_{90, z} / R_{50, z}$ value ranges from 1.9 to 2.8 (10 to 90$^{th}$ percentiles) in our sample, the influence of bulges on our results should not be severe.

\subsection{Tidal strength for the HI sample} \label{sec:S}
We use the tidal parameter suggested by theoretical studies to quantify the instantaneous tidal perturbation experienced by a galaxy. One of the commonly used parameters is the dimensionless tidal parameter $S_0 = (\frac{M_p}{M_g})(\frac{R_g}{d_{peri}})^3(\frac{\Delta t}{T})$ \citep{Oh2008}, where $M_p$ is the mass of the perturbing object, $M_g$ is the mass of the galaxy of interest, $R_g$ is the radius of the galaxy, and $d_{peri}$ is the pericenter distance between the perturbing object and the galaxy. $\Delta t$ is the time elapsed for the perturbing object to move over 1 radian near the pericenter, and $T \equiv \sqrt\frac{R_g^3}{G M_g}$ is the time taken by a test mass at $R = R_g$ to rotate 1 radian about the galaxy center. One obvious limitation with the $S_0$ parameter is that it does not apply to mergers in the coalescence stage where the gravitational effect is strong but the perturber is blended with the galaxy of interest. However, from inspecting the optical and $\hi$ images of the $\hi$ sample, there is only one system (NGC 1359, ID 69) clearly identified to be in the coalescence stage of a merger. We do not exclude this system, but discuss its potential contamination to results when necessary. 

In order to derive the parameter with observables, we make a few approximations. We use the projected distance ($d_{proj}$) to approximate the pericentric distance, and use $\frac{V_{circ}/R_g}{\Delta v_{rad}/d_{proj}}$ to approximate $\frac{\Delta t}{T}$, where $V_{circ}$ is the circular velocity of the galaxy, and $\Delta v_{rad}$ is the difference in line-of-sight velocity of the two objects. 
In addition, we add a smooth parameter $V_{smooth}$ to $\Delta v_{rad}$ to avoid zero divides. By adding the smooth parameter we artificially set a lower limit to the relative line-of-sight velocity between any two galaxies. We use a fiducial value of $V_{smooth} = V_{circ}$, which ranges from $\sim$50 to $\sim$180 km s$^{-1}$ and has a median of $\sim$80 km s$^{-1}$, but also tried other options including 50 km s$^{-1}$, 100 km s$^{-1}$ and 200 km s$^{-1}$. 
We confirm that all the major trends in Section \ref{sec:results} do not significantly change unless $V_{smooth}$ is set to unrealistically small values like 10 km s$^{-1}$. 

The derived dimensionless tidal parameter is thus calculated as 
\begin{equation}
S = \left(\frac{M_p}{M_g}\right) \left(\frac{R_g}{d_{proj}}\right)^2 \left(\frac{V_{circ}}{\sqrt{{(\Delta v_{rad})}^2 + V_{smooth}^2}}\right)
\end{equation}

Such approximations have obvious uncertainties. Consider the case in which $V_{smooth} = 0$. Firstly, the projected quantities $d_{proj}$ and $\Delta v_{rad}$ are the lower limits of the real separation ($d$) and velocity difference ($\Delta v$) between the galaxy and the perturber, so $(\frac{1}{d_{proj}})^2(\frac{1}{\Delta v_{rad}})$ is an over-estimate of $(\frac{1}{d})^2(\frac{1}{\Delta v})$.
On the other hand, the real separation and velocity difference are upper and lower limits of the distance and rotational velocity ($d_{peri}$ and $v_{peri}$, respectively) of the perturber at the pericenter. If we assume the angular momentum to be conserved in the frame centered on the galaxy, $d_{peri}v_{peri}=d\Delta v \equiv const$, then $(\frac{1}{d})^2(\frac{1}{\Delta v})$ is an under-estimate of $(\frac{1}{d_{peri}})^2(\frac{1}{\Delta v_{peri}})$. As a result, depending on circumstances the observationally derived $S$ can either under-estimate or over-estimate the physical $S_0$. So $S$ should only be viewed as a statistically correct indicator of $S_0$, and its relevant analysis should only be interpreted in a statistical sense. 
We roughly quantify the uncertainty of $S$ by using the velocity dispersion of Eridanus supergroup galaxies ($\sigma =$ 265 km s$^{-1}$, \citealt{Willmer1989}) as the relative line-of-sight velocity $\Delta v_{rad}$. We estimate a typical error of 0.3 dex for each perturber-perturbed object pair.

In addition, the infall history of the galaxies also introduces physical uncertainty. The tidal effect is strongest when galaxies are at pericenter. For galaxies that have already gone past that, the instantaneous measure of tidal strength, i.e. the tidal parameter, may under-estimate the persistent effect of tidal interaction. Therefore, the tidal parameter may be regarded as a lower limit of the real effect. On the other hand, our sample is biased against galaxies that are at pericenter, since their velocities there are then the highest, and thus the time elapsed is the shortest. But this last limitation generally applies to all randomly selected samples.

Despite these uncertainties, we point out that the estimate of the $S$ parameter is observationally supported by previous findings of tidally induced galactic features (e.g., enhanced central star formation, more perturbed optical disk morphologies) being correlated with the mass ratio, projected distance, and radial velocity offset of galaxy pairs separately \citep{2004MNRAS.352.1081A, Ellison2010, 2018MNRAS.479.3381B, Pan2018}. The use of the $S$ parameter combines these dependent factors in a physically motivated way, and avoids addressing the degeneracy between these factors when investigating the dependence of other galactic properties on the tidal strength. The latter point is particularly important for this study as the sample is not very large.
In the following, we describe the derivation of $V_{circ}$, and our options of $R_g$ and masses ($M_p$ and $M_g$).

\subsubsection{Estimation of total mass}
We calculate the baryonic mass as $M_b = M_* + 1.4 M_{\rm HI}$. The rotational velocity $V_{circ}$ is estimated using the baryonic Tully-Fisher relation of \citet{McGaugh2000}. We do not directly derive $V_{circ}$ from the WALLABY data cube because only a small fraction of the galaxies are spatially resolved. We do not derive $V_{circ}$ from the width of the global $\hi$ profile either, because the integral line width may not well trace $V_{circ}$ when galaxies are perturbed \citep[e.g.,][]{Reynolds2020, 2020MNRAS.492.3672W, 2020MNRAS.499.5205W}.
Furthermore, a significant fraction of galaxies in the $\hi$ sample are dwarf irregular galaxies, which in the optical tend to have thick disks and uncertain axis ratio due to the irregular morphology \citep[e.g.,][]{2010MNRAS.406L..65S, 2019MNRAS.482..821O}. Thus deriving the inclination angle and de-projecting the global line widths are expected to have large uncertainties. 

The total mass enclosed by radius $R$ is calculated as $M = M(R) = V_{circ}^2 R / G$, where $G$ is the gravitational constant, and $R$ should be large enough to be roughly in the regime where the rotation curve reaches $V_{circ}$. At this point, we have three options for $R$: the optical radius $R_{25, g}$, the $\hi$ radius $R_{\rm HI}$, and the baryonic radius $R_b$ which is the larger of $R_{25, g}$ and $R_{\rm HI}$. Directly referring to the theory of tidal interaction, masses (thus tidal parameters) estimated with $R_{25, g}$ ($R_{\rm HI}$) should be more sensitive probes of perturbations to the stellar ($\hi$) disk, while tidal parameters estimated using $R_b$ should be a more general indicator of whether the galaxy is perturbed. 
We take the mass (and thus the tidal parameter) estimated with $R_{25, g}$ as the fiducial measure, for in the analysis later we mainly focus on the effect of tidal interactions on the optical color gradients. But we re-iterate that, if we change to $R_{b}$ or $R_{\rm HI}$, our major conclusions are not affected.

\subsubsection{Tidal strength from different perturbers}
We use the term ``perturber'' to refer to the galaxy that causes tidal perturbation on the galaxy of interest. For each galaxy in the $\hi$ sample, the perturbers come from the superset combining the $\hi$ sample and the optical sample (75 galaxies altogether). We estimate the strength of tidal perturbation caused by each perturber, $S_i$, where $i$ denotes the specific perturber.

Strengths of tidal perturbation caused by perturber(s) are estimated in three ways as follows:
\begin{enumerate}
\item
That which is caused by the nearest perturber ($S_{nearest}$). The nearest perturber is defined as the one that has the smallest projected angular distance to the galaxy.
\item
That which reflects the most severe perturbation caused by any perturber ($S_{strongest}$), i.e. $S_{strongest} = max(S_i)$.
\item
That which reflects the summed effect of perturbation caused by all the perturbers ($S_{sum}$). Mathematically, $S_{sum} = \Sigma_i S_i$, where $i$ denotes different perturber. Referred to as ``the summed tidal parameter'' in the following.
\end{enumerate}

From first principles, tidal forces are vectors. However, when quantifying the cumulative tidal effects on a galaxy, it is not straightforward to treat the observables as vectors. For example, even when the instantaneous tidal forces from several companions cancel out as vectors at a given time, the cumulative tidal effects from these companions do not necessarily do so. This is because the orbits of the companions during one rotation period of the subject galaxy do not always cancel out. Based on such consideration, we consider $S_{sum}$ in addition to $S_{strongest}$ and $S_{nearest}$ in our analysis. Both the scalar sum \citep[e.g.,][]{Argudo-Fernandez2015} and strongest \citep[e.g.,][]{2013AJ....145..101K} tidal strengths were calculated in the literature to quantify the tidal fields. 
We find consistent results in most cases, but in some cases, $S_{sum}$ shows stronger correlation with galactic properties (e.g., $CG_{01}$ and $R_{\rm HI} / R_{25, g}$, see Section \ref{sec:CG} and \ref{sec:HId}), implying support for $S_{sum}$ as a more complete indicator of the total tidal effects (but not tidal forces) felt by a galaxy.

The key results of this study do not significantly change if we use other parameters like $\Theta \equiv \log [(M_* / 10^{11} M_\odot) (d_{proj} / $Mpc$)^{-3}]$ as the measurement of tidal strength \citep[e.g.,][]{2013AJ....145..101K, 2016MNRAS.459.1827P, 2021MNRAS.tmp.2077W}.

\subsubsection{Significant contributors of the summed tidal strength}
We investigate how many galaxies significantly contribute to the summed tidal strength ($S_{sum}$) and their spatial distribution for each galaxy. We use the curve of growth to analyze how the cumulative tidal strength increases as more neighbors are considered. For each galaxy, we first rank its perturbers by order of decreasing tidal strength, and calculate the cumulative sum. Then we normalize the cumulative tidal strengths by the summed tidal strength $S_{sum}$. The number of galaxies that contribute to 80 percent of $S_{sum}$ is then determined by the curves of growth, and denoted as $N_{80}$. To illustrate the spatial distribution of these significant contributors, we further determine $D_{80}$ which is the maximum distance between them and the subject galaxy. We also obtain the distances between the nearest (strongest) perturber and the subject galaxy as $D_{nearest}$ ($D_{strongest}$).
To quantify how important low-mass perturbers are, we also calculate the part of $S_{sum}$ that is contributed by perturbers with stellar mass lower than $10^9 M_\odot$, which we denote as $S_{sum, M_* < 10^9 M_\odot}$.

\begin{figure}[hptb]
    \centering
    \includegraphics[width=0.8\textwidth]{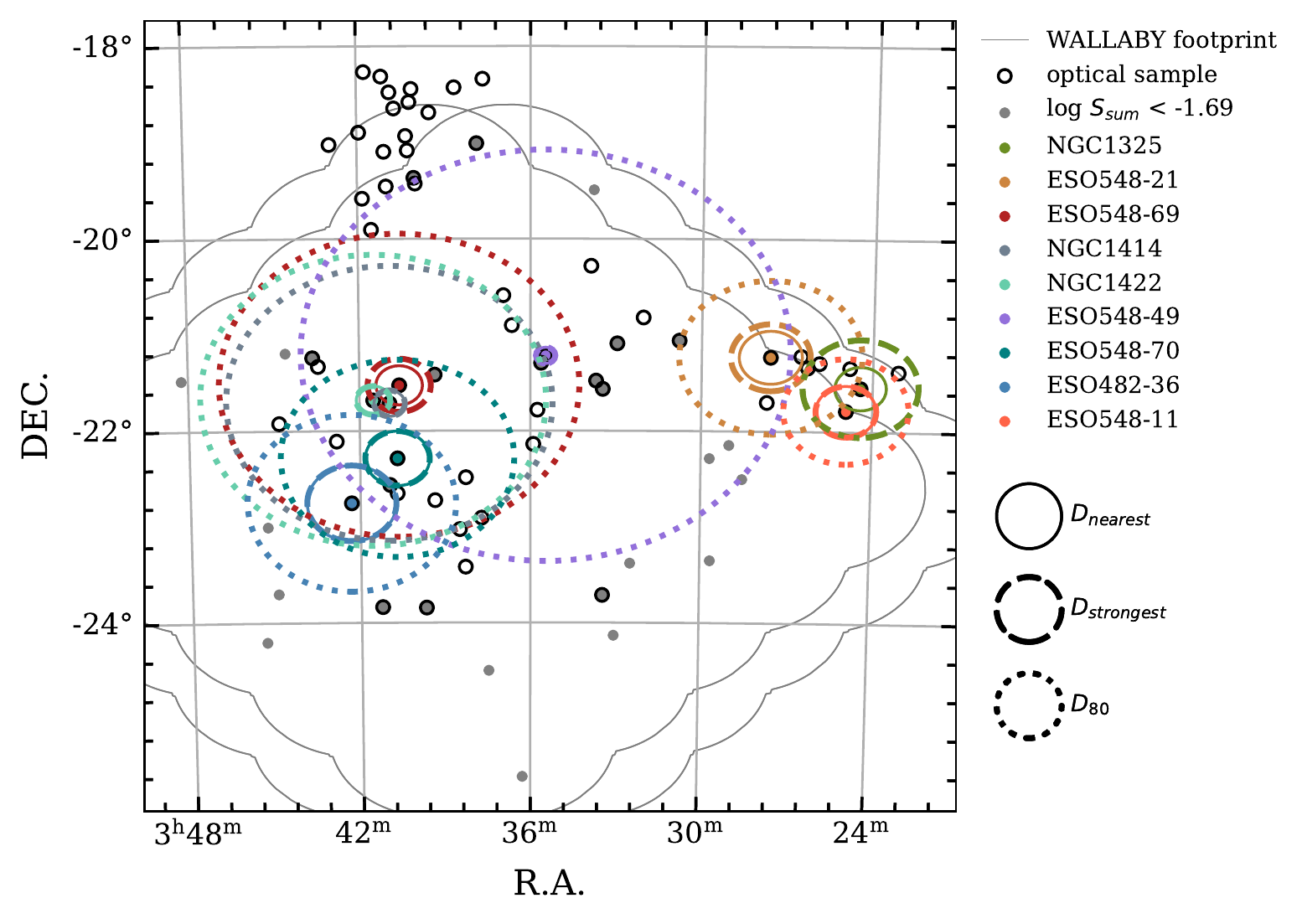}
    \caption{The spatial distribution of galaxies in the $\hi$ and optical samples. 
    Black open circles are galaxies from the optical sample, while filled circles are those from $\hi$ sample. Among the $\hi$ sample galaxies, strongly perturbed galaxies (i.e. those with $\log S_{sum} > -1.69$) are presented in colors (see labels on the right) and weakly perturbed galaxies are plotted in gray. 
    $D_{80}$ (dotted-line ellipses), $D_{strongest}$ (dashed-line ellipses) and $D_{nearest}$ (solid-line ellipses) of nine strongly perturbed galaxies are presented in the same color as the subject galaxy. In many cases, the dashed ellipse overlaps with the solid ellipse. The shape of the two WALLABY Eridanus footprints are shown by light gray lines. Note that NGC 1325 has equal $D_{strongest}$ and $D_{80}$.}
    \label{fig:Scircle}
\end{figure}

\begin{figure}[hptb]
    \centering
    \includegraphics[width=8.5cm]{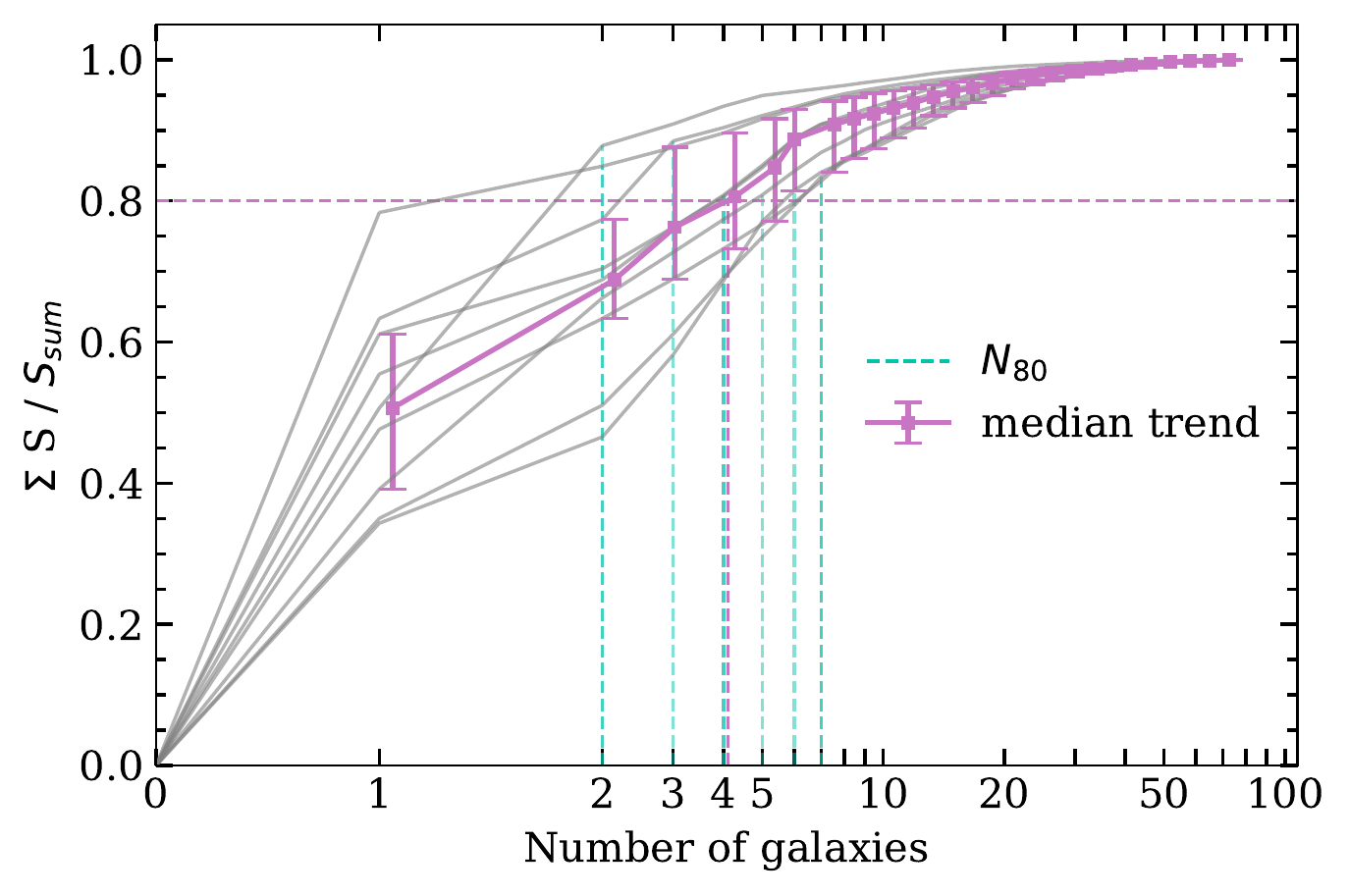}
    \includegraphics[width=8.5cm]{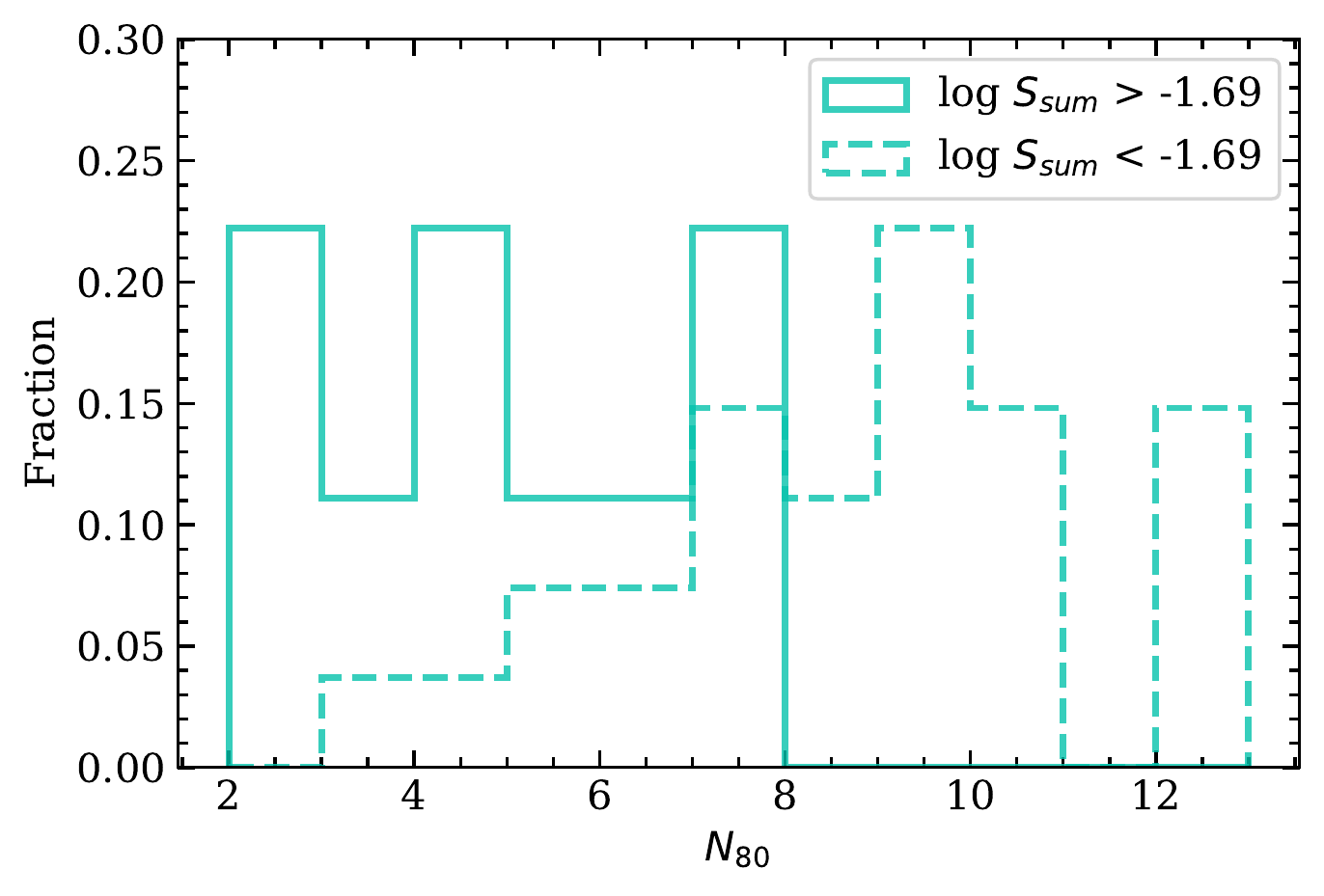}
    \includegraphics[width=8.5cm]{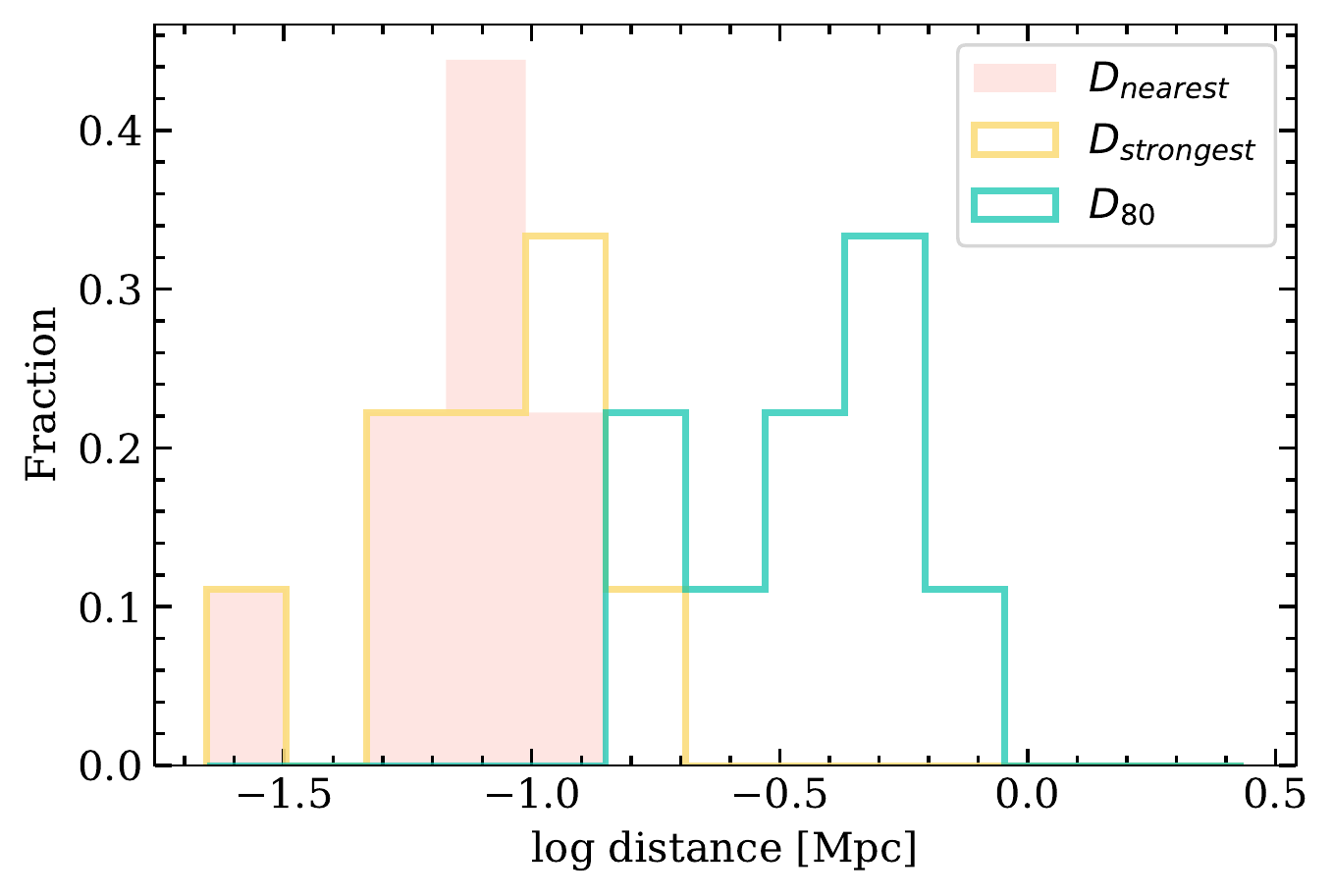}
    \includegraphics[width=8.5cm]{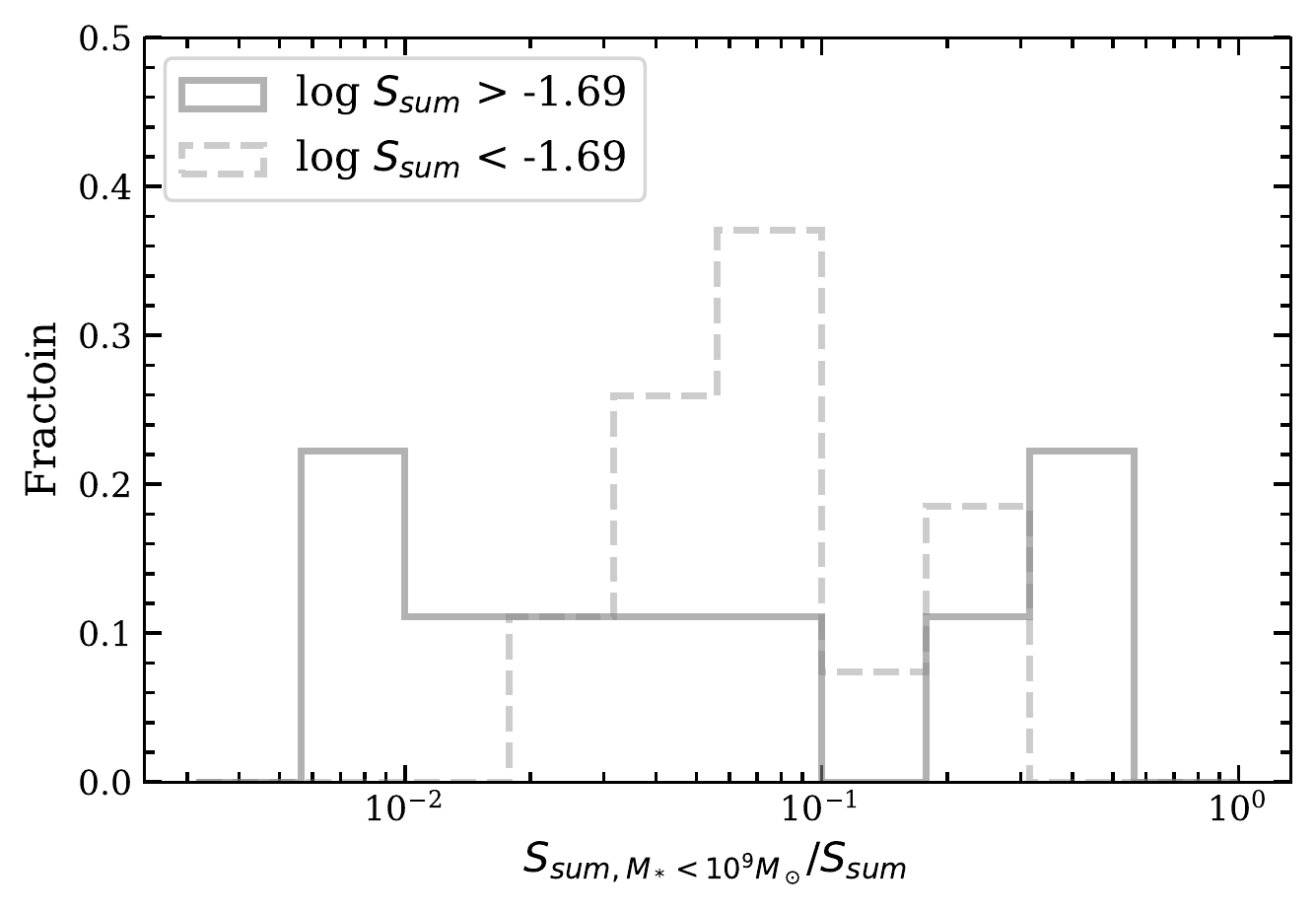}
    \caption{Properties of contributors to $S_{sum}$.
    \textbf{Upper-left:} 
    the curves of growth for individual strongly perturbed galaxies (those with $\log S_{sum} > -1.69$, in grey). The median value and scatter (25 and 75$^{\rm th}$ percentiles) of the grey curves is shown in magenta. The $N_{80}$ of individual (median) curves are labeled by vertical dashed cyan (magenta) lines.
    \textbf{Upper-right:} 
    the distribution of $N_{80}$ for strongly perturbed ($\log S_{sum} > -1.69$, cyan solid) and weakly perturbed ($\log S_{sum} < -1.69$, cyan dashed).
    \textbf{Lower-left:} 
    the distribution of $D_{nearest}$ (pink), $D_{strongest}$ (yellow) and $D_{80}$ (cyan) for nine strongly perturbed galaxies.
    \textbf{Lower-right:} 
    the distribution of the fraction of $S_{sum}$ that is contributed by perturbers with stellar mass smaller than $10^9 M_\odot$ ($S_{sum, M_* < 10^9 M_\odot} / S_{sum}$, see text), for strongly perturbed (gray solid) and weakly perturbed (gray dashed) galaxies.
    }
    \label{fig:Srange}
\end{figure}

We divide the subject galaxies by the 75$^{\rm th}$ percentile value of $\log S_{sum} = -1.69$, to separate them into two groups, hereafter strongly perturbed galaxies and weakly perturbed galaxies. The weakly perturbed galaxies are expected to have larger $D_{80}$ and $N_{80}$ than the strongly perturbed galaxies under similar circumstances, which is confirmed by our results.
As can be seen in Figure \ref{fig:Scircle}, the ellipses of $D_{80}$ are larger than $D_{strongest}$ and $D_{nearest}$ in most cases of strongly perturbed ($\log S_{sum} > -1.69$) galaxies. This suggests that $S_{sum}$ has a significant contribution from perturbers over a relatively large distance range, even though in some cases $D_{80}$ is only $\sim 2$ times $D_{nearest}$. It can also be seen that the strongest perturber is usually the nearest perturber, but not always, as indicated by the frequent overlapping of dashed- and solid-line ellipses.
The difference between $D_{nearest}$, $D_{strongest}$ and $D_{80}$ for strongly perturbed galaxies is more clearly illustrated in the lower-left panel of Figure \ref{fig:Srange}. Most (90\%) of the strongly perturbed galaxies have $D_{80}$ larger than 0.3 Mpc, but $D_{strongest}$ never exceeds this value.

We show in the upper-left panel of Figure \ref{fig:Srange} the curves of growth of tidal strength for individual strongly perturbed ($\log S_{sum} > -1.69$) galaxies and the median trend. In the median, 4 galaxies contribute 80\% of $S_{sum}$. For individual strongly perturbed galaxies, $N_{80}$ ranges from 2 to 7.
In the upper-right panel of Figure \ref{fig:Srange}, we present the distribution of $N_{80}$ for both strongly and weakly perturbed galaxies. Strongly perturbed galaxies do have smaller $N_{80}$ compared to the weakly perturbed galaxies. Most (80\%) galaxies have $N_{80} \geq 5$ and the whole sample has a median $N_{80}$ of 7.

These results consistently suggest that the summed strengths of tidal perturbation experienced by the galaxies in the Eridanus supergroup come from a number ($\gtrsim$ 4) of their neighbor galaxies at larger distances ($\gtrsim$ 0.3 Mpc), rather than only the closest neighbors as is the case in close galaxy pairs or triplet systems in an isolated environment \citep[e.g.,][]{Hibbard2001, Vollmer2005, 2014A&A...564A..94A}. Thus, investigating the tidal effects on group galaxies likely requires a complete sample covering a large enough sky area. Fortunately, the Eridanus supergroup is close-by, and our wide-field WALLABY data helps verify the completeness of redshifts for the gas-rich, low-mass galaxies. 
As we show in the lower-right panel of Figure \ref{fig:Srange}, the low-mass ($M_* < 10^9 M_{\odot}$) galaxies have a median contribution of $\sim$4\% ($\sim$7\% for strongly perturbed galaxies) of $S_{sum}$ and only 6 out of 36 galaxies have a fraction $\gtrsim$20\%. This result implies that $S_{sum}$ is not significantly under-estimated due to the $K$-band flux limit of the either the \citet{Brough2006} catalog or the Cosmicflow-3 catalog.

\subsection{Additional parameters}
In addition to direct measurements of SFR and $\hi$ mass ($M_{\rm HI}$), we also use specific star formation rates (sSFR $\equiv$ SFR/$M_*$), $\hi$ gas fraction ($f_{\rm HI} \equiv M_{\rm HI} / M_*$), deviation of SFR from the star forming main sequence ($\Delta$ SFR $\equiv$ $\log$ SFR - $\log$ SFR$_{\rm SFMS}(M_*)$, where SFR$_{\rm SFMS}(M_*)$ is the SFR expected for a typical star forming galaxy at a given $M_*$, \citealt{Saintonge2016}), and deviation of $\hi$ mass from the median relation of $M_{\rm HI}$ versus $M_*$ ($\Delta M_{\rm HI} \equiv \log M_{\rm HI} - \log M_{\rm HI, med}(M_*)$, where $M_{\rm HI, med}(M_*)$ is the median $M_{\rm HI}$ expected for galaxies of given $M_*$, \citealt{Catinella2018}). 

Another parameter we use is the $\hi$ spectral line asymmetry ($A_{spec}$). It is calculated following the procedure of \citet{Reynolds2020}. $A_{spec} \equiv \Sigma_i|S(i)-S_{flip}(i)| / \Sigma_i|S(i)|$ is the normalized sum of difference in flux intensity between the flipped spectrum and the original one, where $S(i)$ and $S_{flip}(i)$ are the flux intensities in channel $i$ of the original and flipped spectrum, respectively.

\section{Results} \label{sec:results}

Our analysis focuses on the $\hi$ sample. We separate the sample into low-mass and high-mass subsamples by the median value of stellar mass $M_{*} = {10}^{8.95} M_\odot$. The division is chosen to maximize the statistics for both subsamples, and theoretically and observationally it roughly divides two distinct regimes of galaxy formation. A stellar mass of $\sim 10^9 M_{\odot}$ corresponds roughly to a dark matter halo mass of $10^{11} M_{\odot}$ and virial velocity of $\sim$70 km s$^{-1}$ \citep{2019MNRAS.488.3143B}. Thus galaxies with $M_{*}<{10}^9 M_\odot$ tend to be strongly affected by stellar winds and winds launched by supernova which typically have speed of the order of 100 km s$^{-1}$ \citep[e.g.,][]{2005ARA&A..43..769V}. Previous observational studies showed that galaxies with stellar mass above and below this threshold indeed tend to have distinct $g-i$ color gradients \citep{Tortora2010}, disk thickness \citep{Sanchez-Janssen2010}, and slopes of the $\hi$ mass versus stellar mass relations \citep{2015MNRAS.447.1610M}. We point out that, $\sim 10^9 M_\odot$ is also approximately the lower limit of $M_*$ for $M_*$-complete samples selected from SDSS (Sloan Digital Sky Survey, \citealt{York2000}), and the upper limit of $M_*$ for dwarf irregular galaxies studied in the Local Volume in surveys like LITTLE THINGS \citep{Hunter2012}.

We quantify the linear correlation strength with the Pearson R value ($|R| > 0.45$ as significant and $0.3 < |R| < 0.45$ as considerable) and the p value of 5\% significance. The uncertainties in the coefficients are calculated via bootstrapping. Given the sample size, an R value of 0.45 is roughly equivalent to a p value of 0.05 for the correlation.
When SFR is involved, we use the astronomical survival analysis \citep{Feigelson1985, Isobe1986, Lavalley1992} rather than R to account for the lower limits. The python package $pymccorrelation$ and Kendall's-$\tau$ model is used. We also perform a robust linear fit to the data points, and use the deviation of the slope from zero as a measure of the significance of a linear relation. 
We investigate the dependence of galactic properties on the three types of tidal strength parameters ($S_{nearest}$, $S_{strongest}$ and $S_{sum}$). We present Pearson correlation coefficients for dependences on these three tidal parameters separately, but only show figures for trends related to $S_{sum}$. 

\subsection{HI asymmetry}

\begin{figure}[hptb]
    \centering
    \includegraphics[width=0.8\textwidth]{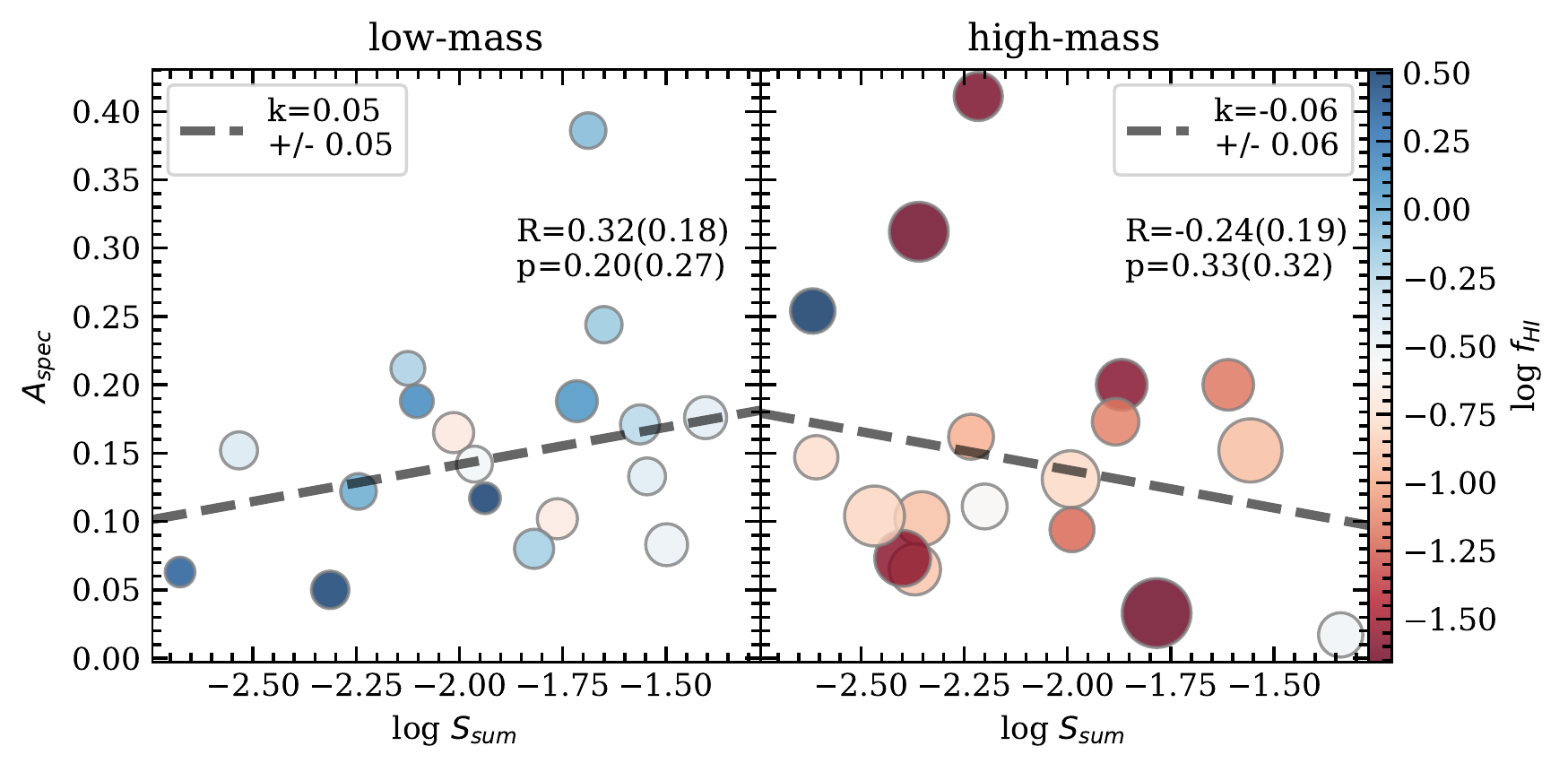}
    \caption{The correlation between $\hi$ spectral line asymmetry and the summed tidal parameter. 
    \textbf{Left:} for low-mass ($\log M_*/M_\odot < 8.95$) galaxies.
    \textbf{Right:} for high-mass ($\log M_*/M_\odot > 8.95$) galaxies.
    Pearson R and p values are shown with bootstrap error in parenthesis. Gray dashed lines are the result of robust linear fit, with the slopes (and error) shown in the corner. All data points are color coded by $\hi$ gas fraction (see colorbar on the right). The size of data points indicates the stellar mass of the galaxy, in the sense that larger data points are used for more massive galaxies.}
    \label{fig:asymmetry}
\end{figure}

The correlations between $\hi$ spectral asymmetries ($A_{spec}$) and the summed tidal parameter ($S_{sum}$) are presented in Figure \ref{fig:asymmetry}.
We do not find statistically significant evidence for a correlation between $\hi$ spectral asymmetries and tidal strength for both low-mass (left panel) and high-mass (right panel) galaxies. It indicates that, if high $S_{sum}$ values are related to stronger perturbations, they are not reflected in the $A_{spec}$ parameter. We remind that $A_{spec}$ are measured from the integral spectra, thus could miss local signatures of perturbation in the $\hi$ disks. It will be interesting to investigate how the multi-dimensional asymmetry is affected when better-resolved $\hi$ images are available in the future.

\subsection{HI disk size} \label{sec:HId}

\begin{figure}[hptb]
    \centering
    \includegraphics[width=0.8\textwidth]{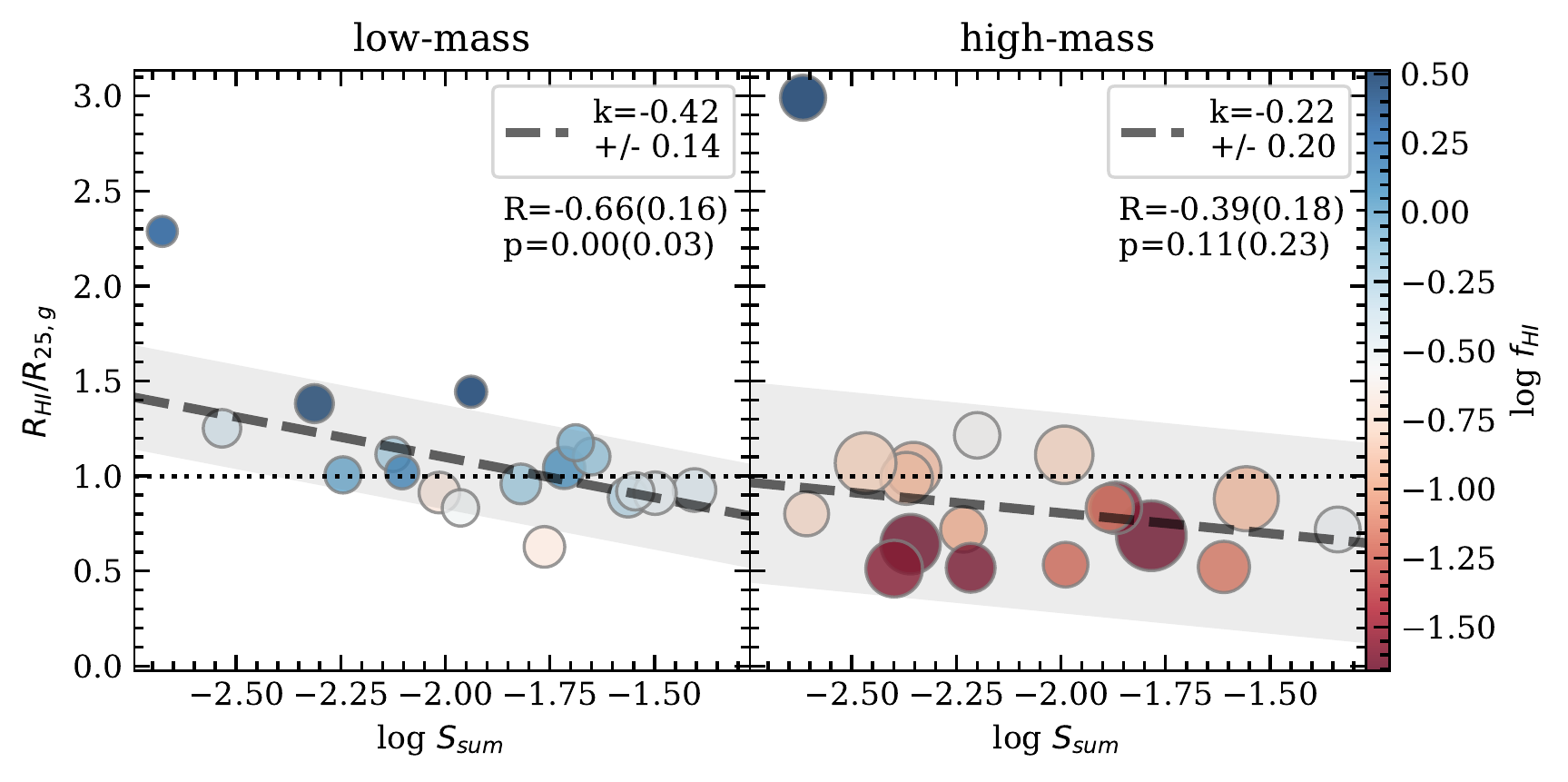}
    \caption{The correlation between the $\hi$-to-optical disk size ratio and the summed tidal parameter.
    The gray shaded area indicates the scatter ($1 \sigma$) of data points about the linear fit. The dotted horizontal line represents $R_{\rm HI} / R_{25, g} = 1$. 
    \textbf{Left:} for low-mass ($\log M_*/M_\odot < 8.95$) galaxies.
    \textbf{Right:} for high-mass ($\log M_*/M_\odot > 8.95$) galaxies.
    The other symbols and text are the same as Figure \ref{fig:asymmetry}: Pearson R and p values are shown with bootstrap error in parenthesis. Gray dashed lines are the result of robust linear fit, with the slopes (and error) shown in the corner. All data points are color coded by $\hi$ gas fraction (see colorbar on the right). The size of data points indicates the stellar mass of the galaxy, in the sense that larger data points are used for more massive galaxies.}
    \label{fig:RHI}
\end{figure}

In Figure \ref{fig:RHI}, we show the connection between the $\hi$-to-optical disk size ratio ($R_{\rm HI} / R_{25, g}$) and $S_{sum}$. Although $R_{\rm HI}$ is derived from $\hi$ mass and $R_{25, g}$ is strongly correlated with stellar mass, using the $\hi$-to-optical disk size ratio instead of the $\hi$-to-stellar mass ratio more directly traces the potential outside-in shrinking of $\hi$ disks with respect to the optical disks as a result of tidal perturbation.

The disk size ratios anti-correlate significantly with tidal strength in the low-mass sample. Most of the low-mass galaxies reside close to the best-fit linear relation except for LEDA 792493 (ID 63, in the upper-left corner) which has the lowest $S_{sum}$ and highest size ratio of the low-mass subsample. LEDA 792493 is a relatively unperturbed, $\hi$-rich dwarf galaxy. If we exclude LEDA 792493 from this analysis, we obtain an R value of $-0.47$ and a p value of 0.06.
For high-mass galaxies, there is no significant correlation between disk size ratio and $S_{sum}$. The best-fit linear relation has a larger scatter than that of low-mass galaxies, implying additional drivers for $\hi$ disk sizes in high-mass galaxies.
The outlier NGC 1359 (ID 69) has the lowest $S_{sum}$ and highest size ratio of the high-mass subsample. NGC 1359 is a merger system in the stage of coalescence. Because the whole system is treated as one galaxy, tidal strength $S_{sum}$ only considers the perturbation from galaxies outside this system. The correlation coefficient between $S_{sum}$ and $R_{\rm HI} / R_{25, g}$ for the high-mass galaxies becomes consistent with 0 if NGC 1359 is excluded from the subsample.

Considering that $R_{25,g}$ may only enclose a small fraction of the total flux for the low-mass galaxies, we test by replacing it with four times the disk scale length ($4 R_d$). We derive $R_d$ by fitting the outer part of the surface brightness profile of the galactic disks in the $g$ band. There is no qualitative difference between the results obtained by adopting $R_{25, g}$ and $4 R_d$ as the disk size estimates (see Figure \ref{fig:Rsl} in appendices). We note that potential systematic uncertainties in the $\hi$ fluxes of the ASKAP data of Eridaunus \citep{2021MNRAS.tmp.2066F} only slightly shift (by maximum $\sim$0.09 dex) but do not affect the tightness of the trend.

\subsection{Color profiles} \label{sec:CP}
\begin{figure}
    \centering
    \includegraphics[width=\textwidth]{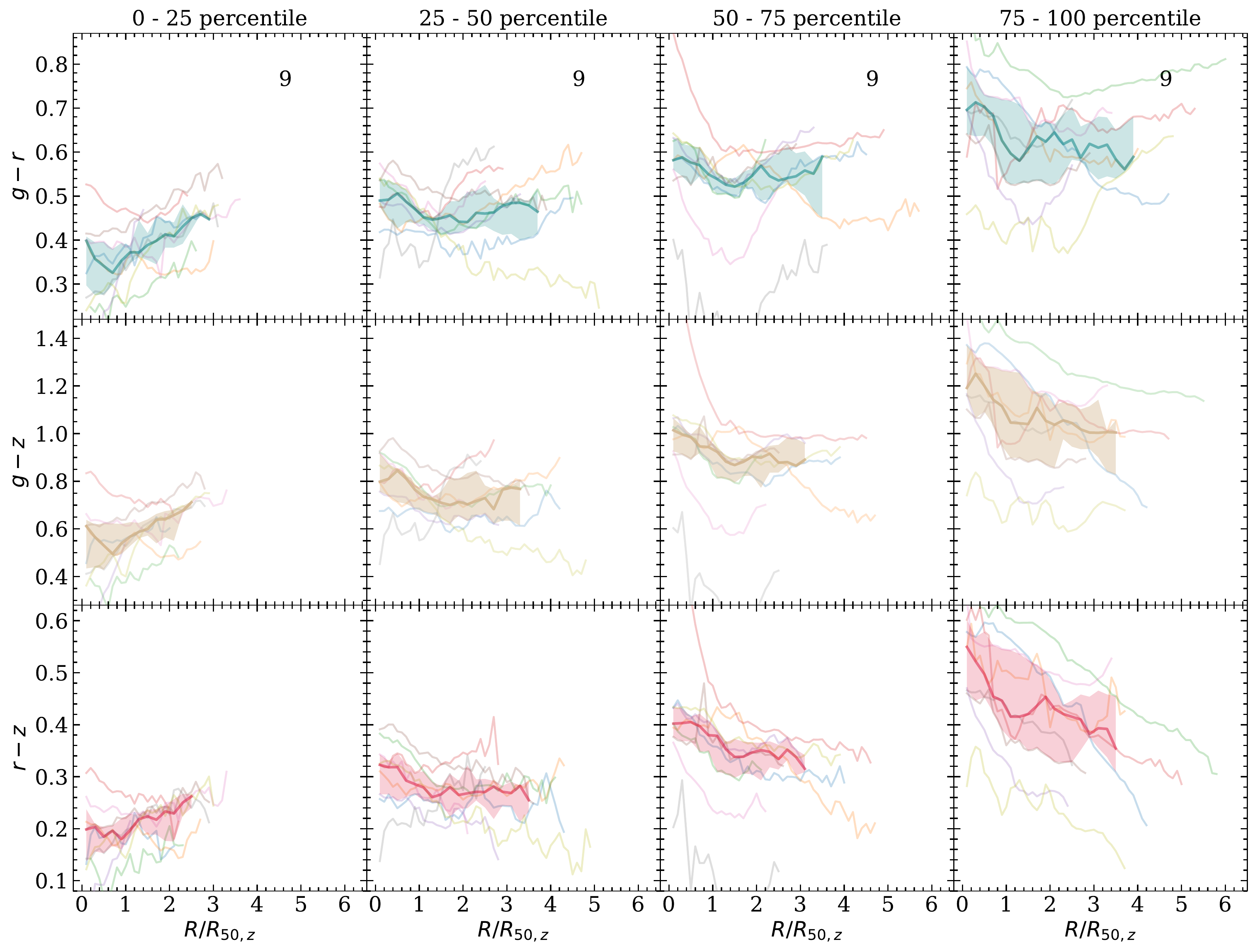}
    \caption{Color profiles in different stellar mass bins. 
    \textbf{From left to right}: $7.5 < \log M_*/M_\odot < 8.43$ , $8.43 < \log M_*/M_\odot < 8.95$, $8.95 < \log M_*/M_\odot < 9.5$ and $9.5 < \log M_*/M_\odot < 10.6$ (corresponding to 0-25$^{\rm th}$, 25-50$^{\rm th}$, 50-75$^{\rm th}$, 75-100$^{\rm th}$ percentile).
    \textbf{From top to bottom}: $g - r$ (green), $g - z$ (yellow) and $r - z$ (red).
    The color profiles for individual galaxies in each stellar mass bin are presented in random light colors. The median value and scatter (25$^{\rm th}$- and 75$^{\rm th}$-percentiles) of the individual profiles is shown in the deep color. The number of galaxies in each stellar mass bin is shown at the top.}
    \label{fig:CPM}
\end{figure}

We investigate the overall shape of color profiles for galaxies of different stellar mass ranges.
We divide the $\hi$ sample evenly into four subsamples by stellar mass. We show three types of color profiles ($g - r$, $g - z$ and $r - z$) in Figure \ref{fig:CPM}. The profiles are radially normalized by the $z$-band half-light radius ($R_{50, z}$). The median profile of each subsample is calculated in the radial range where at least half of the galaxies have color uncertainties lower than 0.1 mag.

For the least-massive galaxies ($7.5 < \log M_*/M_\odot < 8.43$, 0-25$^{\rm th}$ percentile), the colors become redder almost monotonically towards large radius.
For galaxies that have intermediate stellar masses ($8.43 < \log M_*/M_\odot < 8.95$ and $8.95 < \log M_*/M_\odot < 9.5$, 25-50$^{\rm th}$ and 50-75$^{\rm th}$ percentile), the median profiles are almost flat and sometimes show a ``U''-shape. 
For the most massive galaxies in our sample ($9.5 < \log M_*/M_\odot < 10.6$, 75-100$^{\rm th}$ percentile), the color profiles generally become bluer toward large radius.
We checked the profiles after excluding mergers, i.e. NGC 1359 and NGC 1385 (ID 69 and ID 70), but see no significant change compared to that presented above. 

\subsection{Color gradients} \label{sec:CG}
We study the relation between color gradients and $S_{sum}$. Color gradients are commonly used to indicate whether stellar disks grow or stop growing inside-out or outside-in \citep[e.g.,][]{2007ApJ...658.1006M, Wang2011, 2015ApJ...804L..42P, 2016ApJ...819...91P}. We focus on the $g - r$ color hereafter as the other two colors show similar patterns. Regarding the non-trivial shape of color profiles shown in Figure \ref{fig:CPM}, we do not derive one single global color gradient for each galaxy. 
We consider color gradients calculated in different radius ranges, $0 < R < R_{50, z}$ and $R_{50, z} < R < 2 R_{50, z}$ respectively to better probe the variations in response to tidal perturbation.
As before, the analysis is performed for the low-mass ($\log M_*/M_\odot < 8.95$) and high-mass ($\log M_*/M_\odot > 8.95$) galaxies separately.

\subsubsection{Dependence of color gradients and tidal strengths on galaxy properties} \label{sec:dep}
\begin{figure}
    \centering
    \includegraphics[width=\textwidth]{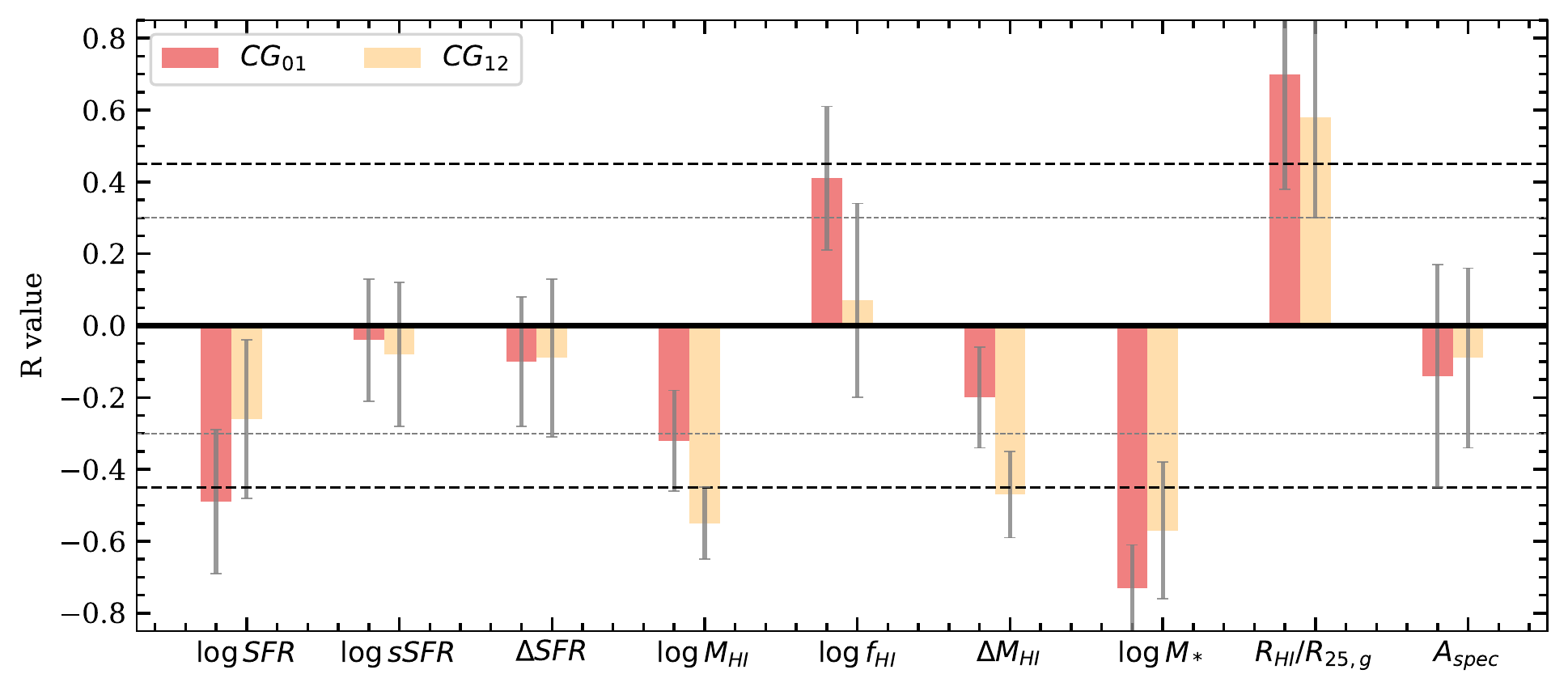}
    \includegraphics[width=\textwidth]{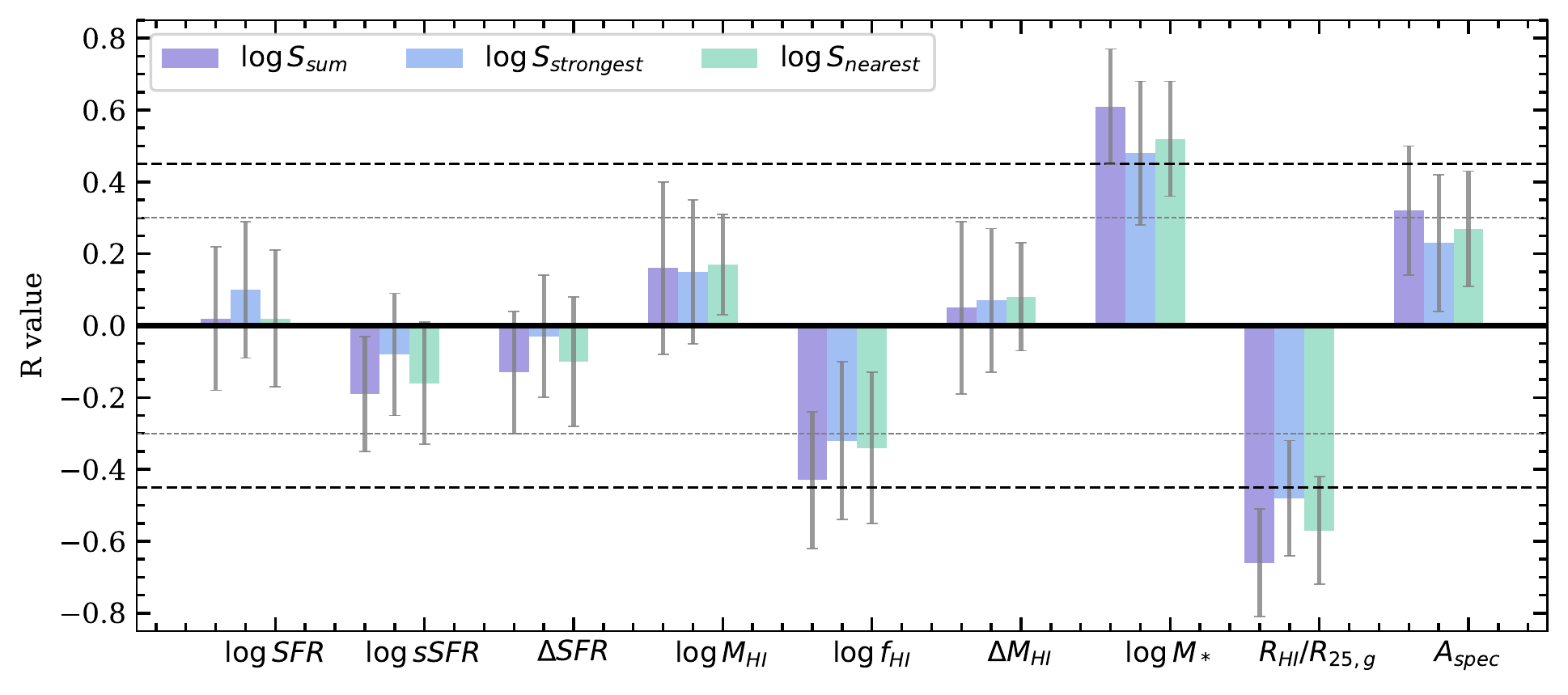}
    \caption{Pearson correlation coefficients for low-mass ($\log M_*/M_\odot < 8.95$) galaxies.
        \textbf{Upper:} between color gradients ($CG_{01}$ and $CG_{12}$) and galaxy properties.
        \textbf{Lower:} between tidal parameters ($S_{sum}$, $S_{strongest}$ and $S_{nearest}$) and galaxy properties.
        The dashed black (gray) horizontal line represents for $|R| = 0.45(0.3)$, which we regard as the criterion for a significant (considerable) (anti-)correlation. The error in R is derived by bootstrap re-sampling.}
    \label{fig:dependence}
\end{figure}

The major goal of this section is to identify the parameters that drive variations in the color gradients, and in the tidal strengths separately, so that we can control for these major driving parameters when investigating correlations between color gradients and tidal strengths later. 
The galaxy properties considered here include SFR related measurements such as SFR, sSFR and $\Delta SFR$, $\hi$ related measurements such as $M_{\rm HI}$, $f_{\rm HI}$ and $\Delta M_{\rm HI}$, the stellar mass $M_*$, the $\hi$-to-optical disk size ratio $R_{\rm HI} / R_{25, g}$ and the $\hi$ spectral line asymmetry $A_{spec}$. These parameters have been described in Section \ref{sec:analysis}. In Figure \ref{fig:dependence} and Table \ref{tab:dependence} we show the Pearson correlation coefficients (R) for the relation between color gradients and these parameters, and between tidal strengths and these parameters for low-mass galaxies. 

\begin{figure}[hptb]
    \centering
    \includegraphics[width=8.5cm]{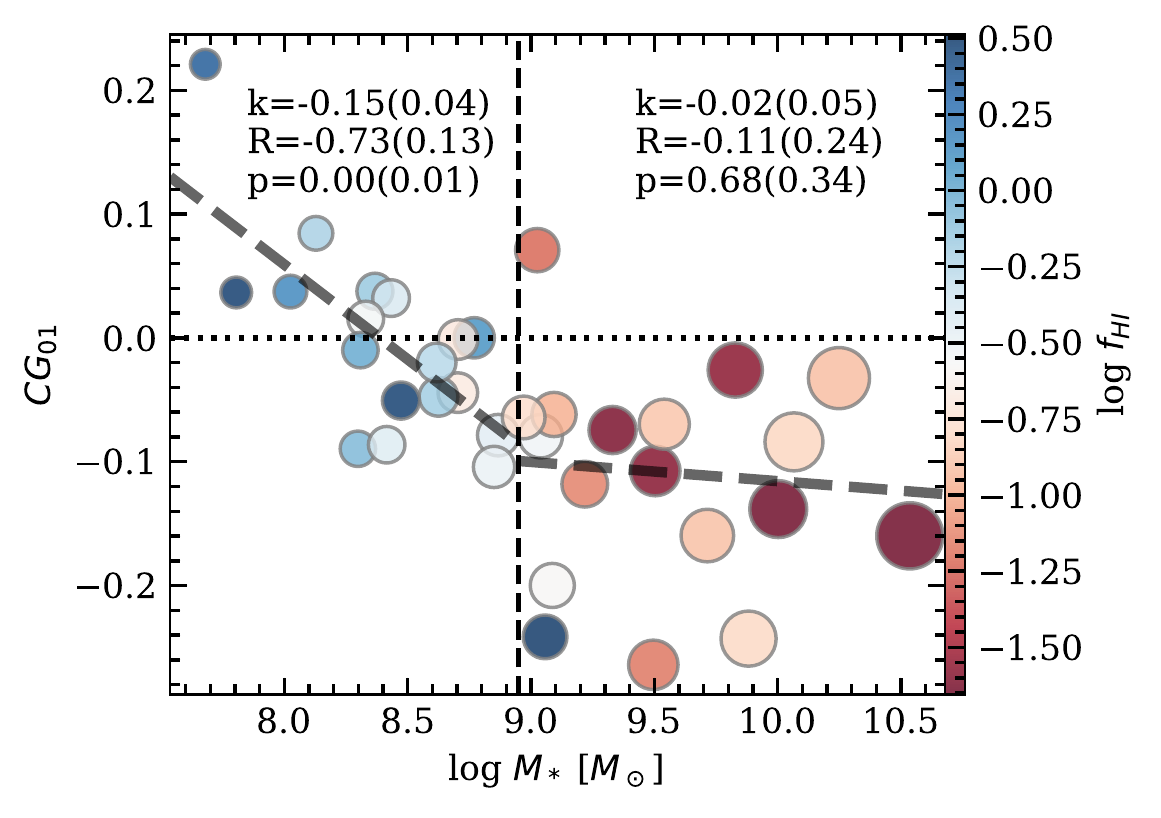}
    \includegraphics[width=8.5cm]{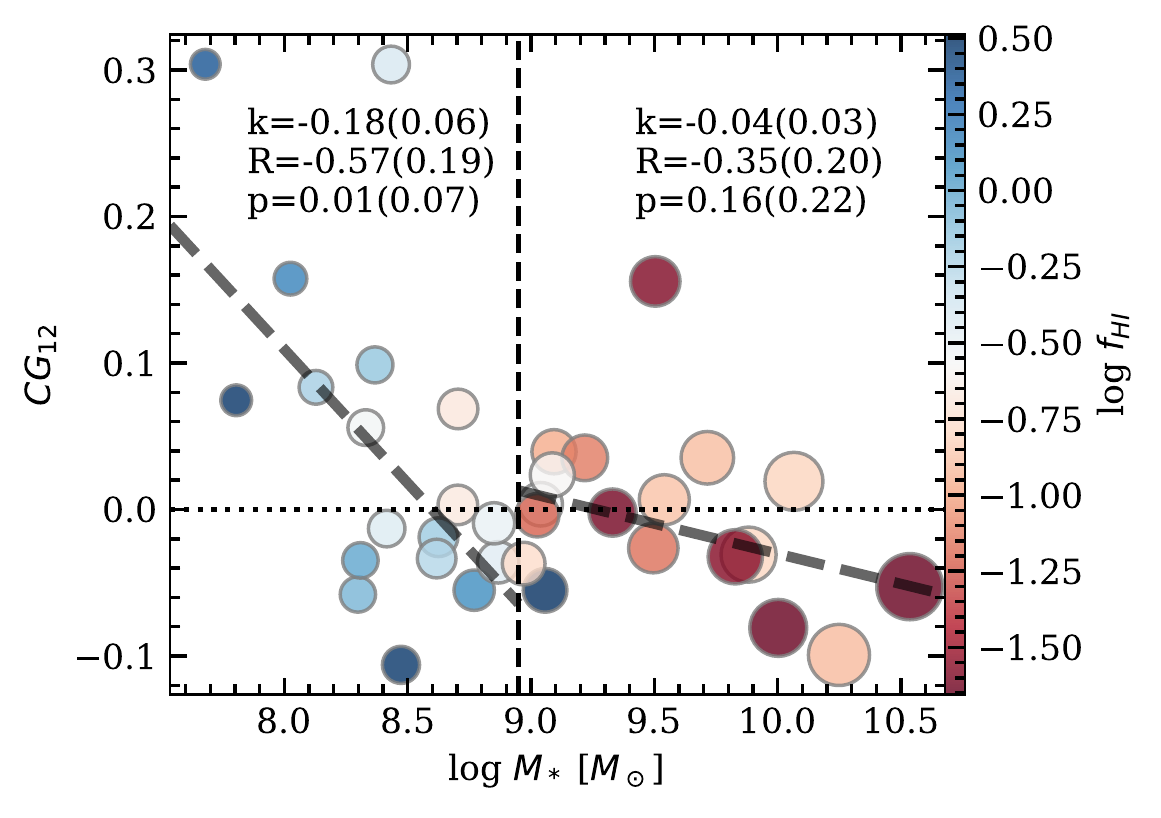}
    \includegraphics[width=8.5cm]{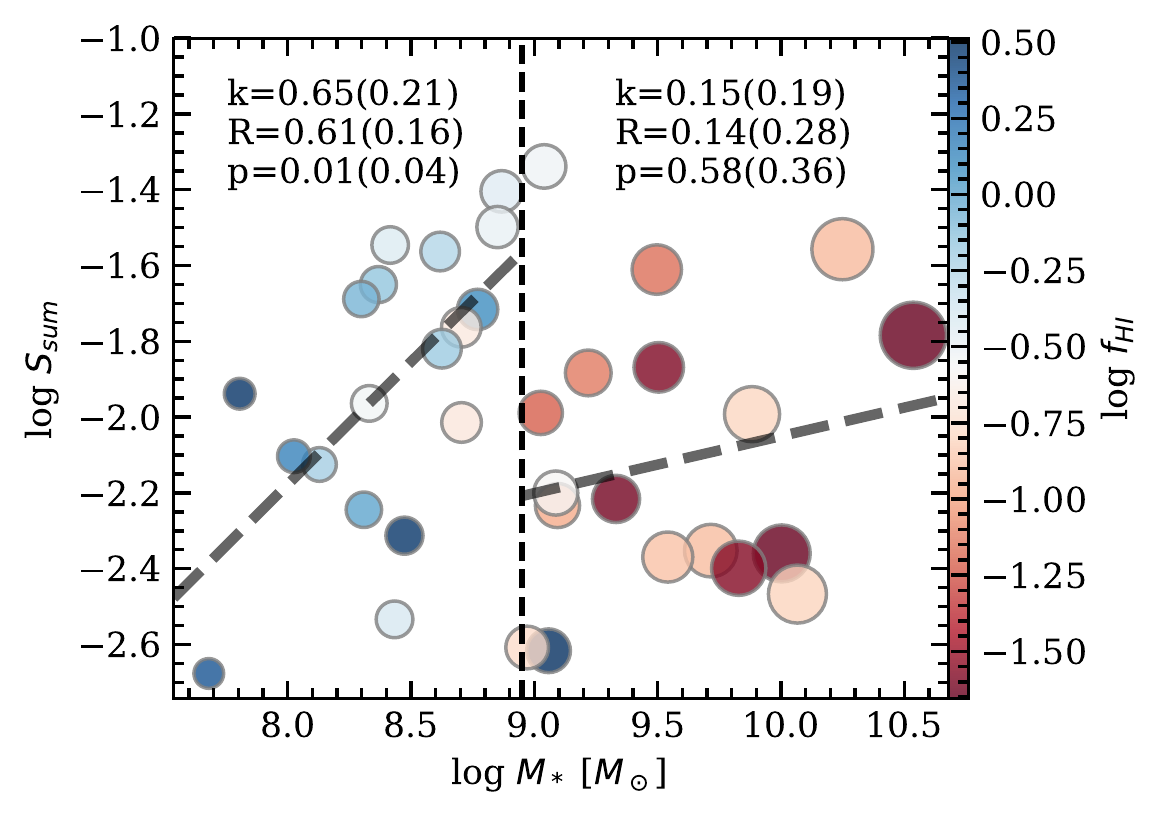}
    \caption{The dependence of color gradients and the summed tidal parameter on stellar mass. The stellar mass division ($M_* = 10^{8.95} M_\odot$) of low- and high-mass subsamples is presented by vertical black dashed lines.
    \textbf{Upper-left:} $CG_{01}$. 
    \textbf{Upper-right:} $CG_{12}$.
    \textbf{Lower:} $S_{sum}$.
    Pearson R and p values are shown with bootstrap error in parenthesis for both low- ($\log M_*/M_\odot < 8.95$, left) and high-mass ($\log M_*/M_\odot > 8.95$, right) galaxies separately. Gray dashed lines are the result of robust linear fit, with the slopes (and error) shown above the R and p value. All data points are color coded by $\hi$ gas fraction (see colorbar on the right). The size of data points indicates the stellar mass of the galaxy, in the sense that larger data points are used for more massive galaxies.}
    \label{fig:low-int}
\end{figure}

As can be seen from Table \ref{tab:dependence}, $CG_{01}$ shows significant ($|R| > 0.45$, shown in bold) anti-correlation with star formation rate and stellar mass. $CG_{12}$ significantly anti-correlates with $\hi$ mass, $\Delta M_{\rm HI}$ and stellar mass. 
All three types of tidal parameters show significant correlation with stellar mass and considerable anti-correlation with $\hi$ gas fraction. 
As an example to illustrate the (anti-)correlations, the dependence of $CG_{01}$, $CG_{12}$ and $S_{sum}$ on stellar mass is shown in Figure \ref{fig:low-int}.

\begin{table}[hptb]
  \centering
  \caption{Pearson correlation coefficients (R) and p values between color gradients ($CG$), tidal parameters ($S$) and galaxy properties, for low-mass ($\log M_*/M_\odot < 8.95$) galaxies.
  p values are shown in parenthesis. The format of the numbers indicates the significance of the correlation: bold for significant ones ($|R| > 0.45$) and italics for those that are considerable ($0.45 > |R| > 0.3$).}

\begin{tabular}{c|rrrrr}
      & \multicolumn{1}{c}{$CG_{01}$} & \multicolumn{1}{c}{$CG_{12}$} & \multicolumn{1}{c}{$\log S_{sum}$} & \multicolumn{1}{c}{$\log S_{strongest}$} & \multicolumn{1}{c}{$\log S_{nearest}$} \\
\hline
$\log SFR$ & \textbf{-0.49(0.00)} & -0.26(0.13) &  0.02(0.90) &  0.10(0.57) &  0.02(0.90) \\
$\log sSFR$ & -0.04(0.83) & -0.08(0.64) & -0.19(0.28) & -0.08(0.64) & -0.16(0.37) \\
$\Delta SFR$ & -0.10(0.54) & -0.09(0.60) & -0.13(0.44) & -0.03(0.86) & -0.10(0.54) \\
$\log M_{\rm HI}$ & \textit{-0.32(0.19)} & \textbf{-0.55(0.02)} &  0.16(0.53) &  0.15(0.56) &  0.17(0.50) \\
$\log f_{\rm HI}$ & \textit{ 0.41(0.10)} &  0.07(0.78) & \textit{-0.43(0.08)} & \textit{-0.32(0.20)} & \textit{-0.34(0.17)} \\
$\Delta M_{\rm HI}$ & -0.20(0.42) & \textbf{-0.47(0.05)} &  0.05(0.83) &  0.07(0.80) &  0.08(0.75) \\
$\log M_*$ & \textbf{-0.73(0.00)} & \textbf{-0.57(0.01)} & \textbf{ 0.61(0.01)} & \textbf{ 0.48(0.04)} & \textbf{ 0.52(0.03)} \\
$R_{\rm HI}/R_{25, g}$ & \textbf{ 0.70(0.00)} & \textbf{ 0.58(0.02)} & \textbf{-0.66(0.00)} & \textbf{-0.48(0.04)} & \textbf{-0.57(0.01)} \\
$A_{spec}$ & -0.14(0.58) & -0.09(0.71) & \textit{ 0.32(0.20)} &  0.23(0.36) &  0.27(0.28) \\
\end{tabular}%

  \label{tab:dependence}%
\end{table}%

\subsubsection{The relation between color gradients and tidal strengths}
\begin{figure}
    \centering
    \includegraphics[width=0.8\textwidth]{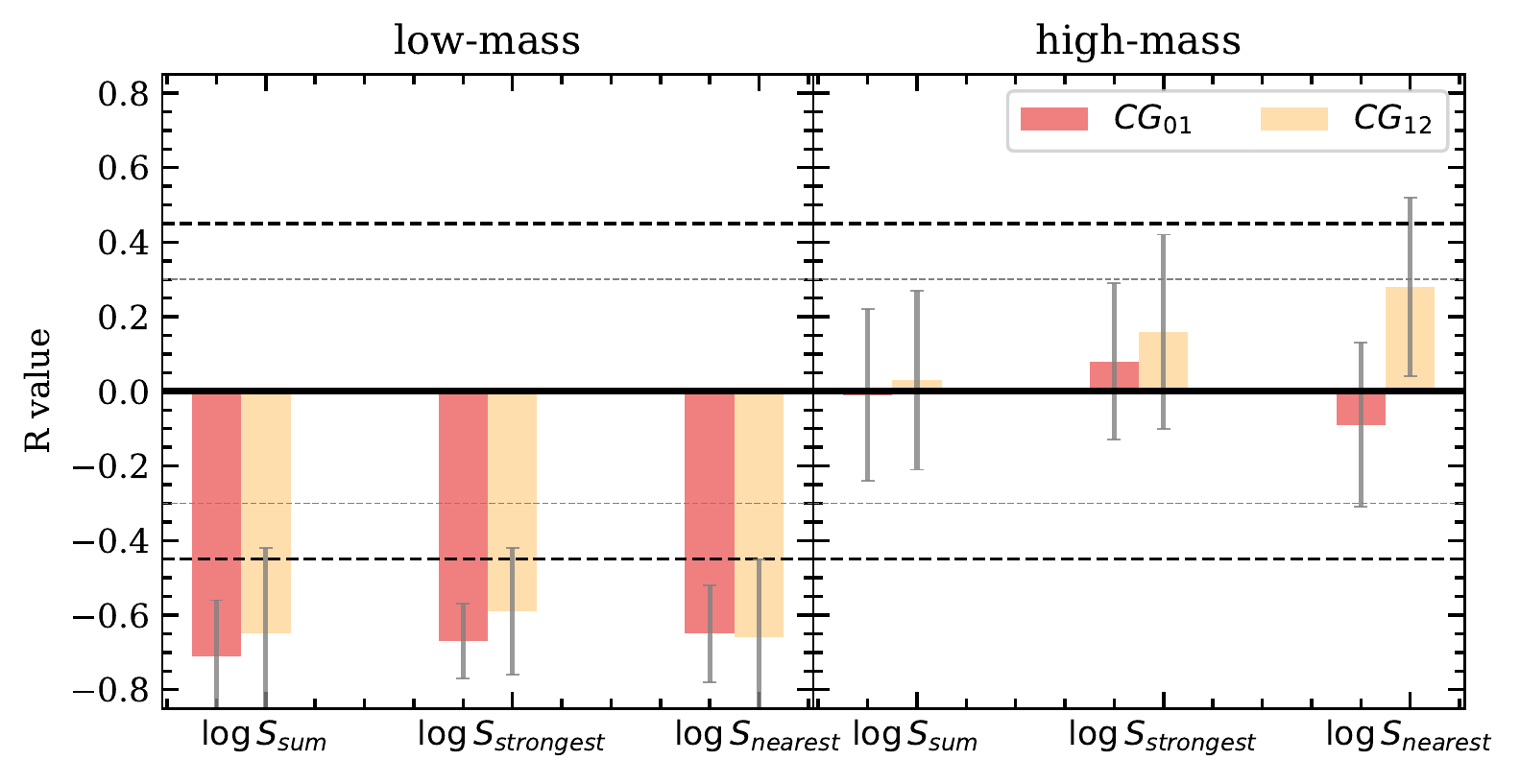}
    \caption{Pearson correlation coefficients between color gradients and tidal parameters. 
        \textbf{Left:} for low-mass ($\log M_*/M_\odot < 8.95$) galaxies.
        \textbf{Right:} for high-mass ($\log M_*/M_\odot > 8.95$) galaxies.
        The dashed black (gray) horizontal line represents $|R| = 0.45(0.3)$, which we regard as the criterion for a significant (considerable) (anti-)correlation. The error in R is derived by bootstrap re-sampling.}
    \label{fig:pearson}
\end{figure}

\begin{figure}
    \centering
    \includegraphics[width=0.8\textwidth]{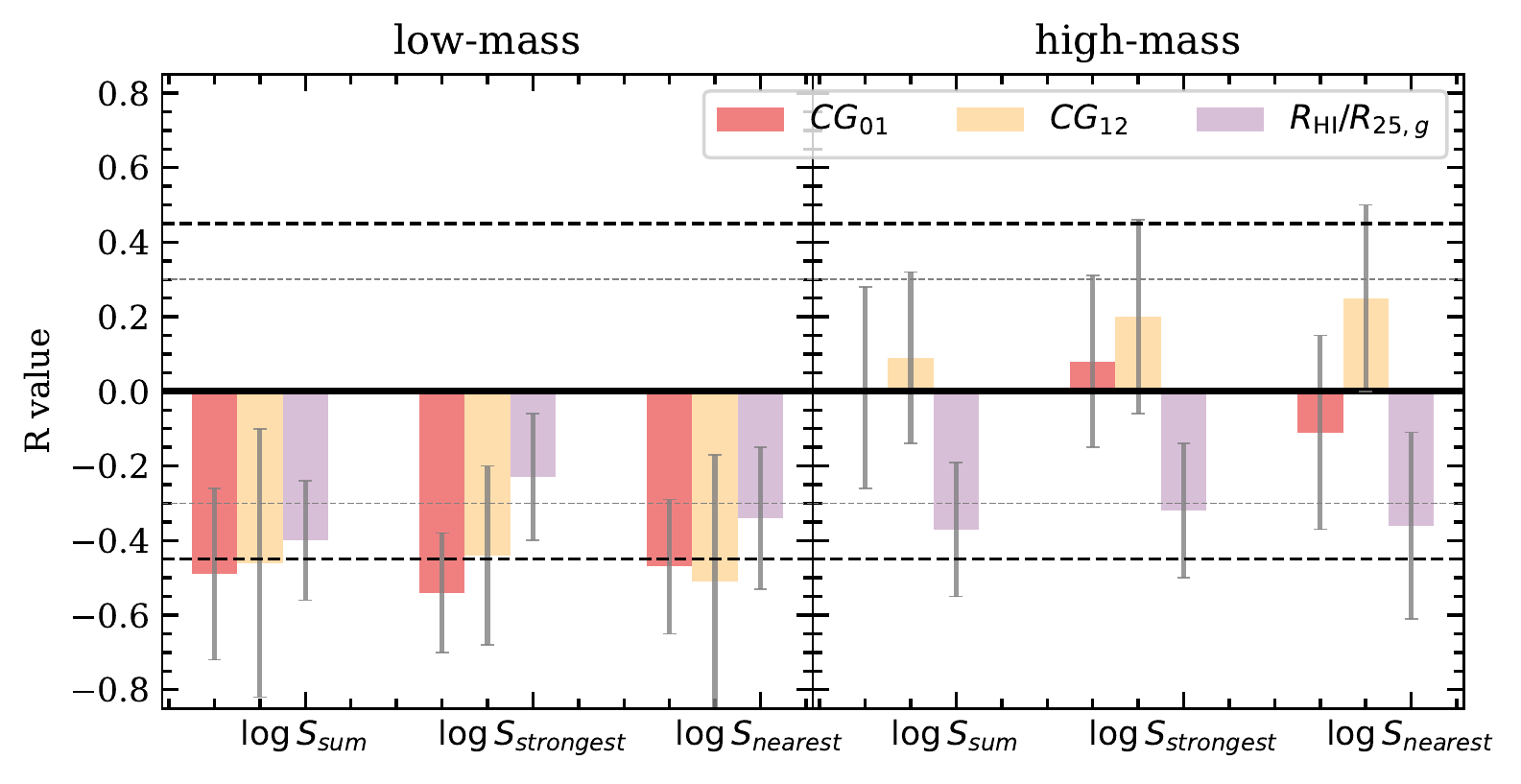}
    \includegraphics[width=0.8\textwidth]{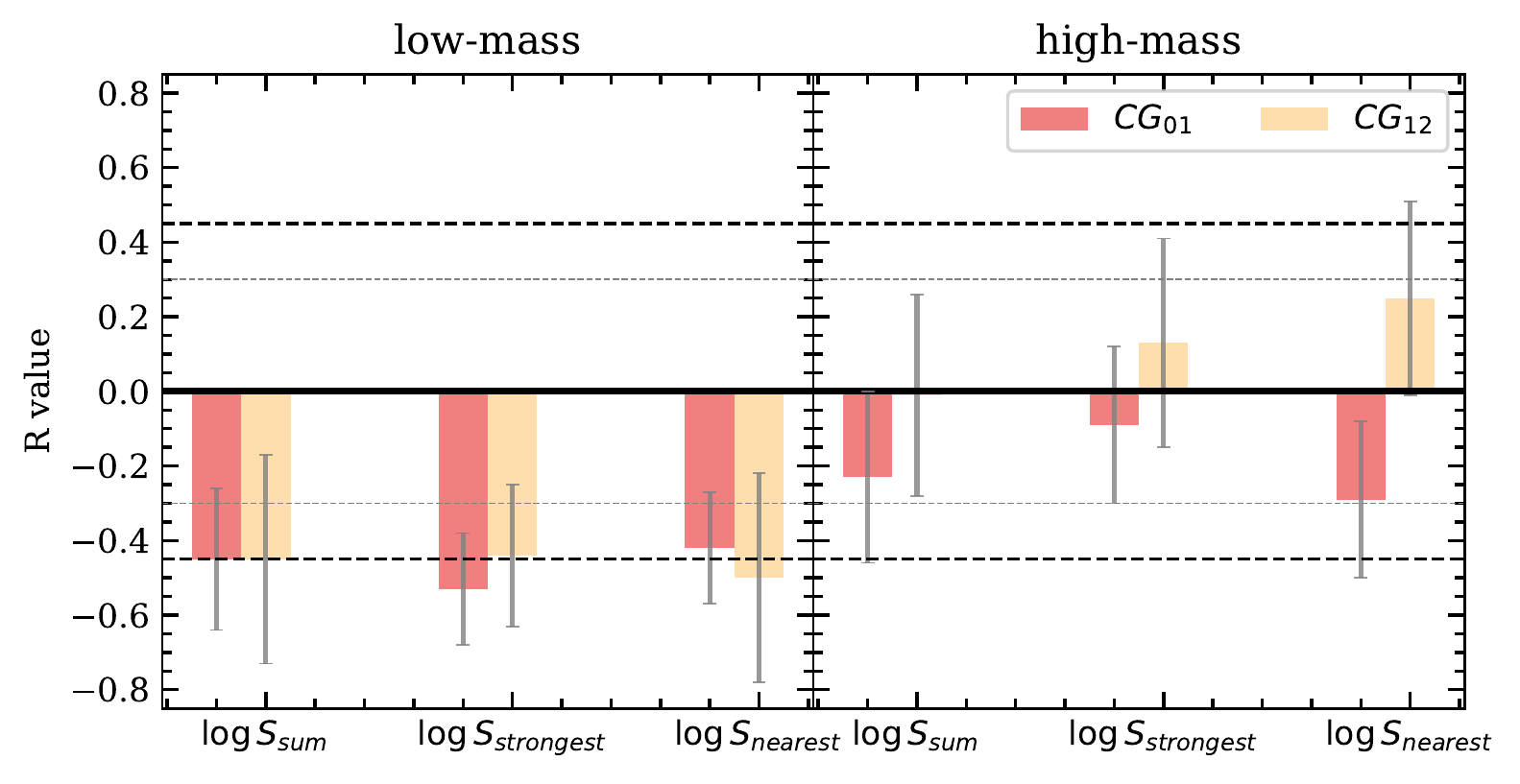}
    \caption{Partial correlation coefficients between color gradients and tidal parameters. 
        \textbf{Upper:} control for stellar mass ($\log M_*$).
        \textbf{Lower:} control for $\hi$-to-optical disk sizes ($R_{\rm HI} / R_{25, g}$).
        \textbf{Left:} for low-mass ($\log M_*/M_\odot < 8.95$) galaxies.
        \textbf{Right:} for high-mass ($\log M_*/M_\odot > 8.95$) galaxies.
        The black(gray) horizontal line represents for $|R| = 0.45(0.3)$, which we regard as the criterion for a significant (considerable) (anti-)correlation. The error in R is derived by bootstrap re-sampling.}
    \label{fig:partial}
\end{figure}

In Figure \ref{fig:pearson} and Table \ref{tab:pearson} we present the Pearson correlation coefficients (R) between color gradients and tidal parameters.
For low-mass galaxies, all three types of tidal parameters show significant anti-correlation with $CG_{01}$ and $CG_{12}$, with the most significant ones being those with $S_{sum}$. 
We need to account for the fact that the correlation between color gradient and tidal strength could be due to a third parameter.
We thus calculate the partial correlation coefficients between color gradients and tidal parameters with the potential third parameter controlled. The results are shown in Figure \ref{fig:partial} and Table \ref{tab:partial}.

\begin{table}[hptb]
  \centering
  \caption{Pearson correlation coefficients (R) and p values between color gradients ($CG$) and tidal parameters ($S$) for low-mass ($\log M_*/M_\odot < 8.95$, left panel) and high-mass ($\log M_*/M_\odot > 8.95$, right panel) galaxies. 
  Symbols are the same as Table \ref{tab:dependence}: p values are shown in parenthesis. The format of the numbers indicates the significance of the correlation: bold for significant ones ($|R| > 0.45$) and italics for those that are considerable ($0.45 > |R| > 0.3$).}

\begin{tabular}{l|rrr|rrr}
\multicolumn{1}{r}{} & \multicolumn{3}{c}{low-mass} & \multicolumn{3}{c}{high-mass} \\
      & \multicolumn{1}{l}{$\log S_{sum}$} & \multicolumn{1}{l}{$\log S_{strongest}$} & \multicolumn{1}{l|}{$\log S_{nearest}$} & \multicolumn{1}{l}{$\log S_{sum}$} & \multicolumn{1}{l}{$\log S_{strongest}$} & \multicolumn{1}{l}{$\log S_{nearest}$} \\
\hline
$CG_{01}$ & \textbf{-0.71(0.00)} & \textbf{-0.67(0.00)} & \textbf{-0.65(0.00)} & -0.01(0.98) &  0.08(0.76) & -0.09(0.71) \\
$CG_{12}$ & \textbf{-0.65(0.00)} & \textbf{-0.59(0.01)} & \textbf{-0.66(0.00)} &  0.03(0.90) &  0.16(0.52) &  0.28(0.27) \\
\end{tabular}%

  \label{tab:pearson}%
\end{table}%

\begin{figure}[hptb]
    \centering
    \includegraphics[width=0.8\textwidth]{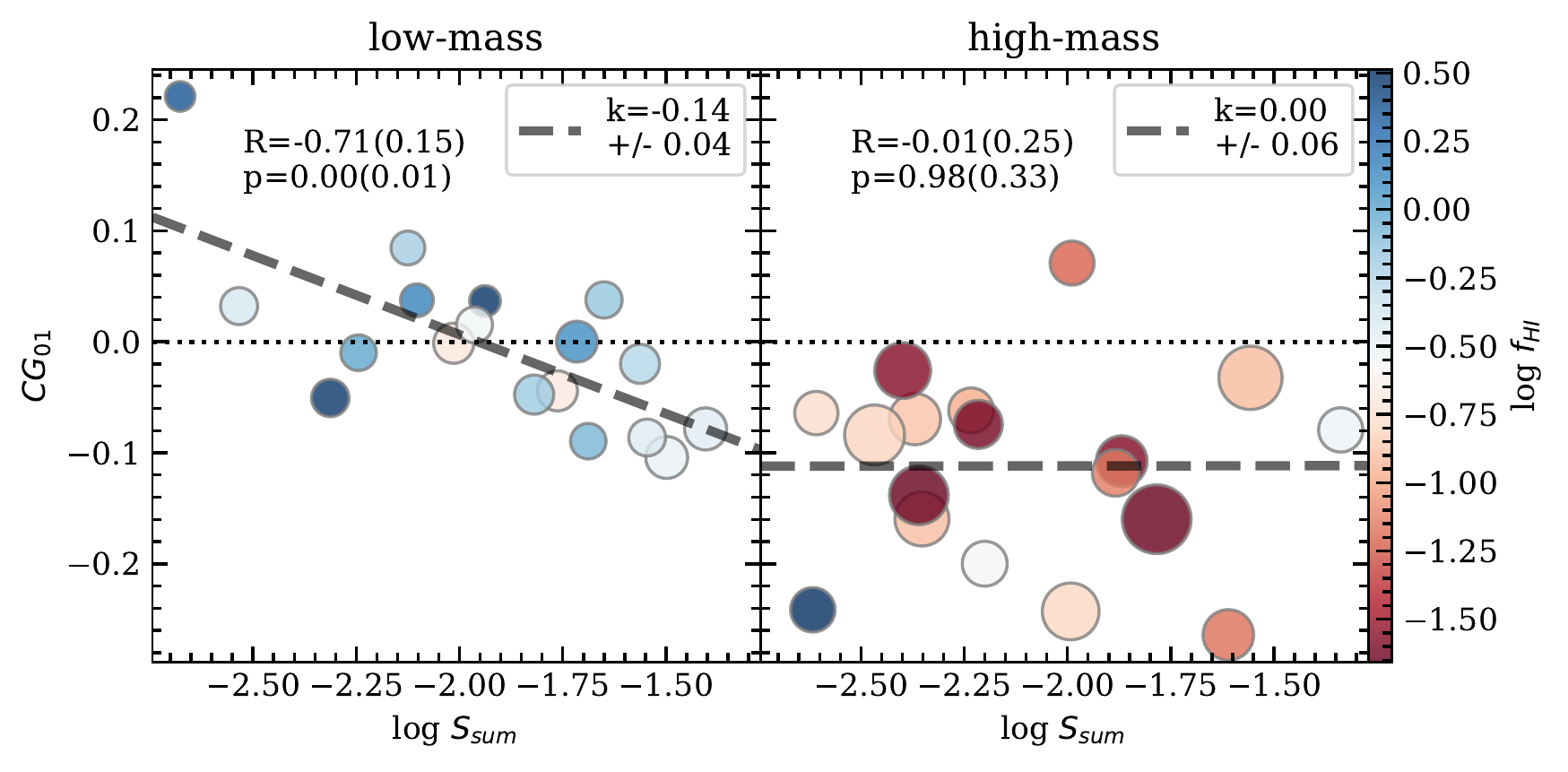}
    \includegraphics[width=0.8\textwidth]{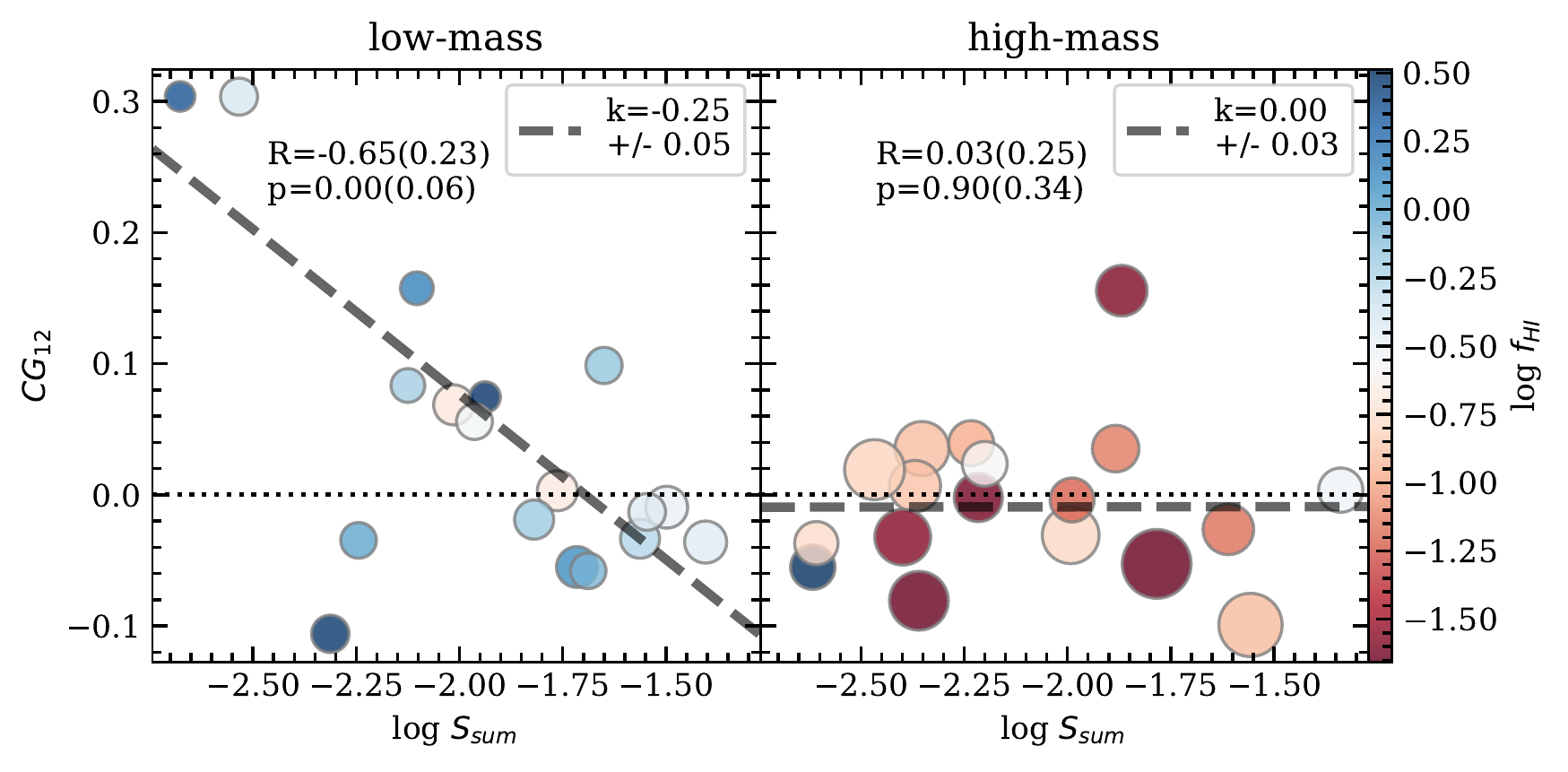}
    \caption{The correlation between color gradients and the summed tidal parameters. The dotted horizontal lines represent $CG_{01} = 0$ and $CG_{12} = 0$, respectively. 
    \textbf{Left:} for low-mass ($\log M_*/M_\odot < 8.95$) galaxies.
    \textbf{Right:} for high-mass ($\log M_*/M_\odot > 8.95$) galaxies.
    The other symbols and text are the same as Figure \ref{fig:asymmetry}: Pearson R and p values are shown with bootstrap error in parenthesis. Gray dashed lines are the result of robust linear fit, with the slopes (and error) shown in the corner. All data points are color coded by $\hi$ gas fraction (see colorbar on the right). The size of data points indicates the stellar mass of the galaxy, in the sense that larger data points are used for more massive galaxies.}
    \label{fig:CGS}
\end{figure}

We find that for the low-mass galaxies the anti-correlation between $CG_{01}$ and $CG_{12}$ and tidal parameters ($S_{sum}$) is still significant when stellar mass is controlled. Similar conclusions are reached when the $\hi$ mass or $\hi$ gas fraction is controlled. 
The similarly significant correlations with $S_{nearest}$ and $S_{strongest}$ suggest that tidal effects produced by the nearest and/or the strongest perturber may dominate the effects of tidal interaction on color gradients. Recall that the situation is different when considering the correlations of tidal strengths with the $\hi$ spectral asymmetry parameters.
We show the relation between $S_{sum}$ and $CG_{12}$, $CG_{01}$ in Figure \ref{fig:CGS}, with data points color coded by $\hi$ gas fraction and their sizes indicating stellar mass. 

\begin{table}[hptb]
  \centering
  \caption{Partial correlation coefficients (R) and p values between color gradients ($CG$), $\hi$-to-optical disk size ratios ($R_{\rm HI} / R_{25, g}$) and tidal parameters ($S$) for low-mass ($\log M_*/M_\odot < 8.95$, left panel) and high-mass ($\log M_*/M_\odot > 8.95$, right panel) galaxies, with stellar masses ($M_*$, upper part) or $\hi$-to-optical disk size ratios ($R_{\rm HI} / R_{25, g}$, lower part) controlled. 
  Symbols are as Table \ref{tab:dependence}: p values are shown in parenthesis. The format of the numbers indicates the significance of the correlation: bold for significant ones ($|R| > 0.45$) and italics for those that are considerable ($0.45 > |R| > 0.3$).}

\begin{tabular}{c|rrr|rrr}
\multicolumn{1}{r}{} & \multicolumn{3}{c}{low-mass} & \multicolumn{3}{c}{high-mass} \\
      & \multicolumn{1}{l}{$\log S_{sum}$} & \multicolumn{1}{l}{$\log S_{strongest}$} & \multicolumn{1}{l|}{$\log S_{nearest}$} & \multicolumn{1}{l}{$\log S_{sum}$} & \multicolumn{1}{l}{$\log S_{strongest}$} & \multicolumn{1}{l}{$\log S_{nearest}$} \\
\hline
$CG_{01}$ & \textbf{-0.49(0.05)} & \textbf{-0.54(0.03)} & \textbf{-0.47(0.06)} &  0.01(0.98) &  0.08(0.75) & -0.11(0.68) \\
$CG_{12}$ & \textbf{-0.46(0.06)} & \textit{-0.44(0.08)} & \textbf{-0.51(0.04)} &  0.09(0.74) &  0.20(0.44) &  0.25(0.33) \\
\multicolumn{1}{c|}{$R_{\rm HI}/R_{25, g}$} & \textit{-0.40(0.11)} & -0.23(0.37) & \textit{-0.34(0.18)} & \textit{-0.37(0.14)} & \textit{-0.32(0.21)} & \textit{-0.36(0.16)} \\
\multicolumn{1}{r}{} &       &       & \multicolumn{1}{r}{} &       &       &  \\
      & \multicolumn{1}{l}{$\log S_{sum}$} & \multicolumn{1}{l}{$\log S_{strongest}$} & \multicolumn{1}{l|}{$\log S_{nearest}$} & \multicolumn{1}{l}{$\log S_{sum}$} & \multicolumn{1}{l}{$\log S_{strongest}$} & \multicolumn{1}{l}{$\log S_{nearest}$} \\
\hline
$CG_{01}$ & \textbf{-0.45(0.07)} & \textbf{-0.53(0.03)} & \textit{-0.42(0.09)} & -0.23(0.37) & -0.09(0.72) & -0.29(0.25) \\
$CG_{12}$ & \textbf{-0.45(0.07)} & \textit{-0.44(0.08)} & \textbf{-0.50(0.04)} & -0.01(0.96) &  0.13(0.61) &  0.25(0.33) \\
\end{tabular}%

  \label{tab:partial}%
\end{table}%

\section{Discussion} \label{sec:Dis}
\subsection{Tidal effects on the distribution of HI}

The observation of the $\hi$ component in the Eridanus supergroup galaxies from WALLABY enables us to probe the most subtle yet direct effects produced by tidal interactions that we might expect. The anti-correlation between $S_{sum}$ and $R_{\rm HI} / R_{25, g}$ among low-mass galaxies (Figure \ref{fig:RHI}) suggests the possible existence of stripping effects \citep[e.g.,][]{1999MNRAS.304..465M} from tidal interactions. The outer $\hi$ disks are preferentially stripped as the gravity is weaker at a larger radius. The stripped $\hi$ may be accreted by more massive galaxies, or disperse in the hot intra-group medium (IGM). Theoretically tidal interactions may cause gas inflow \citep[e.g.,][]{1995ApJ...448...41H, 2001ApJ...547L.123M}, enhance central star formation and accelerate the consumption of $\hi$, which together may also lead to a shrinkage of the $\hi$ disk. However this scenario is unlikely for our low-mass galaxies as the inner color is not bluer but redder (Figure \ref{fig:C12R}). 

Previous studies have shown that image asymmetries in the $\hi$ can be caused by galaxy interactions \citep[e.g.,][]{2009MNRAS.400.1749K, 2010AJ....139..102E, 2011MNRAS.416.2401H}.
The insignificant correlation between $\hi$ spectral line asymmetry and tidal strength we find for Eridanus supergroup galaxies, however, may be due to the incapability of projected line asymmetry to reflect the 3-dimensional true asymmetry caused by perturbations. The weak relation between line asymmetry and environment has been noticed before \citep{2011A&A...532A.117E, 2014A&A...567A..56S, Reynolds2020}. 
Although there have been supportive results based on observed and simulated data on the link between enhanced $\hi$ line asymmetry, gas loss and environment density \citep{2020MNRAS.492.3672W, 2021arXiv210911214M}, a direct causality by galactic tidal interaction has not been clearly established.

Our findings are thus consistent with the idea that the tidal interaction, as an effective mechanism of external perturbation, can strip $\hi$ gas from galaxies and cause the $\hi$ disks to shrink. They are consistent with previous findings that ram pressure stripping is unlikely the primary driver for galactic $\hi$ deficiency in groups with similar mass as Eridanus \citep{2009MNRAS.400.1962K}.

\subsection{Color gradients as a tracer of star formation distributions} \label{subsec:dust}
The color of a stellar disk is dependent not only on stellar age, but also on stellar metallicity and dust extinction. Previous studies \citep{DeJong1996, Bell2000} found that the radial color gradients in disk and spiral galaxies are consistent predominantly with the effects of stellar age gradients. \citet{Gadotti2001} also claimed that dust is unlikely to play a fundamental role in global color gradients in late-type spiral galaxies. It is thus reasonable to assume that the color gradients represent stellar age gradients and the disk assembly history of our galaxies. This method was used by \citet{Wang2011} to study the connection between $\hi$ and disk assembly, and also used by many studies to investigate the connection between environmental effects and radial star formation enhancement and quenching \citep[e.g.,][]{2006MNRAS.372.1161W, Pan2018, 2019ApJ...881..119P, 2020MNRAS.494.4969P}. 

The dependence of color gradients on stellar mass has been extensively studied in the literature. It was found that more massive galaxies tend to show negative or flat color gradients while low-mass galaxies tend to show positive color gradients \citep[e.g.,][]{Tortora2010, Gonzalez-Perez2011, Cibinel2013}. Figure \ref{fig:CPM} confirms this transition with stellar mass. It was also found that relatively less massive galaxies tend to show positive color gradients when they are in the green valley and/or of early-type (\citealt{2015ApJ...804L..42P, 2016ApJ...819...91P, Belfiore2018}; see also \citealt{Cibinel2013}). However, our sample of galaxies (both low- and high-mass subsets) are biased against green-valley galaxies and early-type galaxies (see \citealt{2021MNRAS.tmp.2066F}), thus our results can not be directly compared to these trends.

In the following, we discuss the influence of tidal interactions on the color gradients for the low- and high-mass galaxies separately.

\subsubsection{Tidal effects on color distribution in low-mass galaxies}
Based on a sample of 34 Local Volume low-mass galaxies (all with $M_* < 10^9 M_\odot$ except for one galaxy with $M_* = 1.2 \times 10^9 M_\odot$), \citet{Zhang2012} found that, contrary to the high-mass galaxies, these low-mass galaxies typically show positive color gradients (i.e. blue cores). They considered in-situ star formation, secular redistribution, external influence (e.g., ram pressure stripping and tidal interaction), regulation of star formation through stellar feedback and gas pressure supported dynamics as potential causes, but, possibly due to the limited sample size and selection effects, they did not find conclusive observational evidence that showed why blue cores were dominant. Later models were proposed to explain the radial distribution of stellar age in these dwarf irregular galaxies as a consequence of fountain driven accretion \citep{2014ApJ...796..110E}, and stellar feedback driven, age-dependent stellar migration \citep[e.g.,][]{El-Badry2016}. More recently, the local SFR was found to follow the volumetric star formation law of more massive galaxies, where the volumetric density was derived assuming a hydrostatic quasi-equilibrium between the gravitational potential and the kinetic energy of the gas \citep{2020A&A...641A..70B}. As the color is correlated with sSFR, the radial distribution of the color may be a natural consequence of the balanced radial distribution of the SFR, the gas mass, and the stellar mass. Common features of these models are that the relevant low-mass galaxies are relatively unperturbed by the environment, and their star formation is fueled in a relatively quasi-equilibrium state.

In our study, the overall color gradients of the low-mass galaxies are much more positive than those of the high-mass galaxies, consistent with the findings of \citet{Zhang2012}. Moreover, we find a clear trend of more negative color gradients within $2 R_{50, z}$ under stronger external tidal interaction, highlighting the important role of external tidal effects in modifying the star formation distribution in the low-mass galaxies. 

\begin{figure}
    \centering
    \includegraphics[width=0.8\textwidth]{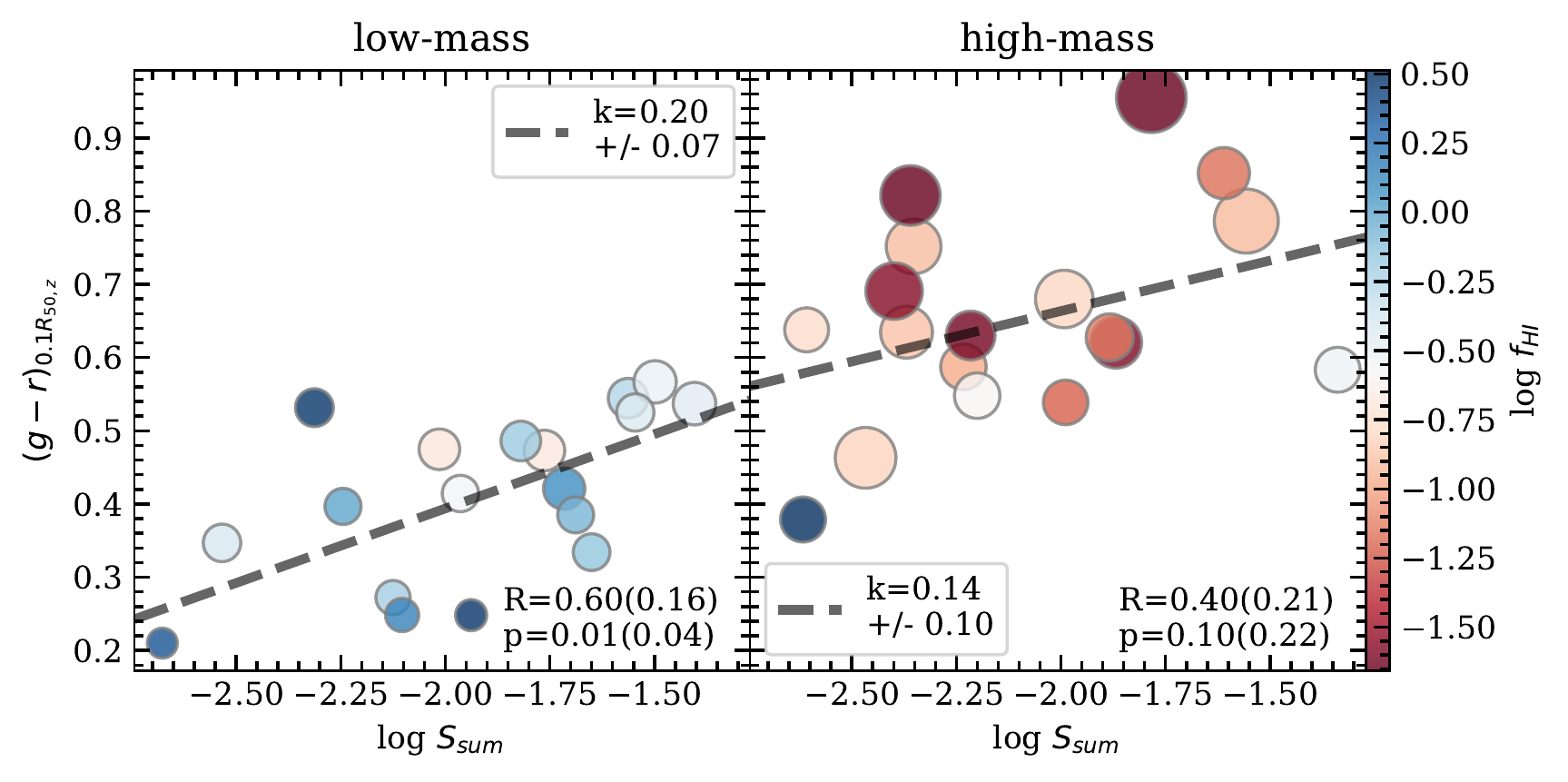}
    \includegraphics[width=0.8\textwidth]{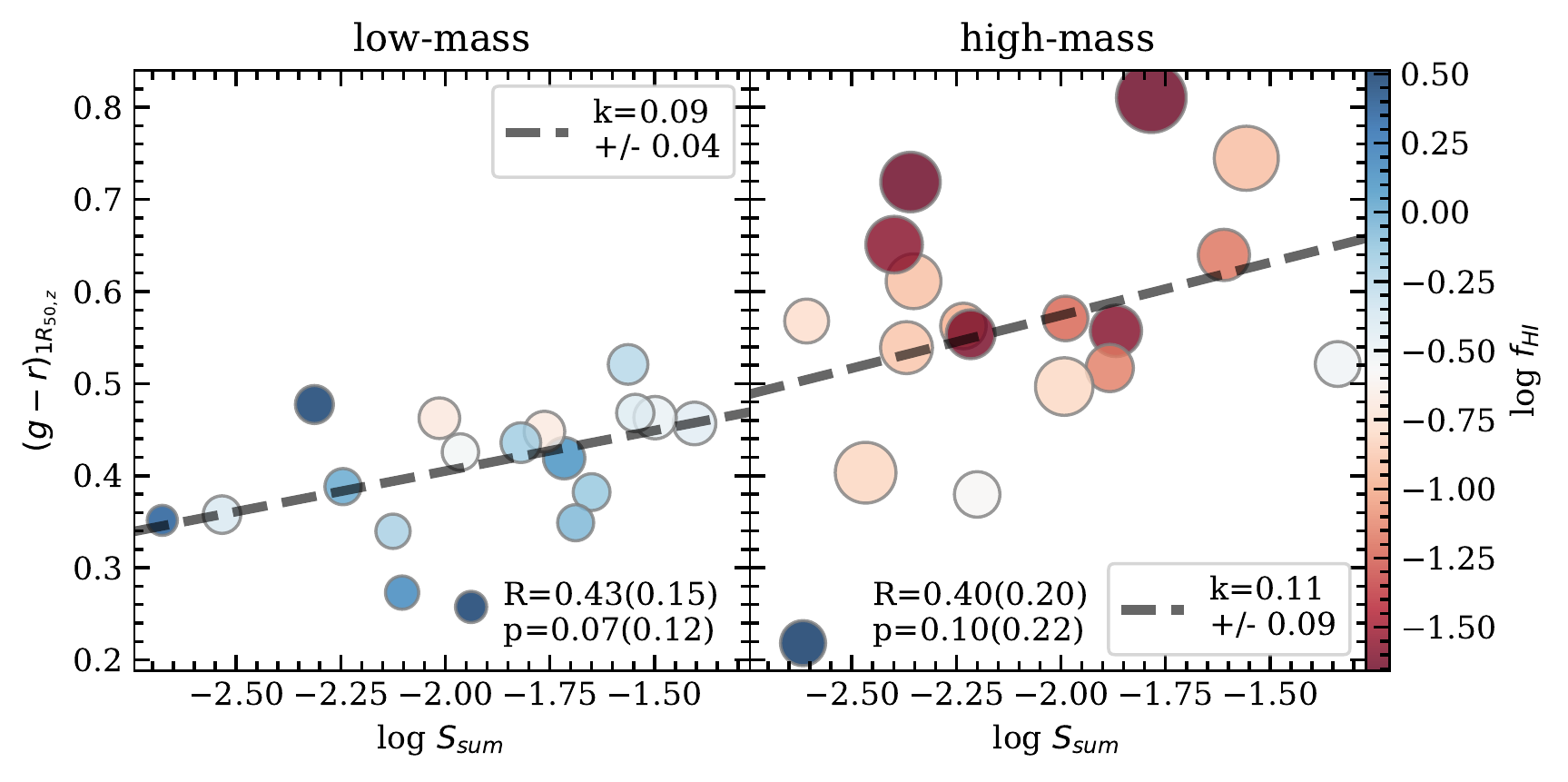}
    \includegraphics[width=0.8\textwidth]{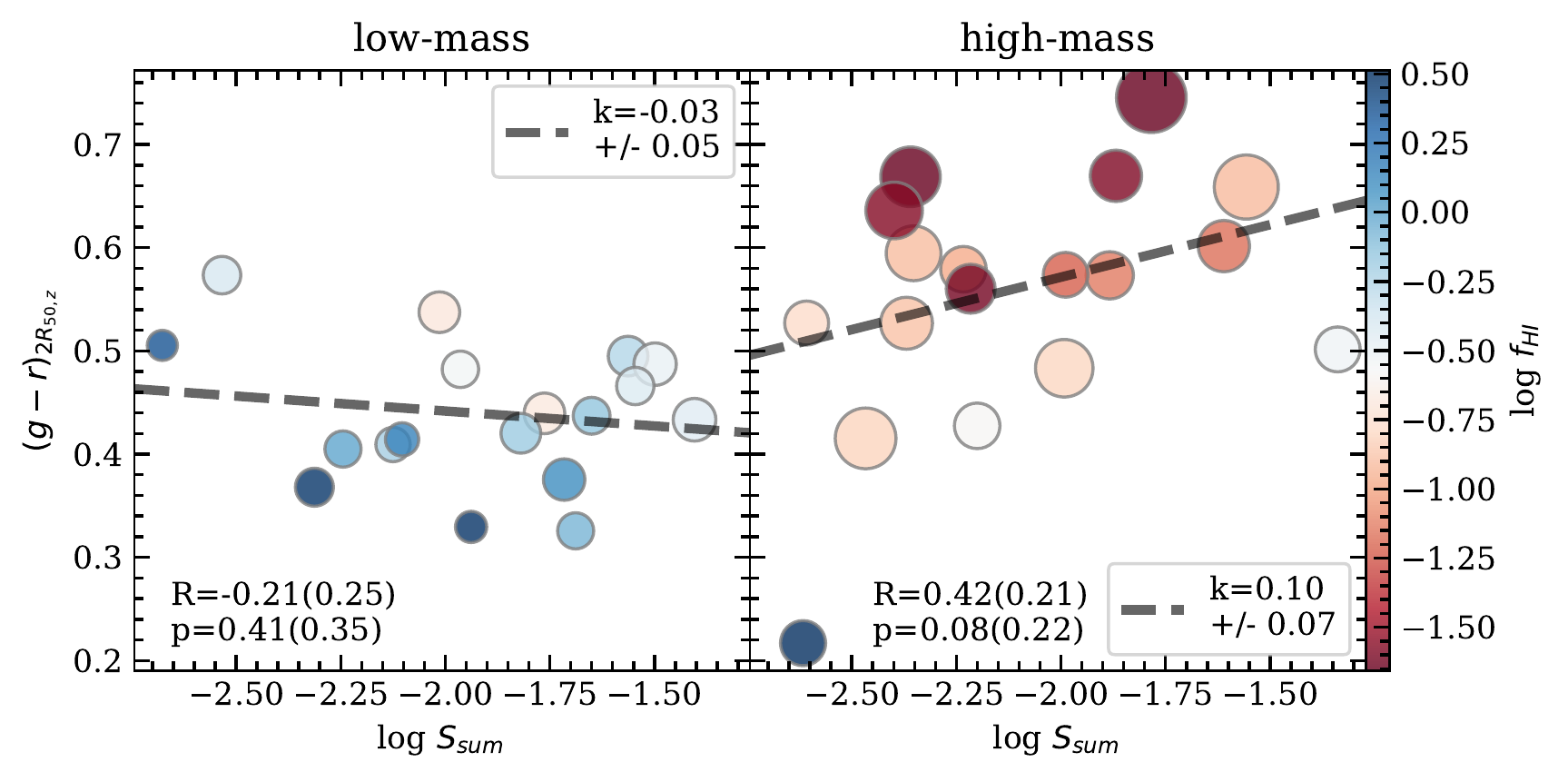}
    \caption{The correlation between $g - r$ color at different radius and the summed tidal parameters. 
    \textbf{Top:} at $R \sim 0.1R_{50, z}$.
    \textbf{Middle:} at $R \sim R_{50, z}$.
    \textbf{Bottom:} at $R \sim 2R_{50, z}$.
    The symbols and text are the same as Figure \ref{fig:asymmetry}: Pearson R and p values are shown with bootstrap error in parenthesis. Gray dashed lines are the result of robust linear fit, with the slopes (and error) shown in the corner. All data points are color coded by $\hi$ gas fraction (see colorbar on the right). The size of data points indicates the stellar mass of the galaxy, in the sense that larger data points are used for more massive galaxies.}
    \label{fig:C12R}
\end{figure}

There are two observational possibilities to explain why the low-mass galaxies have more negative $CG_{01}$ and $CG_{12}$ when $S_{sum}$ increases: either the inner color becomes redder, or the outer color becomes bluer. We distinguish these two possibilities by investigating the dependence of the $g - r$ colors close to $0.1 R_{50, z}$, $R_{50, z}$ and $2R_{50, z}$ on the $S_{sum}$ parameter (Figure \ref{fig:C12R}). We find that the $g - r$ color at the center, i.e. $(g - r)_{0.1 R_{50, z}}$, shows the most significant correlation with $S_{sum}$, the correlation for $(g - r)_{1 R_{50, z}}$ is weaker but still noticeable, while $(g - r)_{2 R_{50, z}}$ only shows a tentative anti-correlation with $S_{sum}$.
Therefore, the drop of the color gradients in these low-mass galaxies is mainly because the inner regions are redder and not because the outer regions are bluer. The star formation is not strongly enhanced in the outer disk, although simulations predict and observations partially support that this could happen as a result of enhanced local gas densities and instabilities \citep[e.g.,][]{2019ApJ...881..119P, Moreno2021}. Instead, the star formation, which could have been concentrated in the inner disks and show a positive color gradient if these low-mass galaxies were in an isolated environment, is likely suppressed by the tidal perturbations. 

\begin{figure}[hptb]
    \centering
    \includegraphics[width=0.5\textwidth]{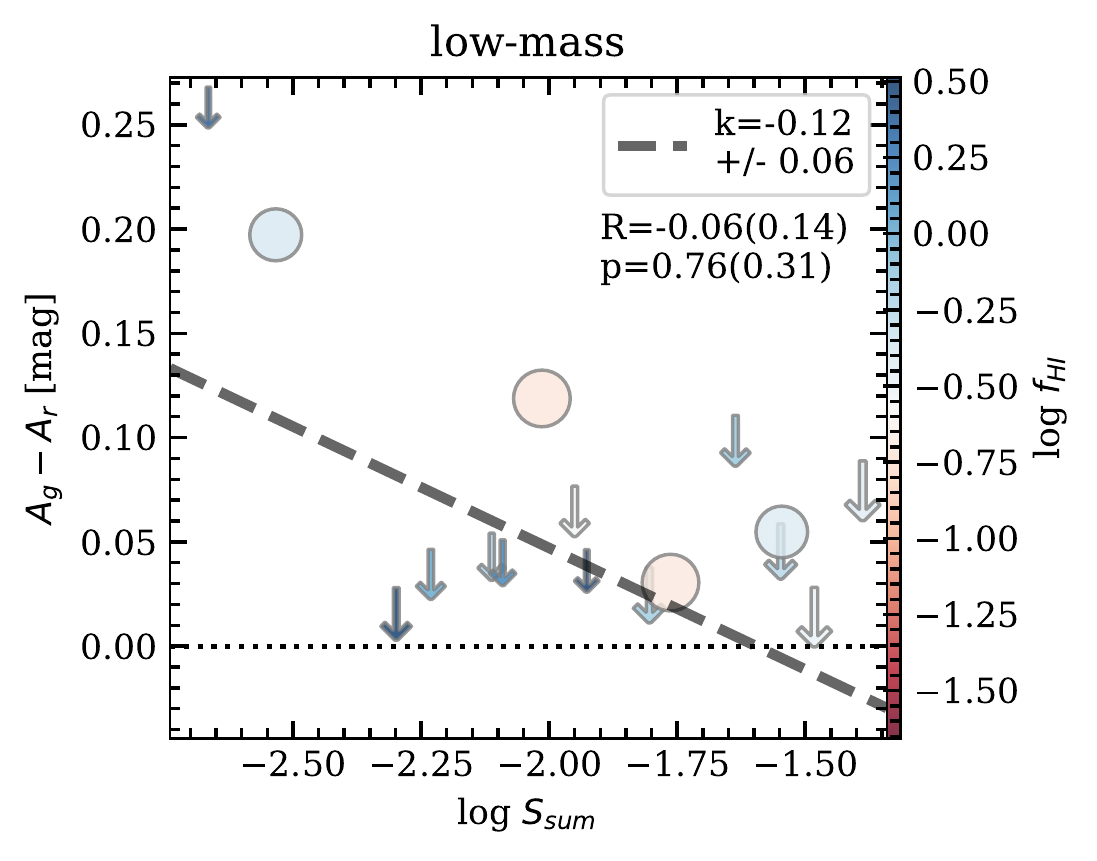}
    \caption{The correlation between color excess and the summed tidal parameters for low-mass galaxies. The dotted horizontal lines represent $A_g - A_r = 0$. The downward arrows are upper limits of the color excess.
    The other symbols and text are the same as Figure \ref{fig:asymmetry}: Pearson R and p value are shown with bootstrap error in parenthesis. Gray dashed line is the result of linear fit (upper limits taken into account), with the slope (and error) shown in the corner. All data points are color coded by $\hi$ gas fraction (see colorbar on the right). The size of data points indicates the stellar mass of the galaxy, in the sense that larger data points are used for more massive galaxies.
    }\label{fig:dust}
\end{figure}

One worry however may arise that the central colors can be redder because of a higher level of dust attenuation, caused by a large amount of centrally concentrated dust, transported there with gas inflows driven by tidal interactions \citep[e.g.,][]{1995ApJ...448...41H, 2001ApJ...547L.123M}.
We thus conduct the following test. We take $SFR_{W4}$ as the dust attenuated part of $SFR$ (see Section \ref{sec:data}), and use $SFR_{W4} / SFR$ to infer $A_{FUV}$ or $A_{NUV}$, depending on which of the two ultraviolet bands is used in estimating $SFR$. When W4 fluxes are not detected, we use the 5 percentile of the W4 flux distribution in our sample as the upper limit. We use the extinction curve of \citet{2007ApJS..173..293W} for the ultraviolet bands, where $A_{FUV} = 8.24 E(B-V)$ and $A_{NUV} = 8.2 E(B-V)$, and the extinction curve of \citet{Calzetti2000} for the optical bands where $A_g - A_r = 1.16 E(B-V)$. Hence $A_g - A_r = 0.14 A_{FUV} = 0.14 A_{NUV}$. Limited by the large PSF of GALEX and WISE images, we are unable to directly trace the attenuation near $0.1R_{50}$, which is typically below 5 arcsec in the low-mass subsample. Although $A_g - A_r$ is a measure for the global level of attenuation, it should be biased toward the condition in the galactic center if the centrally concentrated dust (and hence gas and starburst) dominates the reddening of the central color. In Figure \ref{fig:dust}, we show the relation between $A_g - A_r$ and $S_{sum}$ for the low-mass subsample. There is no significant correlation. More importantly, there should have been a trend for $A_g - A_r$ to increase with $S_{sum}$, if the reddening of the central $g-r$ color had been caused by centrally concentrated dust, 
but there is no evidence for such a trend. We thus conclude that the reddening of the central color with $S_{sum}$ should be more likely associated with old stellar population than with high dust attenuation. 

In addition to the reddening of inner disks, we find on average a decrease in relative size ($R_{\rm HI} / R_{25, g}$) with increasing $S_{sum}$. 
On average, $\hi$ disk size becomes smaller than that of the optical disk (Figure \ref{fig:RHI}) and $CG_{01}$ becomes negative (upper panel of Figure \ref{fig:CGS}) simultaneously at the characteristic $S_{sum}$ $\sim$0.01.
The suppression of the SFR is thus likely linked to the removal of the gas reservoir, probably as well as the decreased star forming efficiency, when the increase in velocity dispersion stabilizes the gas against gravitational collapse and/or radial inflow \citep[e.g.,][]{2008AJ....136.2846B, 2008AJ....136.2782L}. 
It is interesting to point out that the characteristic $S_{sum}$ of $\sim 0.01$ is much lower than the critical value of 0.07 for stellar disks to be strongly perturbed as predicted in previous stellar-only N-body simulations \citep{Oh2008}. It highlights the $\hi$ gas and star formation as more sensitive tracers to tidal perturbation than the morphology and mass of the stellar disks.

\subsubsection{Tidal effects on color distribution in high-mass galaxies}
High-mass galaxies typically show negative color gradients (i.e. red cores), which is consistent with a scenario where the galactic disks form inside-out, driven by the cosmic gas accretion \citep{1998MNRAS.295..319M}. Such a scenario is confirmed by the observed dependence of color gradients on $\hi$ gas fractions at a given stellar mass in high-mass galaxies \citep{Wang2011}. Our results confirm the dependence of color gradients on $\hi$ mass or $\hi$ excess with the high-mass subset, and find that the trend holds in the low-mass galaxies, supporting the important role of $\hi$ abundance in shaping the stellar disks. Large-scale cosmic gas accretion onto low-mass satellite galaxies was hinted before by the observational conformity phenomenon \citep{2013MNRAS.430.1447K, 2015MNRAS.454.1840K, 2015MNRAS.449.2010W}.

On top of the general behaviors of galaxies, tidal interactions have been found to be an important factor significantly affecting the color gradients of massive galaxies. High-mass galaxies with close companions were found to have on average $\sim 0.2$ mag bluer bulges and $\sim 0.1$ mag redder disks than the isolated control galaxies \citep{Ellison2010}, consistent with a scenario where tidal interactions induce gas inflows, either through bar instability \citep[e.g.,][]{1996ApJ...471..115B, 2001ApJ...547L.123M}, or through gravitational torques \citep[e.g.,][]{1995ApJ...448...41H}, that boost the gas density \citep[e.g.,][]{1996AJ....111..655H, 2010ApJ...710L.156R, 2014MNRAS.438.1870D, 2015MNRAS.450.2327Z, Chown2019} and thus significantly elevate the SFR in the center \citep[e.g.,][]{2003ApJ...582..668B, 2006AJ....131.2004K, 2008AJ....135.1877E}. 
\citet{2014ApJ...788L..39F} also found that, in a sample of highly isolated massive galaxies, most bulges are as red as E galaxies, but the subsample of bluer bulges is more likely to be located in galaxies with higher likelihood of (minor) tidal perturbations.
But the previous studies also found that the link between elevated central SFR and a close companion disappears when the galaxy pairs are in high-density environments ($\log \Sigma > 0.15$, where $\log \Sigma \equiv \frac{1}{2} \log (\frac{4}{\pi d_4^2}) + \frac{1}{2} \log (\frac{4}{\pi d_5^2})$ and $d_4$ and $d_5$ are the projected distances to the fourth and fifth nearest neighbours within 1000 km s$^{-1}$) or have large separations ($r_p > 30 h_{70}^{-1}$ kpc) at intermediate densities ($-0.55 < \log \Sigma < 0.15$), which was speculatively attributed to lower gas fractions in such environments \citep{Ellison2010, 2004MNRAS.352.1081A}. We adopt the same magnitude cut as that of \citet{Ellison2010} and derive an averaged $\log \Sigma$ of $-0.14$ and $D_{nearest}$ of $\gtrsim$500 kpc for the Eridanus supergroup region. Thus the small dependence of the color gradients on the $S$ parameters for our high-mass subset is consistent with previous findings. 

Since the $S$ parameter of the high-mass galaxies has a similar range as that of low-mass galaxies, the weak trend in their color gradients is unlikely due to their stronger gravity. We note that the $S$ parameters adopted in this paper principally probe the tidal strength at the edge of the optical disks, and $R_{\rm HI}$ are smaller than $R_{25}$ in most of the high-mass galaxies (see Figure \ref{fig:RHI}). It is thus likely that only a small fraction of these truncated $\hi$ disks suffer significantly from the effect of tidal interactions. Since the $\hi$ is an important intermediate step in fueling star formation \citep{Wang2020}, little influence on the $\hi$ mass may have led to low gas inflow rate, and barely any enhancement in the central SFR.

On the other hand, we do observe a correlation between the colors throughout the disks (at $0.1R_{50, z}$, $1R_{50, z}$, and $2R_{50, z}$) and tidal strengths despite the relatively large scatter (see right column of Figure \ref{fig:C12R}). It implies that the tidal stripping does contribute to accelerating the SFR quenching of these massive galaxies, but the averaged pattern is not inside-out as it would be in the low-mass galaxies of this study, or in general high-mass galaxies \citep{Ellison2018a}. Unlike in the low-mass galaxies, the general $\hi$-rich high-mass galaxies tend to have higher specific SFR and bluer colors in the outer disks than in the inner disks when they are unperturbed \citep{Wang2011}, which might have resulted in the outer disks colors being more sensitive to the stripping of $\hi$ than it would be in low-mass galaxies.

It is also interesting to point out that, for both low- and high-mass galaxies in the $\hi$ sample, the correlation between tidal strength and color gradient seems stronger than that between tidal strength and $\hi$ content (i.e. $f_{\rm HI}$ and $M_{\rm HI}$). \citet{2021MNRAS.tmp.2066F} also found that the global SFR of the $\hi$ sample is not strongly suppressed in galaxies of Eridanus supergroup. Such results suggest that the tidal perturbation as quantified in this paper is likely to more efficiently affect the radial distribution of star formation (and likely also $\hi$) instead of the total amount. Thus we see stronger correlations between tidal strengths and color gradients than between tidal strengths and the amount of $\hi$ content. It is also hinted that tidal interactions which are not able to strip gas from the galaxy may have significant effects on the distribution of gas and star formation.

\subsection{Other environmental effects} \label{sec:other}
We consider two additional environmental effects which may play a role in the galaxies of this paper.
Firstly, we consider the tidal force from the group halo, the mass of which is dominated by the dark matter halo \citep[e.g.,][]{1993ApJ...408...57V, 1996ApJ...459...82H, Fujita1998}. Unlike galaxy-galaxy tidal interactions, interaction with the group halo is more likely to drive gas inflows than to strip the gas. The strength of the effect can be quantified as $P_{gc} = (M_{cluster}/M_{gal})(R/r_{gal})^{-3}$ \citep{Byrd1990}. A threshold value above which significant disk instabilities and thus gas inflows can be triggered, is predicted to range from 0.006 to 0.1 depending on the ratio of dark matter halo mass over the disk mass \citep{Byrd1990}. We estimate that the sum of tidal forces from the groups (Eridanus, the NGC1407 group and the NGC1332 group) enforced on each galaxy ranges from $3 \times 10^{-6}$ to $0.1$, with five (four) high-mass (low-mass) galaxies having tidal force from any of the groups larger than $0.006$. Thus, tidal interaction with the group halo is unlikely to be the dominating effect among the whole sample, but may affect a few galaxies close to group centers. However, we do not observe peculiar color distributions in those galaxies with $P_{gc} > 0.006$, possibly because the interaction with the group halo is not strong enough to counteract the stripping effect from other galaxies, or because projection effects have led to large uncertainties in the estimate of $P_{gc}$ due to inaccurate estimate of the group-centric distances. 

A second effect we consider is ram pressure stripping due to the hot IGM. We follow the method of \citet{2020ApJ...903..103W} and \citet{Wang2021} to compare the ram pressure level with disc restoring forces on the $\hi$ gas. We estimate the IGM density at the location of each galaxy by interpolating (extrapolating for galaxies in the NGC1332 group and the Eridanus group) the gas density profile of the NGC1407 group, because extended X-ray emission is only detected in the core region of the NGC1407 group. Most of the galaxies in the $\hi$ sample have projected distances to NGC1407 in the range of $1.25 R_{virial}$ to $3.5 R_{virial}$. It is thus natural to expect a weak level of ram pressure stripping effects on galaxies in the Eridanus supergroup. And indeed, there are only three, i.e. NGC 1390 (ID 21), ESO 548-65 (ID 23) and NGC 1359 (ID 69), out of 36 galaxies are identified to be candidates for ram pressure stripping, which experience higher ram pressure than the gravitational restoring forces at $R_{\rm HI}$.
Among these three candidates, NGC 1359 is a merging pair, thus the ram pressure stripping could be assisted by tidal effects when the $\hi$ is tidally shifted to regions of low restoring forces. 
Thus, ram pressure stripping should not play an important role in setting the statistical behavior of this sample. But we point out that the density of the IGM has been assumed to be distributed smoothly following the standard beta-model as in \citet{2020ApJ...903..103W} and \citet{Wang2021}, while cold fronts due to shocks are found to be prevalent in merging clusters \citep{2007PhR...443....1M}. A more detailed analysis awaits modeling based on deeper X-ray images, possibly when eROSITA \citep{2012arXiv1209.3114M} results become available in the near future.

\subsection{Caveats and future perspective}
We emphasize that the systematic uncertainty due to projection effects (both in distance and velocity) and the crude estimation of the duration of the tidal encounter inevitably limit the use of the tidal parameter. Thus it should be considered in a relative sense and is only valid for a statistically meaningful sample. As mentioned in Section \ref{sec:S} and Section \ref{sec:HId}, the tidal strength parameter has uncertainties contributed by the orbital history of the galaxies, and is particularly not suitable for describing gravitational effects at the coalescence stage of mergers. This type of systems are few in our $\hi$ sample, but an alternative way of quantifying the physical effects in the coalescence stage should be considered in the future.

As the Eridanus supergroup is in a distinct stage of cluster or supergroup assembly, it is likely that tidal interactions are enhanced. It will be meaningful to apply the technique used in this paper to more general groups in the future, when the WALLABY survey has covered a much larger sky region. Finally, comparing the newly observed results with hydrodynamic simulations using a consistent parametrization will be useful to understand the systematic uncertainties, and derive further insight into complex physical processes which is difficult to address with observations alone, in particular in helping to disentangle the combined effects of tidal interaction and ram pressure stripping, and the cosmological background of group assembly which sets the initial conditions of galaxies upon infall.

It is worth mentioning that the truncated $\hi$ disks of high-mass galaxies indicate pre-processing by environment outside the current field-of-view of the Eridanus supergroup. Alternatively, evolution of high-mass galaxies can be more strongly driven by internal structures or masses than by the environments \citep[e.g.,][]{Peng2010}. To clarify the evolutionary track of high-mass galaxies, we will need to trace them back to larger group-centric radius, or consider smaller galaxy groups. In future work we will address this question for other clusters and groups, partly by exploiting further WALLABY data.

\section{Summary and conclusion} \label{sec:con}
To summarize, we conducted photometric measurements in $g$, $r$ and $z$ bands for 36 WALLABY $\hi$-detected galaxies in the Eridanus supergroup, and derived color gradients and tidal parameters based on an optical sample of Eridanus supergroup member galaxies. 
We confirm that the tidal parameters are capable of reflecting the tidal disturbance experienced by, at least, the $\hi$ disks, in the sense that the shrinkage of $\hi$ disks are connected to large tidal parameters. We do not find clear evidence for the asymmetry of the integral $\hi$ spectra to increase with the tidal parameter, possibly because the spatial information is lost in the integral $\hi$ spectra. It is worth noting that the tidal perturbation in the Eridanus supergroup is contributed by a significant number of neighboring galaxies, which emphasizes the importance of adopting a complete sample to study tidal interaction in galaxy groups.

We show that the color profiles of galaxies in the Eridanus supergroup are strongly dependent on stellar mass. The color profiles show transition from a generally positive gradient (redder towards large radii, an outside-in scheme) at the low-mass end ($\log M_*/M_\odot \lesssim 8.5$), to a tentative ``U''-shape at intermediate mass ($8.5 \lesssim \log M_*/M_\odot \lesssim 9.5$), and finally to a negative gradient (bluer towards large radii, an inside-out scheme) at the high-mass end ($\log M_*/M_\odot \gtrsim 9.5$), which is consistent with previous findings.

We find that tidal interactions play an important role in determining the color gradients within $2 R_{50, z}$ of low-mass galaxies in groups. More negative color gradients (i.e. redder cores and/or bluer outer regions) are clearly related to stronger tidal strengths. Further investigation reveals that the anti-correlation between color gradients and tidal strengths largely originates from the reddening of the central regions rather than bluer outer regions.
High-mass galaxies, on the other hand, do not show a clear correlation between color gradients and tidal strengths. The colors at all three radii ($0.1R_{50, z}$, $1R_{50, z}$, and $2R_{50, z}$) do correlate with tidal strengths, although with large scatter. These results suggest that the quenching in high-mass galaxies can be assisted by tidal perturbation, but not in a normally expected inside-out way.

Combining the results above, we conclude that tidal interaction in the Eridanus supergroup serves as a major mechanism for star formation quenching in low-mass galaxies. It works by stripping the $\hi$ gas from the extended $\hi$ disk; as the reservoir of $\hi$ shrinks, the inner disk where star formation is concentrated is less fuelled than it would be in an unperturbed state. 
On the other hand, the smaller $\hi$ reservoir leaves less space for the same mechanism to work in high-mass galaxies of the Eridanus supergroup. Their SFR throughout the disk are suppressed, possibly due to the halt of their normal inside-out formation as a result of shrinking $\hi$ reservoir at the outskirts of the disk.

\begin{acknowledgements}

We thank the referee for the very constructive, helpful comments.

JW acknowledges support from the National Science Foundation of China (12073002, 11721303), and the science research grants from the China Manned Space Project with NO. CMS-CSST-2021-B02. Parts of this research were supported by High-performance Computing Platform of Peking University.

FB acknowledges funding from the European Research Council (ERC) under the European Union’s Horizon 2020 research and innovation programme (grant agreement No.726384/Empire)).

AB acknowledges support from the Centre National d'Etudes Spatiales (CNES), France.

Parts of this research were supported by the Australian Research Council Centre of Excellence for All Sky Astrophysics in 3 Dimensions (ASTRO 3D), through project number CE170100013. 

This project has received funding from the European Research Council (ERC) under the European Union’s Horizon 2020 research and innovation programme (grant agreement no. 679627; project name FORNAX) 

The Australian SKA Pathfinder is part of the Australia Telescope National Facility which is managed by CSIRO. Operation of ASKAP is funded by the Australian Government with support from the National Collaborative Research Infrastructure Strategy. ASKAP uses the resources of the Pawsey Supercomputing Centre. Establishment of ASKAP, the Murchison Radio-astronomy Observatory and the Pawsey Supercomputing Centre are initiatives of the Australian Government, with support from the Government of Western Australia and the Science and Industry Endowment Fund. We acknowledge the Wajarri Yamatji people as the traditional owners of the Observatory site.

This research has made use of the NASA/IPAC Extragalactic Database, which is funded by the National Aeronautics and Space Administration and operated by the California Institute of Technology.

The Legacy Surveys consist of three individual and complementary projects: the Dark Energy Camera Legacy Survey (DECaLS; Proposal ID \#2014B-0404; PIs: David Schlegel and Arjun Dey), the Beijing-Arizona Sky Survey (BASS; NOAO Prop. ID \#2015A-0801; PIs: Zhou Xu and Xiaohui Fan), and the Mayall z-band Legacy Survey (MzLS; Prop. ID \#2016A-0453; PI: Arjun Dey). DECaLS, BASS and MzLS together include data obtained, respectively, at the Blanco telescope, Cerro Tololo Inter-American Observatory, NSF’s NOIRLab; the Bok telescope, Steward Observatory, University of Arizona; and the Mayall telescope, Kitt Peak National Observatory, NOIRLab. The Legacy Surveys project is honored to be permitted to conduct astronomical research on Iolkam Du’ag (Kitt Peak), a mountain with particular significance to the Tohono O’odham Nation.

NOIRLab is operated by the Association of Universities for Research in Astronomy (AURA) under a cooperative agreement with the National Science Foundation.

This project used data obtained with the Dark Energy Camera (DECam), which was constructed by the Dark Energy Survey (DES) collaboration. Funding for the DES Projects has been provided by the U.S. Department of Energy, the U.S. National Science Foundation, the Ministry of Science and Education of Spain, the Science and Technology Facilities Council of the United Kingdom, the Higher Education Funding Council for England, the National Center for Supercomputing Applications at the University of Illinois at Urbana-Champaign, the Kavli Institute of Cosmological Physics at the University of Chicago, Center for Cosmology and Astro-Particle Physics at the Ohio State University, the Mitchell Institute for Fundamental Physics and Astronomy at Texas A\&M University, Financiadora de Estudos e Projetos, Fundacao Carlos Chagas Filho de Amparo, Financiadora de Estudos e Projetos, Fundacao Carlos Chagas Filho de Amparo a Pesquisa do Estado do Rio de Janeiro, Conselho Nacional de Desenvolvimento Cientifico e Tecnologico and the Ministerio da Ciencia, Tecnologia e Inovacao, the Deutsche Forschungsgemeinschaft and the Collaborating Institutions in the Dark Energy Survey. The Collaborating Institutions are Argonne National Laboratory, the University of California at Santa Cruz, the University of Cambridge, Centro de Investigaciones Energeticas, Medioambientales y Tecnologicas-Madrid, the University of Chicago, University College London, the DES-Brazil Consortium, the University of Edinburgh, the Eidgenossische Technische Hochschule (ETH) Zurich, Fermi National Accelerator Laboratory, the University of Illinois at Urbana-Champaign, the Institut de Ciencies de l’Espai (IEEC/CSIC), the Institut de Fisica d’Altes Energies, Lawrence Berkeley National Laboratory, the Ludwig Maximilians Universitat Munchen and the associated Excellence Cluster Universe, the University of Michigan, NSF’s NOIRLab, the University of Nottingham, the Ohio State University, the University of Pennsylvania, the University of Portsmouth, SLAC National Accelerator Laboratory, Stanford University, the University of Sussex, and Texas A\&M University.

BASS is a key project of the Telescope Access Program (TAP), which has been funded by the National Astronomical Observatories of China, the Chinese Academy of Sciences (the Strategic Priority Research Program “The Emergence of Cosmological Structures” Grant \# XDB09000000), and the Special Fund for Astronomy from the Ministry of Finance. The BASS is also supported by the External Cooperation Program of Chinese Academy of Sciences (Grant \# 114A11KYSB20160057), and Chinese National Natural Science Foundation (Grant \# 11433005).

The Legacy Survey team makes use of data products from the Near-Earth Object Wide-field Infrared Survey Explorer (NEOWISE), which is a project of the Jet Propulsion Laboratory/California Institute of Technology. NEOWISE is funded by the National Aeronautics and Space Administration.

The Legacy Surveys imaging of the DESI footprint is supported by the Director, Office of Science, Office of High Energy Physics of the U.S. Department of Energy under Contract No. DE-AC02-05CH1123, by the National Energy Research Scientific Computing Center, a DOE Office of Science User Facility under the same contract; and by the U.S. National Science Foundation, Division of Astronomical Sciences under Contract No. AST-0950945 to NOAO.

\end{acknowledgements}

\appendix
\section{Measurements of disk sizes}
\subsection{Disk scale length as the measure of optical disk sizes}
\begin{figure}[hptb]
    \centering
    \includegraphics[width=0.8\textwidth]{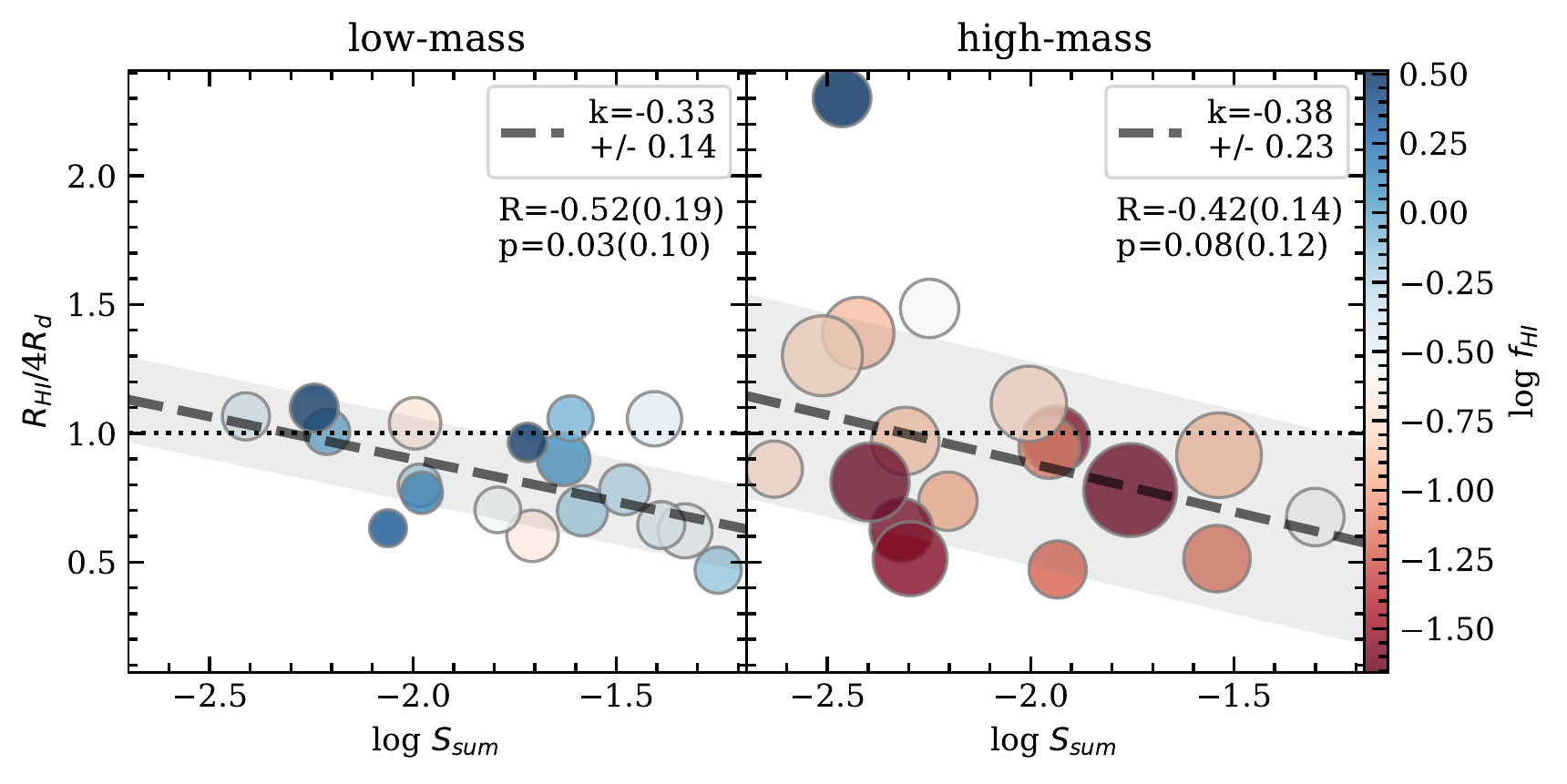}
    \caption{The correlation between the $\hi$-to-optical disk size ratio (adopting $4 R_d$ as the optical disk sizes) and the summed tidal parameter.
    The gray shaded area indicates the scatter ($1 \sigma$) of data points about the linear fit. The dotted horizontal line represents $R_{\rm HI} / 4 R_d = 1$.
    \textbf{Left:} for low-mass ($\log M_*/M_\odot < 8.95$) galaxies.
    \textbf{Right:} for high-mass ($\log M_*/M_\odot > 8.95$) galaxies.
    The other symbols and text are the same as Figure \ref{fig:asymmetry}: Pearson R and p values are shown with bootstrap error in parenthesis. Gray dashed lines are the result of robust linear fit, with the slopes (and error) shown in the corner. All data points are color coded by $\hi$ gas fraction (see colorbar on the right). The size of data points indicates the stellar mass of the galaxy, in the sense that larger data points are used for more massive galaxies.}
    \label{fig:Rsl}
\end{figure}

The majority of the galaxies in the $\hi$ sample are faint dwarf galaxies. Thus it may be a concern that a surface brightness cut at 25 mag arcsec$^{-2}$ does not enclose the bulk of the galaxy light, so that $R_{25, g}$ is not an ideal quantification of the galaxy size in this study. One of the size estimations that is less affected by the faintness is the disk scale length (assume exponential disks). We derive the disk scale length ($R_d$) by fitting the outer disk surface brightness profile in $g$ band and use 4 $R_d$ as the estimates of disk size (enclose $\sim$90\% of light assuming exponential disks). We confirm that the results are not qualitatively different from those obtained by using $R_{25, g}$ as the estimates of disk size. We show the relation between $R_{\rm HI} / 4 R_d$ and $S_{sum}$ in Figure \ref{fig:Rsl}, which is quite similar to that shown in Figure \ref{fig:RHI}.

\subsection{Correction for projection effect in measuring optical disk sizes}
As we have mentioned in the main text, the de-projection on dwarf irregular galaxies based on photometrically derived inclinations and axis ratios could introduce large uncertainties. Previous studies also showed that the observed apparent diameter based on surface brightness isophote is not sensitive to inclination (\citealt{1972MNRAS.159P..35T, 1991Natur.353..515B, 1991MNRAS.250..486C, 1995A&A...296...64B}; but see \citealt{1985ApJS...58...67T}).

\begin{figure}[hptb]
    \centering
    \includegraphics[width=0.8\textwidth]{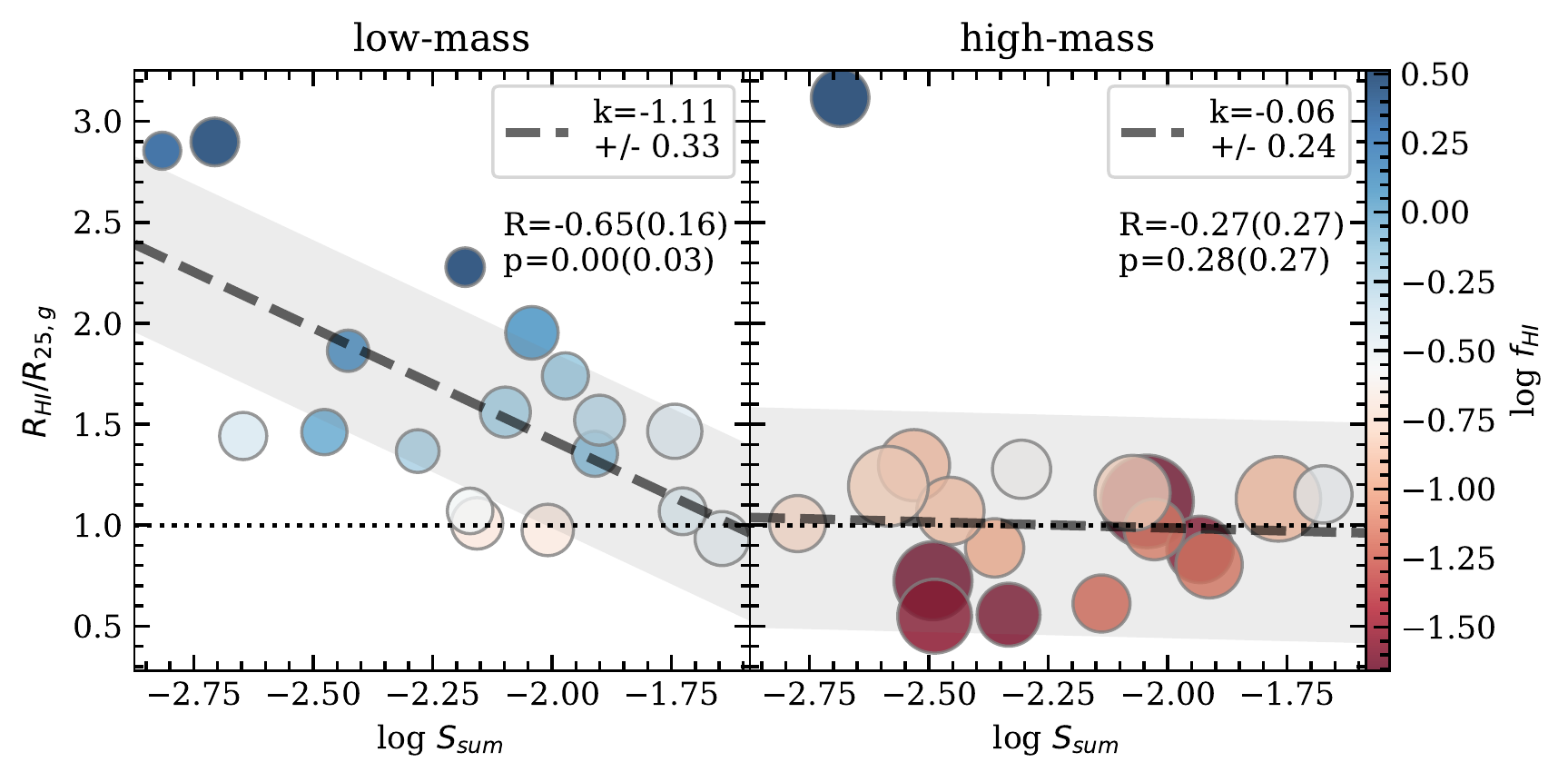}
    \caption{The correlation between the $\hi$-to-optical disk size ratio (de-projection performed) and the summed tidal parameter.
    The gray shaded area indicates the scatter ($1 \sigma$) of data points about the linear fit. The dotted horizontal line represents $R_{\rm HI} / R_{25, g} = 1$.
    \textbf{Left:} for low-mass ($\log M_*/M_\odot < 8.95$) galaxies.
    \textbf{Right:} for high-mass ($\log M_*/M_\odot > 8.95$) galaxies.
    The other symbols and text are the same as Figure \ref{fig:asymmetry}: Pearson R and p values are shown with bootstrap error in parenthesis. Gray dashed lines are the result of robust linear fit, with the slopes (and error) shown in the corner. All data points are color coded by $\hi$ gas fraction (see colorbar on the right). The size of data points indicates the stellar mass of the galaxy, in the sense that larger data points are used for more massive galaxies.}
    \label{fig:deproj}
\end{figure}

We confirm that most of our major results are robust against the treatment of de-projection. When de-projection is performed when deriving $R_{25, g}$, as shown in Figure \ref{fig:deproj}, the $\hi$-to-optical disk size ratios of high-mass galaxies are insensitive to the summed tidal parameter. All high-mass galaxies except for the merging pair (NGC 1359, ID 69) have disk size ratios close to 1, with significant scatter though.

\subsection{Measure HI disk sizes}
\begin{figure}[hptb]
    \centering
    \includegraphics[width=0.8\textwidth]{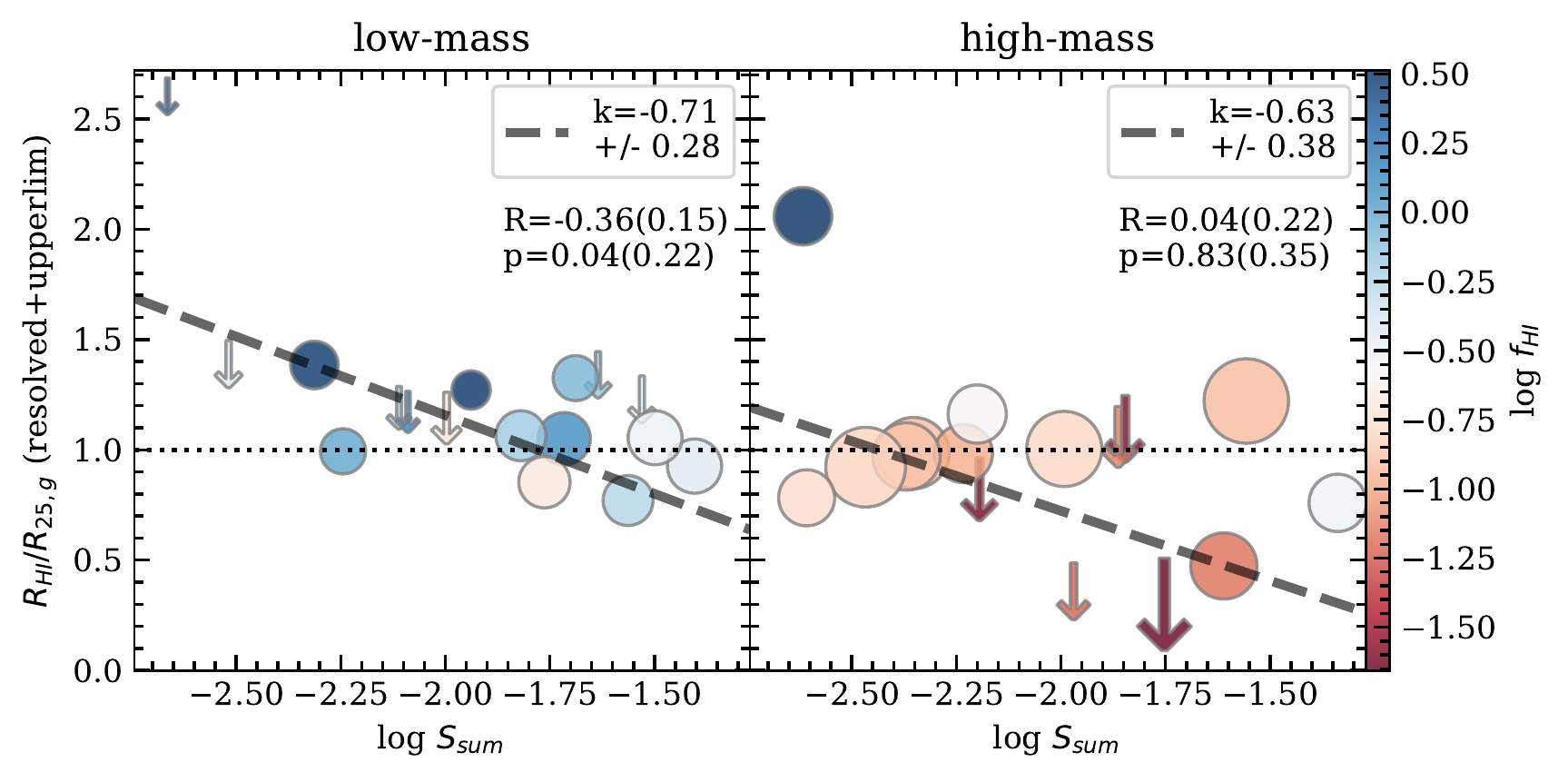}
    \caption{The correlation between the $\hi$-to-optical disk size ratio (adopt directly measured $\hi$ disk sizes) and the summed tidal parameter.
    The dotted horizontal line represents $R_{\rm HI} / R_{25, g} = 1$. The downward arrows are upper limits of $\hi$-to-optical disk size ratios.
    \textbf{Left:} for low-mass ($\log M_*/M_\odot < 8.95$) galaxies.
    \textbf{Right:} for high-mass ($\log M_*/M_\odot > 8.95$) galaxies.
    The other symbols and text are the same as Figure \ref{fig:asymmetry}: Kendall R and p values are shown with bootstrap error in parenthesis. Gray dashed lines are the result of linear fit invoking survival analysis, with the slopes (and error) shown in the corner. All data points are color coded by $\hi$ gas fraction (see colorbar on the right). The size of data points indicates the stellar mass of the galaxy, in the sense that larger data points are used for more massive galaxies.}
    \label{fig:res}
\end{figure}

We directly measure the characteristic sizes ($R_{\rm HI, 0}$) of the $\hi$ disks in the $\hi$ sample. The projection effect is corrected based on the axis ratios of $\hi$ disks. For 21 resolved $\hi$ disks, i.e. those with $R_{\rm HI, 0}$ larger than 3 $B_{maj}$ (3 $\times$ 30 arcsec), and the beam smearing effect is corrected as $R_{\rm HI} = \sqrt{R_{\rm HI, 0}^2 - B_{maj} \times B_{min}}$ \citep{Wang2016}, where $R_{\rm HI, 0}$ is the uncorrected measurement and $R_{\rm HI}$ is the corrected one. The sizes of the unresolved $\hi$ disks are treated as upper limits and no smearing correction is applied. We check the anti-correlation between the $\hi$-to-optical disk size ratios and tidal parameter using survival analysis (see Figure \ref{fig:res}). If we only include resolved $\hi$ disks for such test, a Pearson R value of $-0.49$ and a p value of 0.15 for the low-mass subsample is obtained (not shown in the figure).

\section{Galactic properties}
We present the basic and derived galactic properties, including Petrosian magnitudes in $g$, $r$ and $z$ band, stellar mass and $R_{25, g}$ in Table \ref{tab:SourceCata}. And color gradients and tidal parameters of galaxies in the $\hi$ sample are provided in Table \ref{tab:HI}.
The uncertainties of $V_{opt}$, $grz$ magnitudes, $R_{25, g}$, color gradients and tidal parameters are shown in the parenthesis following. The typical uncertainty of stellar mass is $\sim$0.11 dex, which is dominated by the scatter of the stellar mass-to-light ratio as a function of $g - r$ color \citep{Zibetti2009}.

\begin{longrotatetable}
\begin{deluxetable*}{cccrccrrrrr}
\tablecaption{Basic information and derived quantities for all galaxies in the optical and $\hi$ sample. \label{tab:SourceCata}}
\tablehead{
\colhead{Object ID}  &  \colhead{$R.A.$}  &  \colhead{$DEC.$}  &  \colhead{$V_{opt}$}  &  \colhead{Reference}  &  \colhead{Detected}  &  \colhead{$g$}  &  \colhead{$r$}  &  \colhead{$z$}  &  \colhead{$\log M_*$}  &  \colhead{$R_{25, g}$}\\
\colhead{}  &  \colhead{(deg)}  &  \colhead{(deg)}  &  \colhead{(km s$^{-1}$)}  &  \colhead{}  &  \colhead{}  &  \colhead{(mag)}  &  \colhead{(mag)}  &  \colhead{(mag)}  &  \colhead{($M_\odot$)}  &  \colhead{(kpc)}\\
\colhead{(1)}  &  \colhead{(2)}  &  \colhead{(3)}  &  \colhead{(4)}  &  \colhead{(5)}  &  \colhead{(6)}  &  \colhead{(7)}  &  \colhead{(8)}  &  \colhead{(9)}  &  \colhead{(10)}  &  \colhead{(11)}
}
\startdata
01    & 51.106541  & -21.544029  & 1588(4) & WALLABY & 1     & 11.894(0.001) & 11.242(0.001) & 10.770(0.001) & 10.2  & 150.42(0.23) \\
02    & 51.571734  & -21.335320  & 1526(29) & NED optical & 0     & 10.707(0.001) & 9.924(0.000) & 9.319(0.000) & 11.0  & 165.07(0.09) \\
03    & 51.897030  & -21.228390  & 1686(4) & WALLABY & 1     & 14.264(0.003) & 13.740(0.002) & 13.415(0.002) & 9.0   & 71.29(0.09) \\
04    & 53.361130  & -23.712780  & 1810(4) & WALLABY & 1     & 13.058(0.002) & 12.439(0.001) & 11.950(0.001) & 9.7   & 69.67(0.13) \\
05    & 53.471198  & -20.282588  & 1177(29) & NED optical & 0     & 13.157(0.002) & 12.453(0.001) & 11.953(0.001) & 9.9   & 49.89(0.19) \\
06    & 53.906212  & -21.294060  & 1802(4) & WALLABY & 1     & 14.341(0.003) & 13.933(0.002) & 13.717(0.002) & 8.8   & 74.27(0.14) \\
07    & 54.623726  & -23.027488  & 1701(19) & NED optical & 0     & 10.364(0.000) & 9.595(0.000) & 8.991(0.000) & 11.1  & 182.06(0.04) \\
08    & 54.878546  & -18.688174  & 590(14) & NED optical & 0     & 11.252(0.001) & 10.490(0.001) & 9.918(0.000) & 10.7  & 103.93(0.12) \\
09    & 55.236919  & -22.564470  & 1552(4) & WALLABY & 1     & 11.936(0.001) & 11.136(0.001) & 10.559(0.001) & 10.5  & 114.55(0.34) \\
10    & 55.704739  & -22.108405  & 1445(6) & NED optical & 0     & 11.840(0.001) & 11.104(0.001) & 10.546(0.001) & 10.4  & 105.82(0.06) \\
11    & 55.457583  & -19.581250  & 1914(...) & NED preferred & 0     & 15.635(0.005) & 15.060(0.004) & 14.728(0.004) & 8.6   & 22.25(0.14) \\
12    & 55.382542  & -19.905139  & 1545(43) & NED optical & 0     & 14.096(0.003) & 13.444(0.002) & 13.009(0.002) & 9.4   & 40.60(0.07) \\
13    & 55.483667  & -18.895167  & 2031(31) & NED optical & 0     & 13.661(0.002) & 12.970(0.002) & 12.500(0.001) & 9.6   & 46.29(0.09) \\
14    & 55.268375  & -19.094444  & 987(45) & NED optical & 0     & 14.885(0.004) & 14.308(0.003) & 13.931(0.003) & 8.9   & 32.48(0.19) \\
15    & 55.066375  & -19.081778  & 1614(45) & NED optical & 0     & 14.875(0.004) & 14.204(0.003) & 13.751(0.002) & 9.1   & 23.60(0.10) \\
16    & 55.180125  & -18.645306  & 1374(45) & NED optical & -1    & 14.791(0.004) & 14.166(0.003) & 13.759(0.002) & 9.0   & 28.58(0.28) \\
17    & 55.000333  & -19.426306  & 1874(33) & NED optical & 0     & 13.949(0.003) & 13.257(0.002) & 12.776(0.001) & 9.5   & 34.69(0.18) \\
18    & 55.219708  & -18.478000  & 1680(...) & NED preferred & -1    & 15.230(0.004) & 14.560(0.003) & 14.109(0.003) & 9.0   & 22.78(0.15) \\
19    & 54.413500  & -18.339472  & 2009(19) & NED optical & -1    & 12.853(0.002) & 12.109(0.001) & 11.547(0.001) & 10.1  & 67.99(0.21) \\
20    & 55.288042  & -18.314139  & 1245(41) & NED optical & -1    & 14.222(0.003) & 13.550(0.002) & 13.090(0.002) & 9.4   & 29.87(0.17) \\
21    & 54.467369  & -19.008360  & 1218(4) & WALLABY & 1     & 14.249(0.003) & 13.702(0.002) & 13.341(0.002) & 9.1   & 44.94(0.20) \\
22    & 54.660750  & -18.427972  & 2127(26) & NED optical & -1    & 12.524(0.001) & 11.785(0.001) & 11.241(0.001) & 10.2  & 71.74(0.10) \\
23    & 55.011250  & -19.366610  & 1216(4) & WALLABY & 1     & 14.835(0.004) & 14.363(0.003) & 14.089(0.003) & 8.7   & 46.01(0.22) \\
24    & 55.029750  & -18.443472  & 1841(30) & NED optical & -1    & 13.444(0.002) & 12.749(0.001) & 12.258(0.001) & 9.7   & 56.13(0.18) \\
25    & 55.049333  & -18.580139  & 1779(14) & NED optical & -1    & 9.966(0.000) & 9.194(0.000) & 8.592(0.000) & 11.3  & 208.52(0.04) \\
26    & 55.079875  & -18.931500  & 1693(26) & NED optical & 0     & 13.456(0.002) & 12.770(0.001) & 12.277(0.001) & 9.7   & 49.15(0.15) \\
27    & 55.251042  & -19.455389  & 2034(...) & NED preferred & -1    & 15.692(0.006) & 15.303(0.004) & 15.089(0.004) & 8.2   & 28.88(0.27) \\
28    & 55.436125  & -18.267000  & 2013(47) & NED optical & -1    & 12.770(0.002) & 12.056(0.001) & 11.549(0.001) & 10.0  & 71.15(0.09) \\
29    & 55.739125  & -19.020806  & 1111(36) & NED optical & -1    & 14.313(0.003) & 13.895(0.002) & 13.663(0.003) & 8.8   & 45.83(0.07) \\
30    & 54.430542  & -22.908194  & 1515(55) & NED optical & 0     & 14.540(0.003) & 13.930(0.002) & 13.511(0.002) & 9.1   & 43.37(0.09) \\
31    & 54.568958  & -22.486500  & 1359(45) & NED optical & 0     & 16.101(0.006) & 15.487(0.004) & 15.115(0.004) & 8.5   & 19.39(0.12) \\
32    & 53.974792  & -22.139722  & 1374(45) & NED optical & 0     & 15.308(0.004) & 14.634(0.003) & 14.181(0.003) & 8.9   & 21.97(0.12) \\
33    & 53.938625  & -21.783111  & 1638(45) & NED optical & 0     & 15.682(0.005) & 15.045(0.004) & 14.668(0.004) & 8.7   & 19.84(0.08) \\
34    & 53.365261  & -21.564659  & 1509(4) & WALLABY & 1     & 13.753(0.002) & 13.102(0.002) & 12.632(0.001) & 9.5   & 30.86(0.31) \\
35    & 53.240120  & -21.089420  & 1665(4) & WALLABY & 1     & 14.012(0.002) & 13.450(0.002) & 13.121(0.002) & 9.2   & 37.07(0.09) \\
36    & 53.012417  & -20.818944  & 1587(10) & NED optical & 0     & 11.674(0.001) & 10.910(0.001) & 10.315(0.000) & 10.6  & 119.80(0.05) \\
37    & 54.236458  & -20.589722  & 1689(31) & NED optical & 0     & 15.053(0.004) & 14.482(0.003) & 14.101(0.002) & 8.8   & 18.25(0.22) \\
38    & 54.162792  & -20.902000  & 1809(30) & NED optical & 0     & 12.970(0.001) & 12.277(0.001) & 11.791(0.001) & 9.9   & 62.42(0.13) \\
39    & 55.150711  & -21.525669  & 1644(4) & WALLABY & 1     & 15.407(0.005) & 14.987(0.004) & 14.908(0.005) & 8.4   & 32.87(0.07) \\
40    & 55.237659  & -21.713140  & 1695(4) & WALLABY & 1     & 14.375(0.003) & 13.913(0.002) & 13.607(0.002) & 8.9   & 50.02(0.15) \\
41    & 55.860250  & -21.328944  & 1711(45) & NED optical & 0     & 16.308(0.007) & 15.705(0.005) & 15.380(0.005) & 8.4   & 16.05(0.11) \\
42    & 55.379459  & -21.681530  & 1644(4) & WALLABY & 1     & 13.678(0.002) & 13.045(0.002) & 12.600(0.001) & 9.5   & 79.05(0.12) \\
43    & 55.173083  & -22.651139  & 1803(34) & NED optical & 0     & 14.849(0.003) & 14.175(0.003) & 13.727(0.002) & 9.1   & 28.87(0.07) \\
44    & 52.696541  & -21.058220  & 1292(4) & WALLABY & 1     & 13.748(0.002) & 13.183(0.002) & 12.811(0.001) & 9.3   & 38.86(0.11) \\
45    & 53.424461  & -21.478640  & 1859(4) & WALLABY & 1     & 12.062(0.001) & 11.557(0.001) & 11.267(0.001) & 9.9   & 88.97(0.37) \\
46    & 53.866959  & -21.217030  & 1518(4) & WALLABY & 1     & 15.095(0.004) & 14.770(0.003) & 14.611(0.003) & 8.3   & 31.32(0.11) \\
47    & 54.573500  & -23.419167  & 1687(37) & NED optical & 0     & 14.127(0.003) & 13.446(0.002) & 12.971(0.002) & 9.4   & 38.73(0.05) \\
48    & 54.839291  & -21.415720  & 1622(4) & WALLABY & 1     & 14.921(0.004) & 14.432(0.003) & 14.106(0.003) & 8.7   & 31.14(0.16) \\
49    & 54.841042  & -22.724694  & 1495(28) & NED optical & 0     & 12.692(0.001) & 11.966(0.001) & 11.413(0.001) & 10.1  & 76.11(0.24) \\
50    & 55.311790  & -23.836941  & 1885(4) & WALLABY & 1     & 13.236(0.002) & 12.668(0.001) & 12.307(0.001) & 9.5   & 60.66(0.28) \\
51    & 55.909050  & -21.237700  & 1612(4) & WALLABY & 1     & 14.933(0.004) & 14.480(0.003) & 14.304(0.003) & 8.6   & 49.50(0.10) \\
52    & 56.208125  & -21.920556  & 1670(10) & NED optical & 0     & 11.642(0.001) & 10.915(0.001) & 10.437(0.001) & 10.5  & 102.29(0.03) \\
53    & 54.921719  & -23.844400  & 1622(4) & WALLABY & 1     & 16.179(0.006) & 15.883(0.006) & 15.784(0.007) & 7.8   & 27.41(0.11) \\
54    & 55.170792  & -22.287060  & 1774(4) & WALLABY & 1     & 15.273(0.004) & 14.757(0.003) & 14.427(0.003) & 8.6   & 48.36(0.12) \\
55    & 55.578350  & -22.752560  & 1569(4) & WALLABY & 1     & 14.635(0.003) & 14.130(0.002) & 13.909(0.003) & 8.9   & 46.92(0.06) \\
56    & 50.777500  & -21.375194  & 1597(26) & NED optical & -1    & 12.872(0.001) & 12.159(0.001) & 11.633(0.001) & 10.0  & 56.13(0.08) \\
57    & 51.469250  & -21.289056  & 1428(45) & NED optical & 0     & 15.064(0.004) & 14.332(0.003) & 13.790(0.002) & 9.1   & 14.16(0.23) \\
58    & 51.925667  & -21.699611  & 1295(...) & NED preferred & 0     & 15.356(0.004) & 14.720(0.003) & 14.287(0.003) & 8.8   & 34.31(0.15) \\
59    & 51.202083  & -21.336528  & 1333(60) & NED optical & 0     & 13.002(0.002) & 12.459(0.001) & 12.105(0.001) & 9.6   & 63.81(0.02) \\
60    & 51.230431  & -21.783890  & 1456(4) & WALLABY & 1     & 15.611(0.005) & 15.128(0.004) & 14.885(0.004) & 8.4   & 30.28(0.07) \\
61    & 51.630583  & -21.216806  & 1548(45) & NED optical & 0     & 14.501(0.003) & 13.776(0.002) & 13.217(0.002) & 9.4   & 22.60(0.23) \\
62    & 56.321789  & -23.002470  & 1546(4) & WALLABY & 1     & 12.751(0.001) & 12.053(0.001) & 11.533(0.001) & 10.0  & 65.52(0.23) \\
63    & 56.344791  & -24.202419  & 1733(4) & WALLABY & 1     & 16.931(0.010) & 16.549(0.008) & 16.441(0.014) & 7.7   & 12.88(0.03) \\
64    & 56.235859  & -23.699699  & 1819(4) & WALLABY & 1     & 15.363(0.004) & 14.920(0.004) & 14.686(0.004) & 8.4   & 23.49(0.13) \\
65    & 57.058670  & -21.474470  & 1586(4) & WALLABY & 1     & 12.962(0.002) & 12.308(0.001) & 11.912(0.001) & 9.8   & 74.14(0.20) \\
66    & 56.145222  & -21.192070  & 1578(4) & WALLABY & 1     & 15.721(0.005) & 15.258(0.004) & 15.038(0.004) & 8.3   & 26.27(0.13) \\
67    & 52.252880  & -22.146580  & 1627(4) & WALLABY & 1     & 14.479(0.003) & 13.920(0.002) & 13.598(0.002) & 9.0   & 41.78(0.12) \\
68    & 52.422920  & -22.285000  & 1755(4) & WALLABY & 1     & 13.540(0.002) & 13.134(0.002) & 12.898(0.002) & 9.1   & 41.44(0.10) \\
69    & 53.448792  & -19.492060  & 1964(4) & WALLABY & 1     & 12.543(0.001) & 12.345(0.001) & 12.303(0.001) & 9.1   & 58.47(0.03) \\
70    & 54.370201  & -24.500299  & 1497(4) & WALLABY & 1     & 11.196(0.001) & 10.769(0.001) & 10.522(0.001) & 10.1  & 113.29(0.19) \\
71    & 52.409611  & -23.350330  & 1657(4) & WALLABY & 1     & 15.426(0.004) & 15.031(0.004) & 14.801(0.004) & 8.3   & 39.80(0.13) \\
72    & 53.117611  & -23.380930  & 1755(4) & WALLABY & 1     & 15.772(0.005) & 15.398(0.004) & 15.203(0.005) & 8.1   & 23.06(0.14) \\
73    & 52.134258  & -22.501329  & 1774(4) & WALLABY & 1     & 16.084(0.006) & 15.699(0.005) & 15.490(0.005) & 8.0   & 33.97(0.16) \\
74    & 53.259392  & -24.132870  & 1915(4) & WALLABY & 1     & 14.938(0.004) & 14.558(0.003) & 14.354(0.003) & 8.5   & 61.93(0.18) \\
75    & 54.072398  & -25.604380  & 1590(4) & WALLABY & 1     & 14.552(0.003) & 14.005(0.002) & 13.636(0.002) & 9.0   & 44.32(0.13) \\
\enddata
\tablecomments{Column (1): object ID. Column (2)-(3): right ascension and declination (J2000). Column (4): optical velocity $V_{opt} = cz$, where $c$ is the speed of light and $z$ is the redshift. Column (5): the reference of $V_{opt}$. Column (6): whether it is detected by WALLABY: -1 = not in the footprint; 0 = not detected; 1 = detected. Column (7)-(9): Petrosian magnitude in $g$, $r$ and $z$ band. Column (10): derived stellar mass, see text in Section \ref{sec:RHI}. Column (11): optical disk size measured in $g$ band at 25 mag arcsec$^{-2}$ isophote, see text in Section \ref{sec:RHI}.}
\end{deluxetable*}
\end{longrotatetable}

\begin{deluxetable*}{cccrccrrrrr} 
\tablecaption{Identifications and derived quantities for galaxies in the $\hi$ sample. \label{tab:HI}}
\tablehead{
\colhead{Object ID}  &  \colhead{WALLABY ID}  &  \colhead{other ID}  &  \colhead{$CG_{01}$}  &  \colhead{$CG_{12}$}  &  \colhead{$S_{sum}$}  &  \colhead{$S_{strongest}$}  &  \colhead{$S_{nearest}$}\\
\colhead{}  &  \colhead{}  &  \colhead{}  &  \colhead{(mag $R_{50, z}^{-1}$)}  &  \colhead{(mag $R_{50, z}^{-1}$)}  &  \colhead{}  &  \colhead{}  &  \colhead{}\\
\colhead{(1)}  &  \colhead{(2)}  &  \colhead{(3)}  &  \colhead{(4)}  &  \colhead{(5)}  &  \colhead{(6)}  &  \colhead{(7)}  &  \colhead{(8)}
}
\startdata
01    & J032425-213233 &  NGC 1325 & -0.03(0.01) & -0.10(0.01) & -1.56(0.16) & -1.76(0.23) & -2.51(0.45) \\
03    & J032735-211339 &  ESO 548-21 & -0.08(0.01) & 0.00(0.01) & -1.34(0.20) & -1.44(0.25) & -3.03(0.33) \\
04    & J033326-234246 &  IC 1952 & -0.16(0.01) & 0.04(0.03) & -2.35(0.11) & -2.75(0.24) & -3.85(0.19) \\
06    & J033537-211742 &  IC 1962 & 0.00(0.01) & -0.06(0.01) & -1.72(0.10) & -2.37(0.16) & -2.45(0.46) \\
09    & J034056-223350 &  NGC 1415 & -0.16(0.05) & -0.05(0.00) & -1.78(0.12) & -2.36(0.31) & -2.38(0.26) \\
21    & J033752-190024 &  NGC 1390 & -0.06(0.01) & 0.04(0.01) & -2.23(0.43) & -2.59(0.86) & -2.98(1.04) \\
23    & J034002-192200 &  ESO 548-65 & -0.04(0.02) & 0.00(0.01) & -1.76(0.47) & -1.98(0.75) & -1.98(0.75) \\
34    & J033327-213352 &  ESO 548-36 & -0.11(0.09) & 0.16(0.02) & -1.87(0.38) & -2.03(0.56) & -2.03(0.56) \\
35    & J033257-210513 &  ESO 548-34 & -0.12(0.02) & 0.04(0.01) & -1.88(0.14) & -2.15(0.24) & -2.15(0.24) \\
39    & J034036-213129 &  ESO 548-69 & 0.04(0.02) & 0.10(0.03) & -1.65(0.05) & -2.11(0.10) & -2.44(0.13) \\
40    & J034057-214245 &  NGC 1414 & -0.08(0.01) & -0.04(0.01) & -1.40(0.09) & -1.62(0.14) & -1.62(0.14) \\
42    & J034131-214051 &  NGC 1422 & -0.26(0.01) & -0.03(0.01) & -1.61(0.08) & -2.07(0.17) & -2.07(0.17) \\
44    & J033047-210333 &  ESO 548-29 & -0.07(0.01) & -0.00(0.01) & -2.22(0.24) & -2.45(0.40) & -2.45(0.40) \\
45    & J033341-212844 &  IC 1953 & -0.24(0.01) & -0.03(0.02) & -1.99(0.23) & -2.38(0.55) & -2.38(0.55) \\
46    & J033527-211302 &  ESO 548-49 & -0.09(0.01) & -0.06(0.01) & -1.69(0.22) & -2.01(0.46) & -2.01(0.46) \\
48    & J033921-212450 &  LEDA 13460 & -0.00(0.02) & 0.07(0.02) & -2.01(0.04) & -2.82(0.16) & -3.18(0.11) \\
50    & J034114-235017 &  ESO 482-35 & -0.07(0.03) & 0.01(0.00) & -2.37(0.18) & -2.66(0.33) & -3.97(0.43) \\
51    & J034337-211418 &  ESO 549-6 & -0.05(0.01) & -0.02(0.01) & -1.82(0.05) & -2.45(0.14) & -2.73(0.19) \\
53    & J033941-235054 &  ESO 482-27 & 0.04(0.04) & 0.07(0.01) & -1.94(0.09) & -2.23(0.15) & -3.14(0.43) \\
54    & J034040-221711 &  ESO 548-70 & -0.02(0.02) & -0.03(0.00) & -1.56(0.16) & -1.97(0.37) & -1.97(0.37) \\
55    & J034219-224520 &  ESO 482-36 & -0.10(0.01) & -0.01(0.01) & -1.50(0.08) & -1.75(0.12) & -1.75(0.12) \\
60    & J032455-214701 &  ESO 548-11 & -0.09(0.01) & -0.01(0.01) & -1.55(0.13) & -1.84(0.23) & -1.84(0.23) \\
62    & J034517-230001 &  NGC 1438 & -0.14(0.03) & -0.08(0.02) & -2.36(0.09) & -3.05(0.19) & -4.68(0.45) \\
63    & J034522-241208 &  LEDA 79249 & 0.22(0.05) & 0.30(0.07) & -2.68(0.05) & -3.02(0.10) & -4.19(0.16) \\
64    & J034456-234158 &  LEDA 13743 & 0.03(0.04) & 0.30(0.03) & -2.53(0.09) & -3.02(0.21) & -4.46(0.17) \\
65    & J034814-212824 &  ESO 549-18 & -0.03(0.01) & -0.03(0.01) & -2.40(0.08) & -2.90(0.21) & -4.41(0.18) \\
66    & J034434-211123 &  LEDA 13511 & 0.02(0.01) & 0.06(0.01) & -1.96(0.05) & -2.56(0.10) & -2.56(0.10) \\
67    & J032900-220851 &  ESO 548-25 & 0.07(0.03) & -0.00(0.02) & -1.99(0.07) & -2.49(0.17) & -2.73(0.23) \\
68    & J032941-221642 &  NGC 1347 & -0.20(0.02) & 0.02(0.02) & -2.20(0.08) & -2.86(0.24) & -2.86(0.24) \\
69    & J033347-192946 &  NGC 1359 & -0.24(0.09) & -0.06(0.03) & -2.62(0.15) & -3.07(0.35) & -4.26(1.20) \\
70    & J033728-243010 &  NGC 1385 & -0.08(0.02) & 0.02(0.03) & -2.47(0.14) & -2.91(0.36) & -4.39(0.27) \\
71    & J032937-232103 &  ESO 481-30 & -0.01(0.01) & -0.03(0.03) & -2.24(0.05) & -2.76(0.12) & -4.20(0.18) \\
72    & J033228-232245 &  ESO 482-3 & 0.08(0.02) & 0.08(0.01) & -2.12(0.06) & -2.59(0.12) & -2.59(0.12) \\
73    & J032831-222957 &  ESO 481-28 & 0.04(0.03) & 0.16(0.02) & -2.10(0.07) & -2.70(0.09) & -2.70(0.09) \\
74    & J033302-240756 &  ESO 482-5 & -0.05(0.01) & -0.11(0.01) & -2.31(0.13) & -2.66(0.21) & -2.66(0.21) \\
75    & J033617-253615 &  ESO 482-11 & -0.06(0.01) & -0.04(0.01) & -2.61(0.07) & -3.18(0.19) & -3.18(0.19) \\
\enddata
\tablecomments{Column (1): object ID. Column (2): WALLABY identifier. Column (3): other identification. Column (4)-(5): color gradients in $R < R_{50, z}$ and $R_{50, z} < R < 2 R_{50, z}$. Column (6)-(8): tidal parameters of the summed, the strongest and the nearest perturber.}
\end{deluxetable*}

\section{atlas} \label{sec:atlas}
We present the optical color images and color profiles for galaxies in the $\hi$ sample in Figure \ref{fig:atlas_1}, \ref{fig:atlas_2}, \ref{fig:atlas_3} and \ref{fig:atlas_4}. The foreground stars are masked. For each galaxy, we show $g - r$, $g - z$ and $r - z$ profiles respectively. $CG_{01}$ and $CG_{12}$ are also shown along with the $g - r$ profile.

\begin{figure}
    \centering
\includegraphics[width=8.5cm]{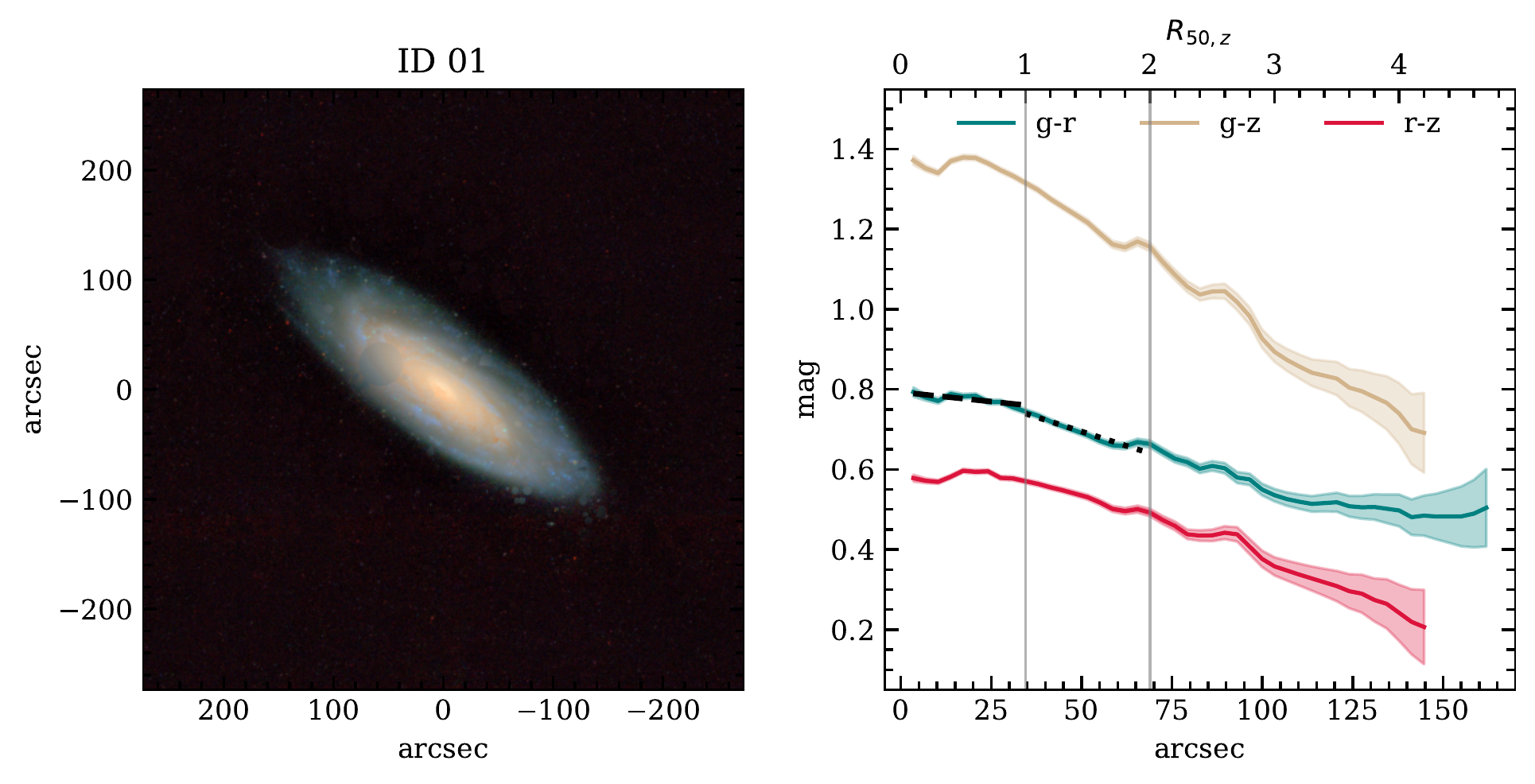}
\includegraphics[width=8.5cm]{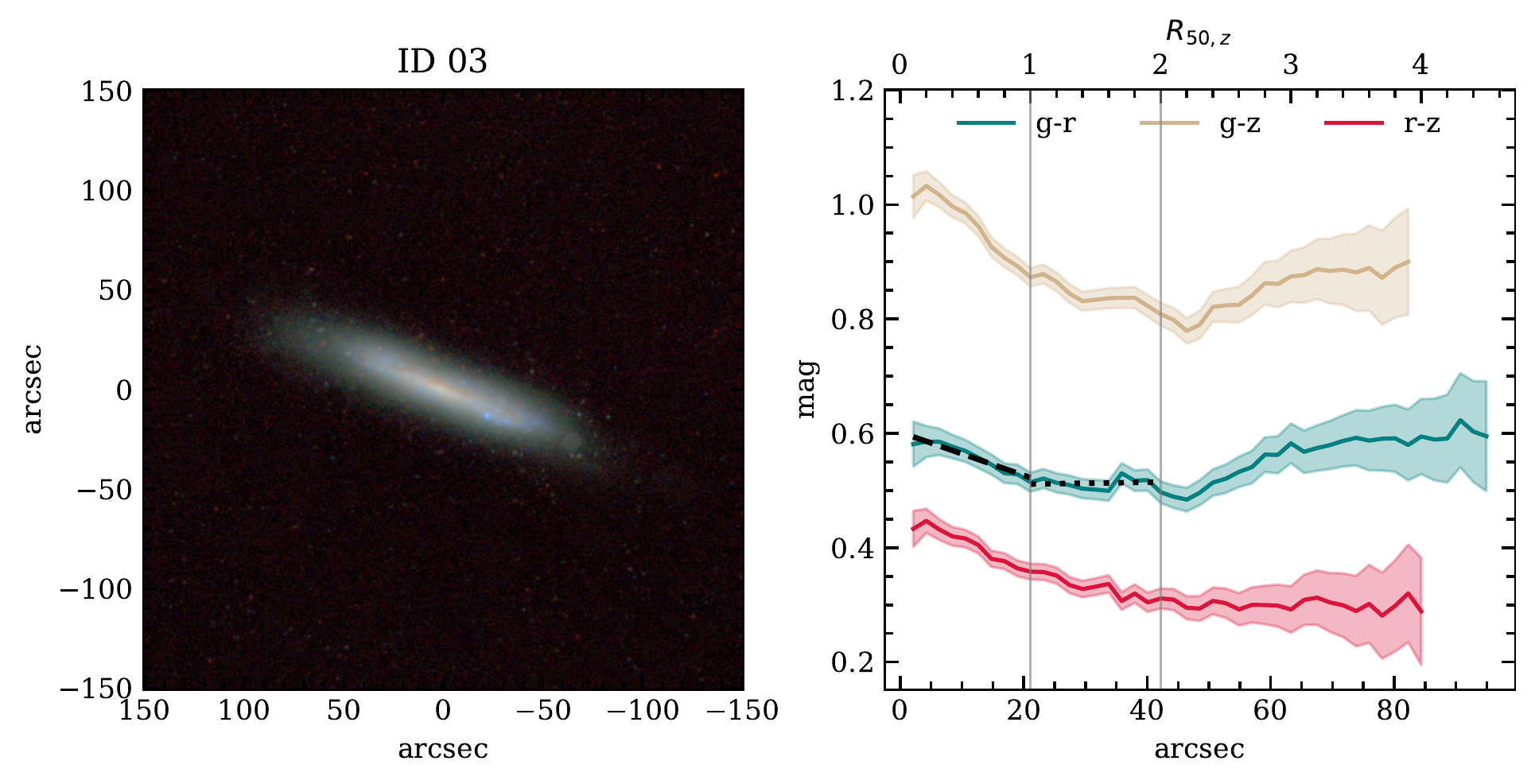}
\includegraphics[width=8.5cm]{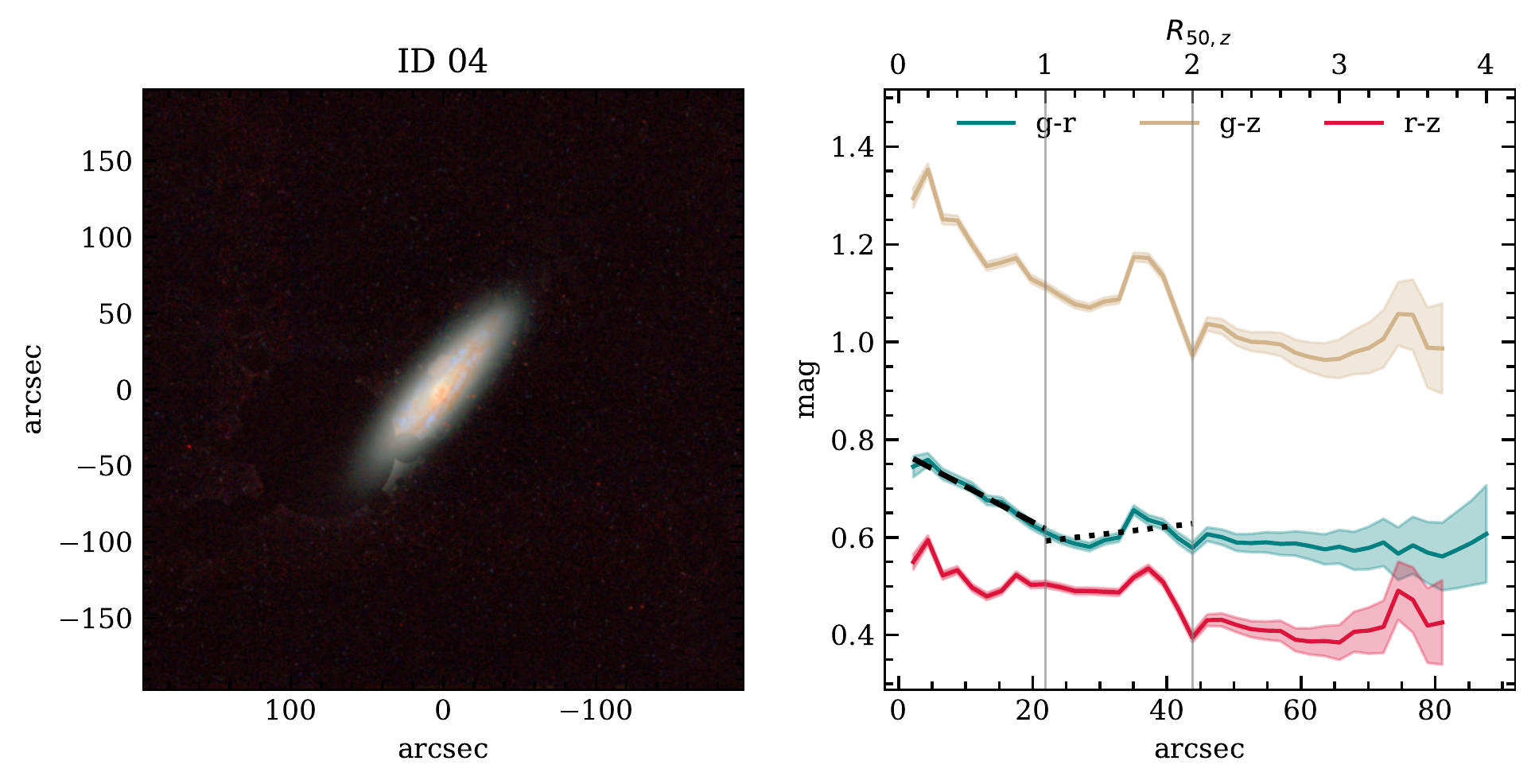}
\includegraphics[width=8.5cm]{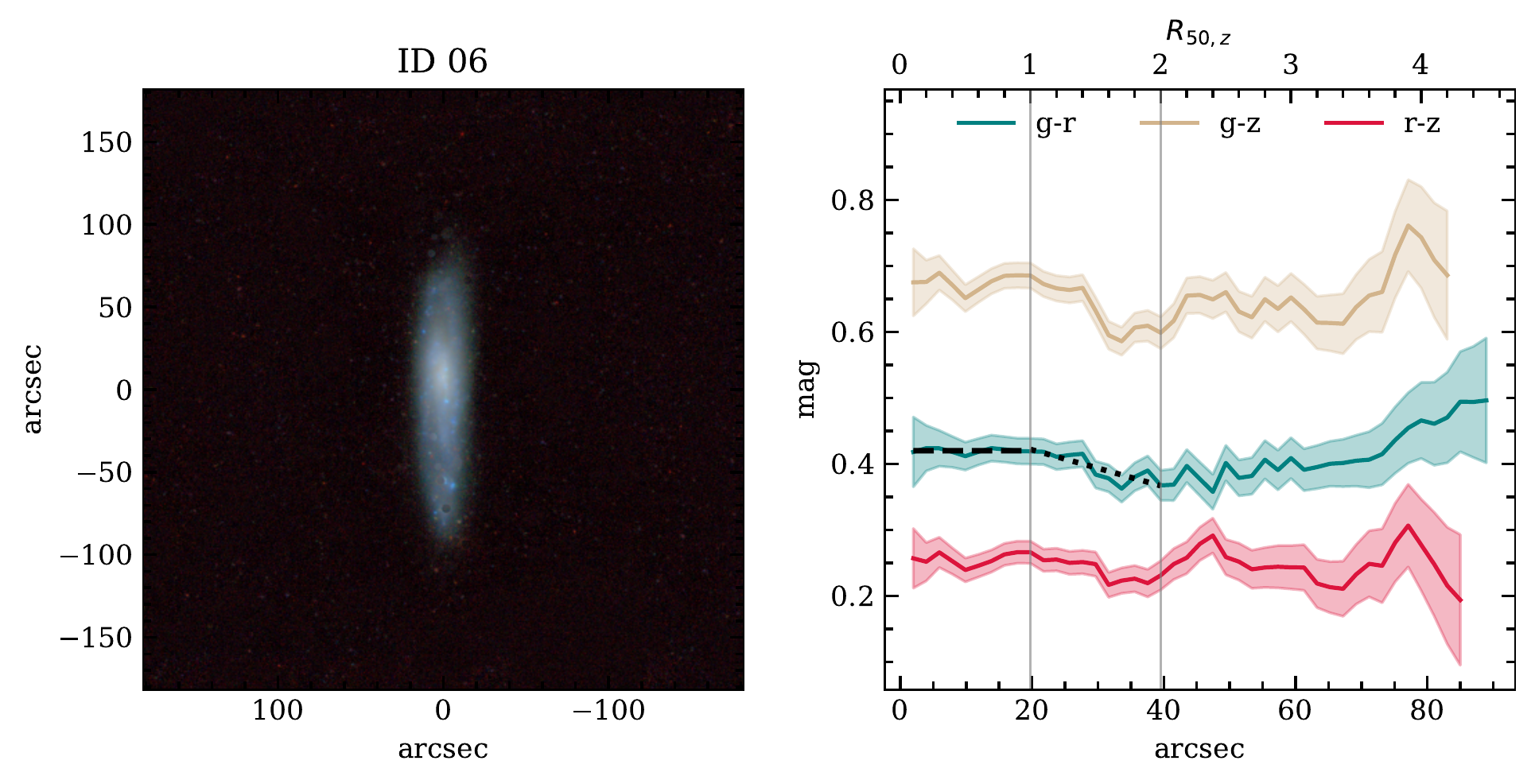}
\includegraphics[width=8.5cm]{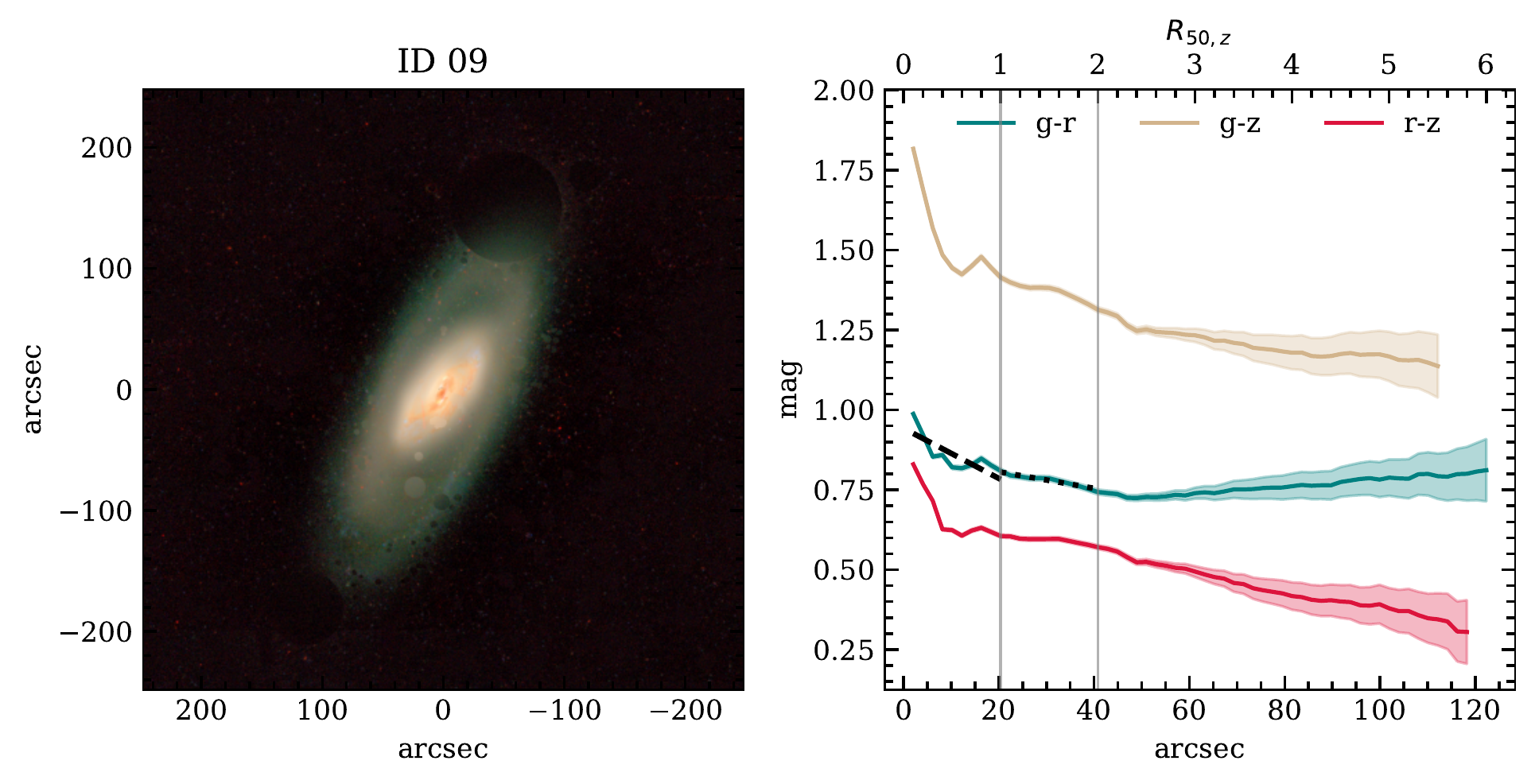}
\includegraphics[width=8.5cm]{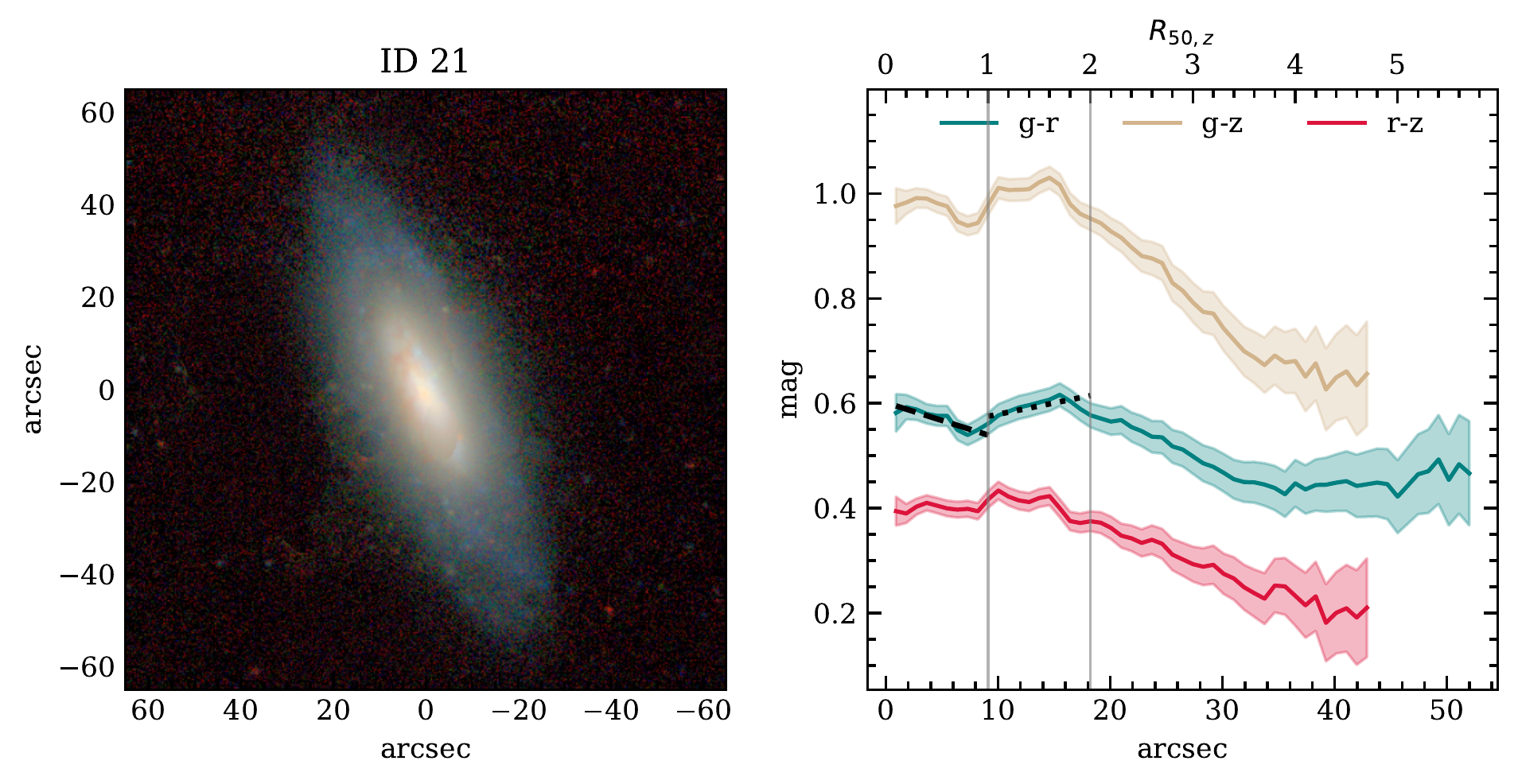}
\includegraphics[width=8.5cm]{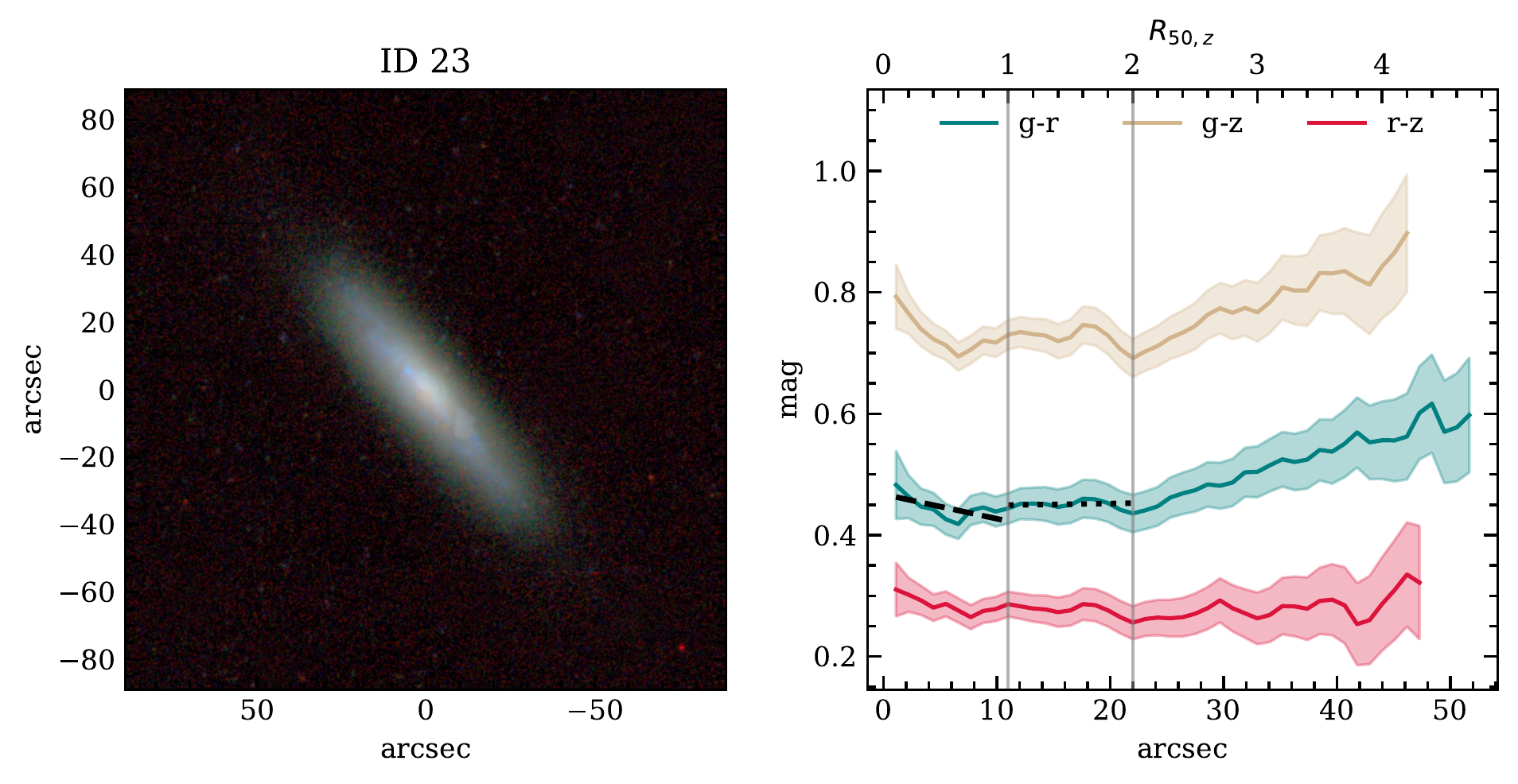}
\includegraphics[width=8.5cm]{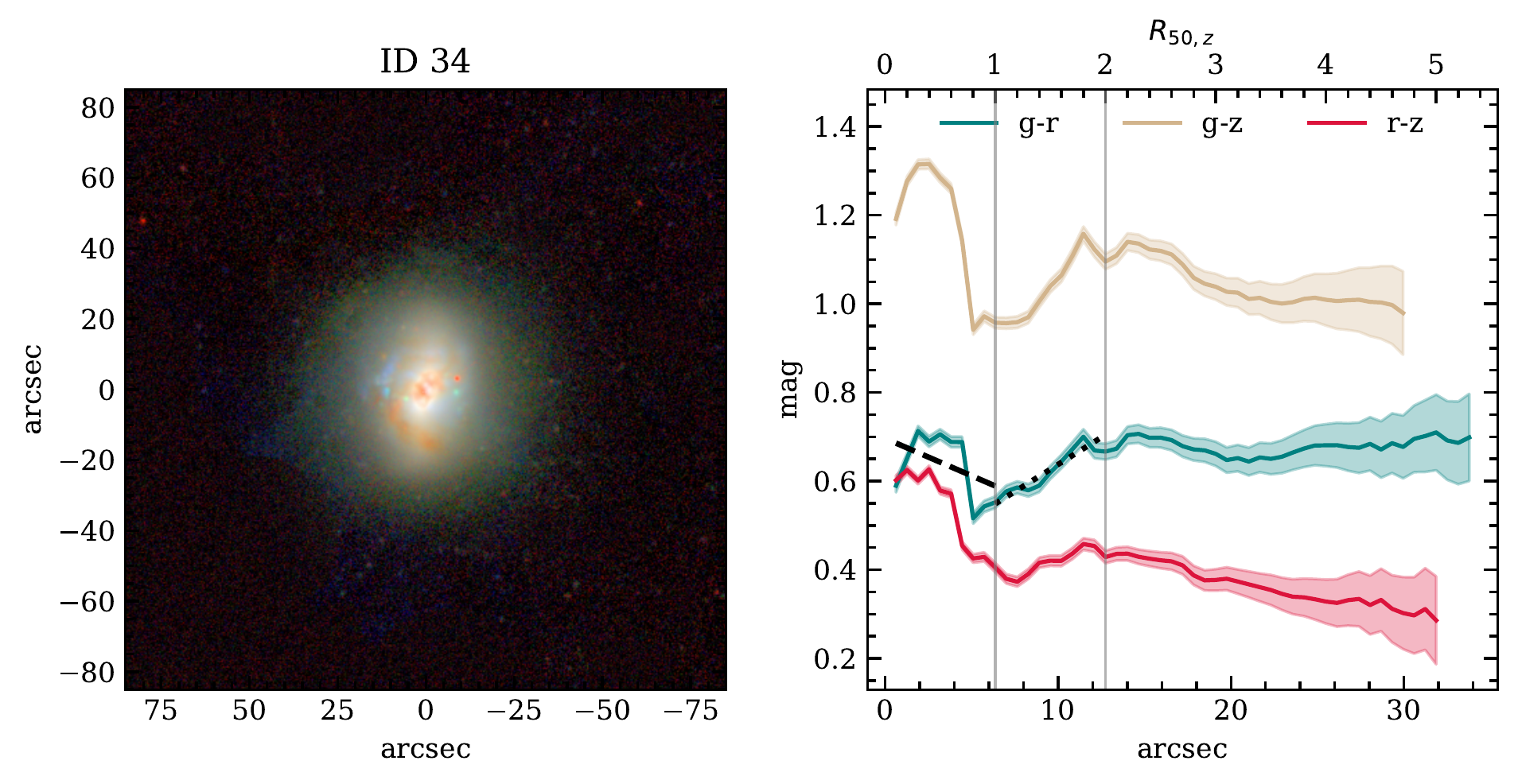}
    \caption{The optical color images (left panel) and color profiles (right panel) for galaxies in the $\hi$ sample. $g - r$, $g - z$ and $r - z$ profiles are shown in green, yellow and red, respectively. The shaded area indicate the uncertainties of the profile. The fitted linear lines with slopes equal to $CG_{01}$ and $CG_{12}$ are shown as black dashed and dotted line overlapping on the $g - r$ profile. The gray vertical lines indicate $1 R_{50, z}$ and $2 R_{50, z}$.}
    \label{fig:atlas_1}
\end{figure}

\begin{figure}
    \centering
\includegraphics[width=8.5cm]{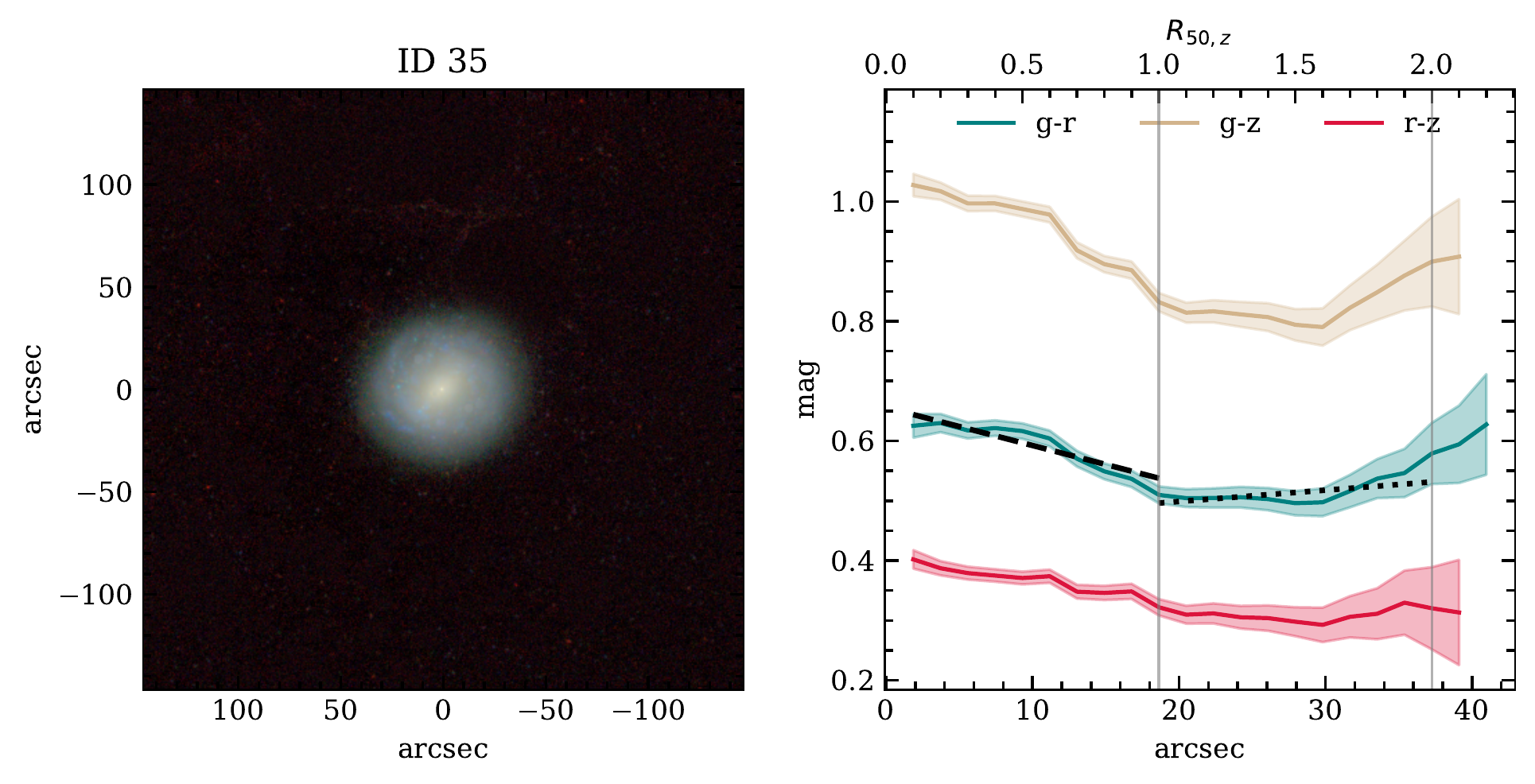}
\includegraphics[width=8.5cm]{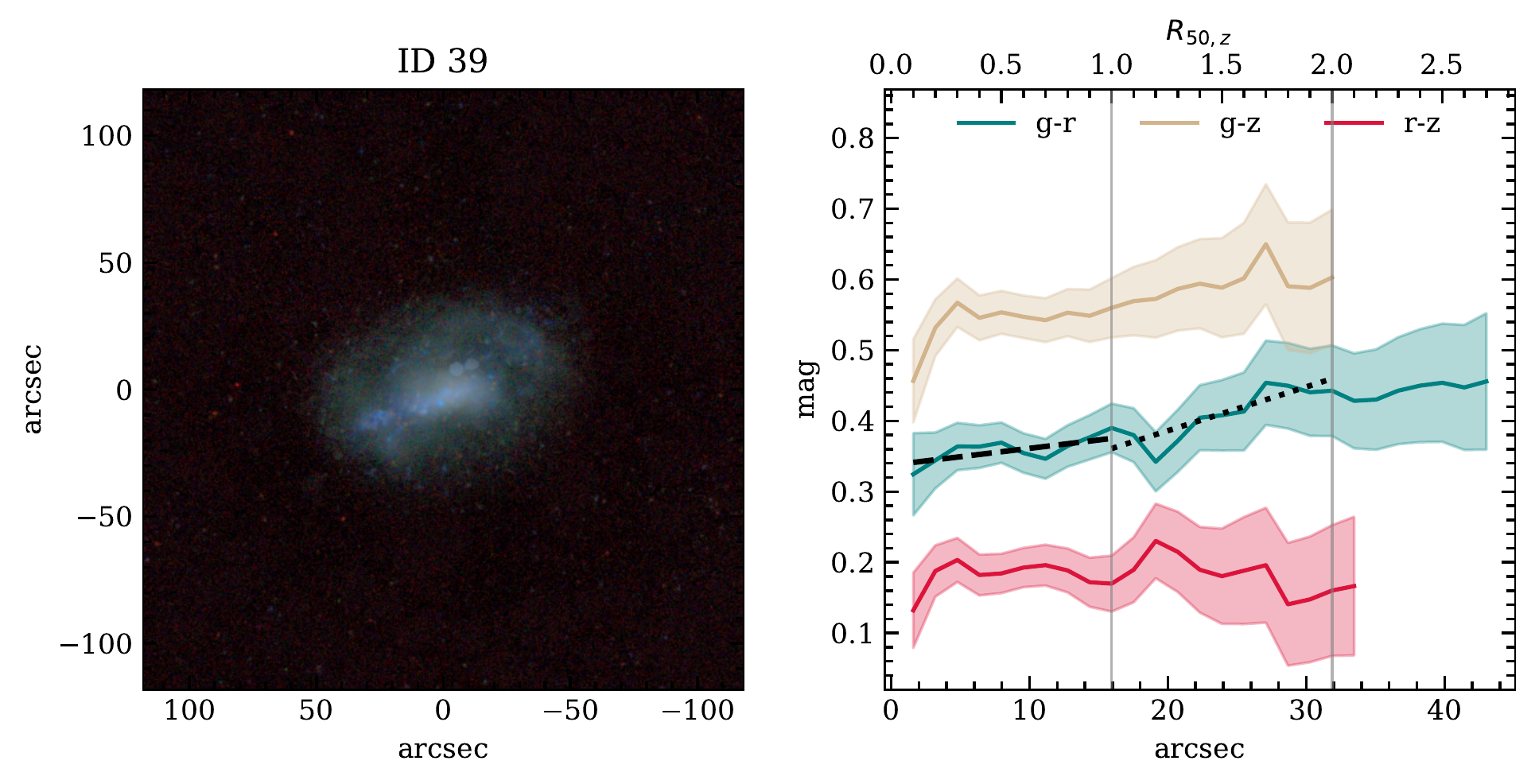}
\includegraphics[width=8.5cm]{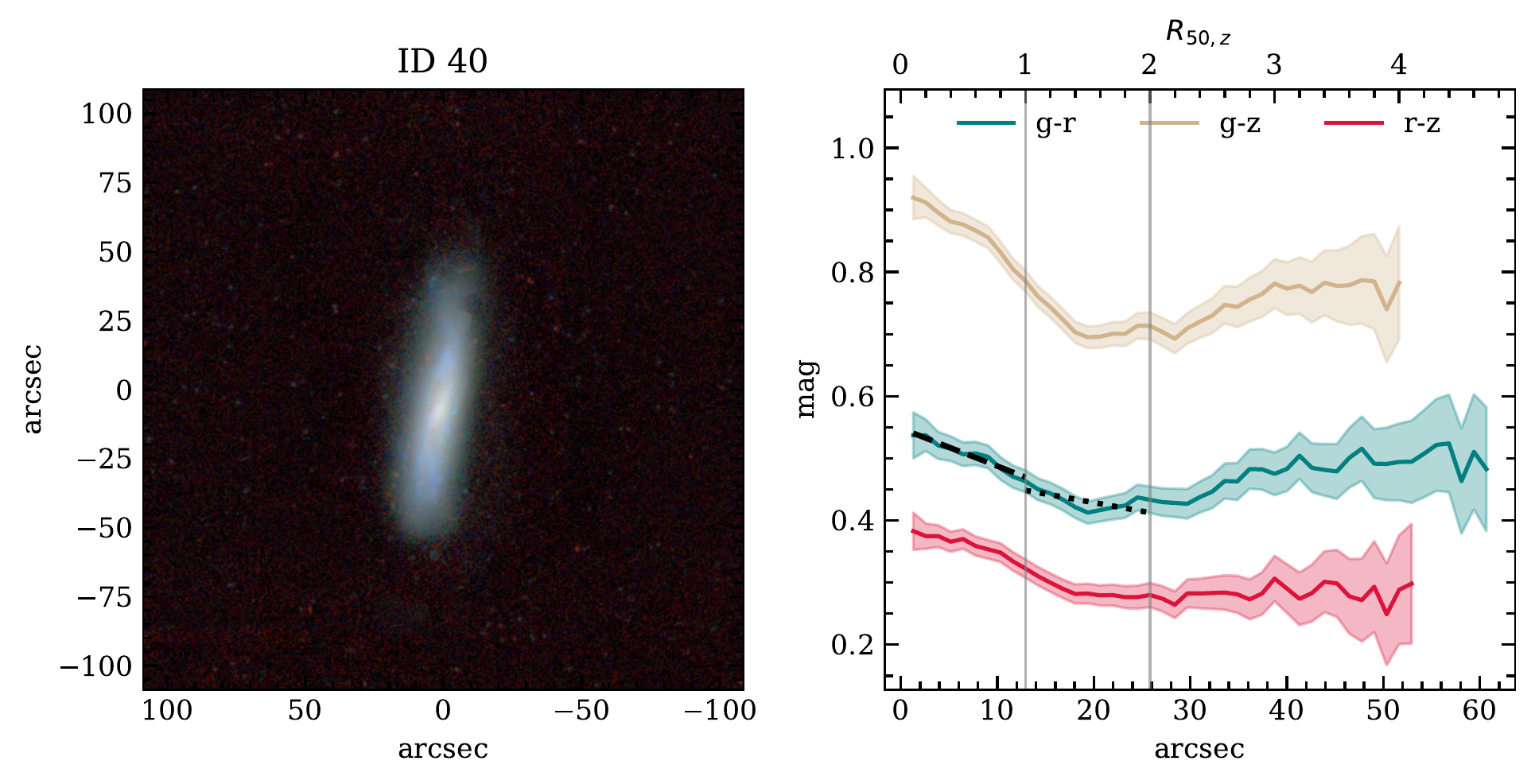}
\includegraphics[width=8.5cm]{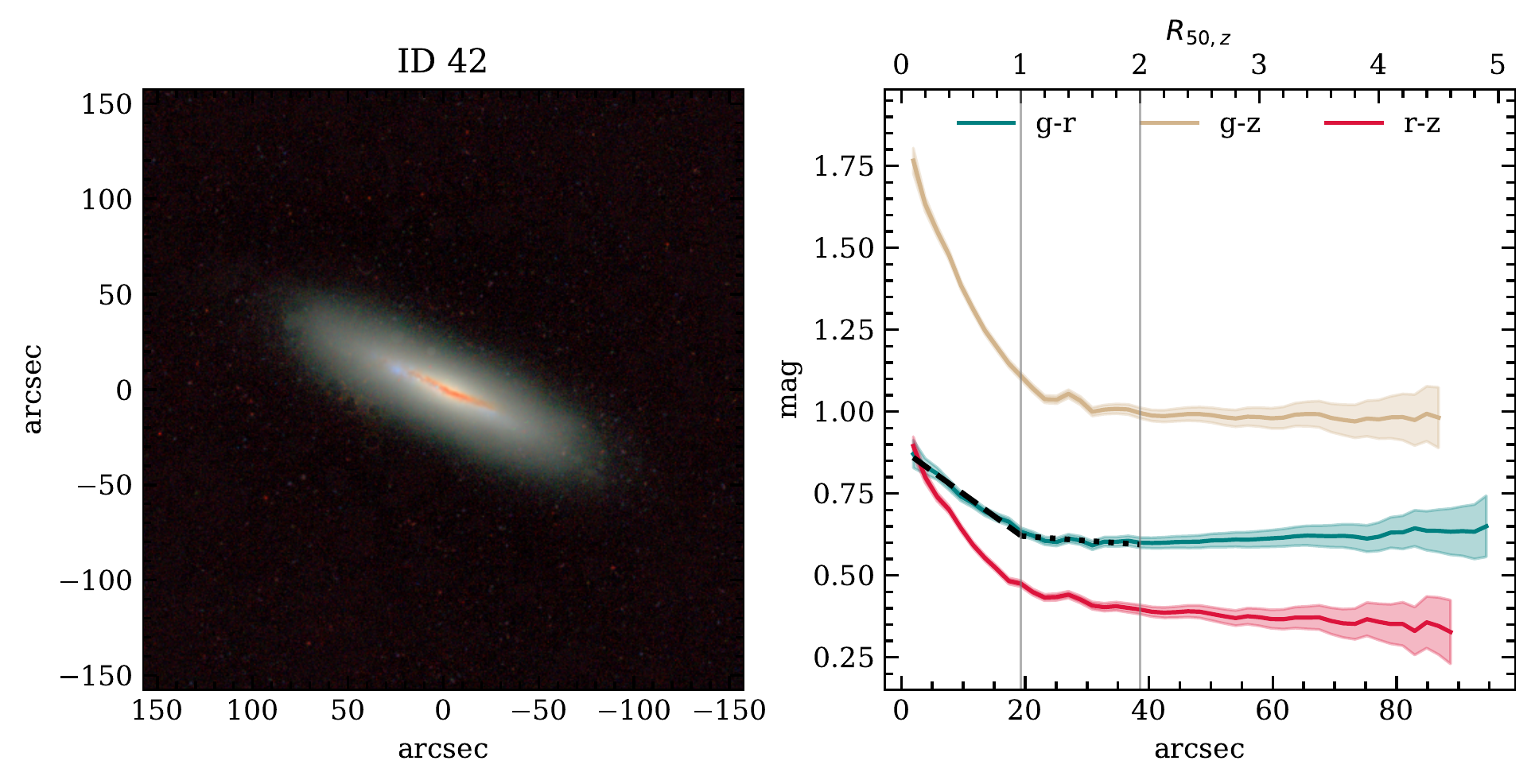}
\includegraphics[width=8.5cm]{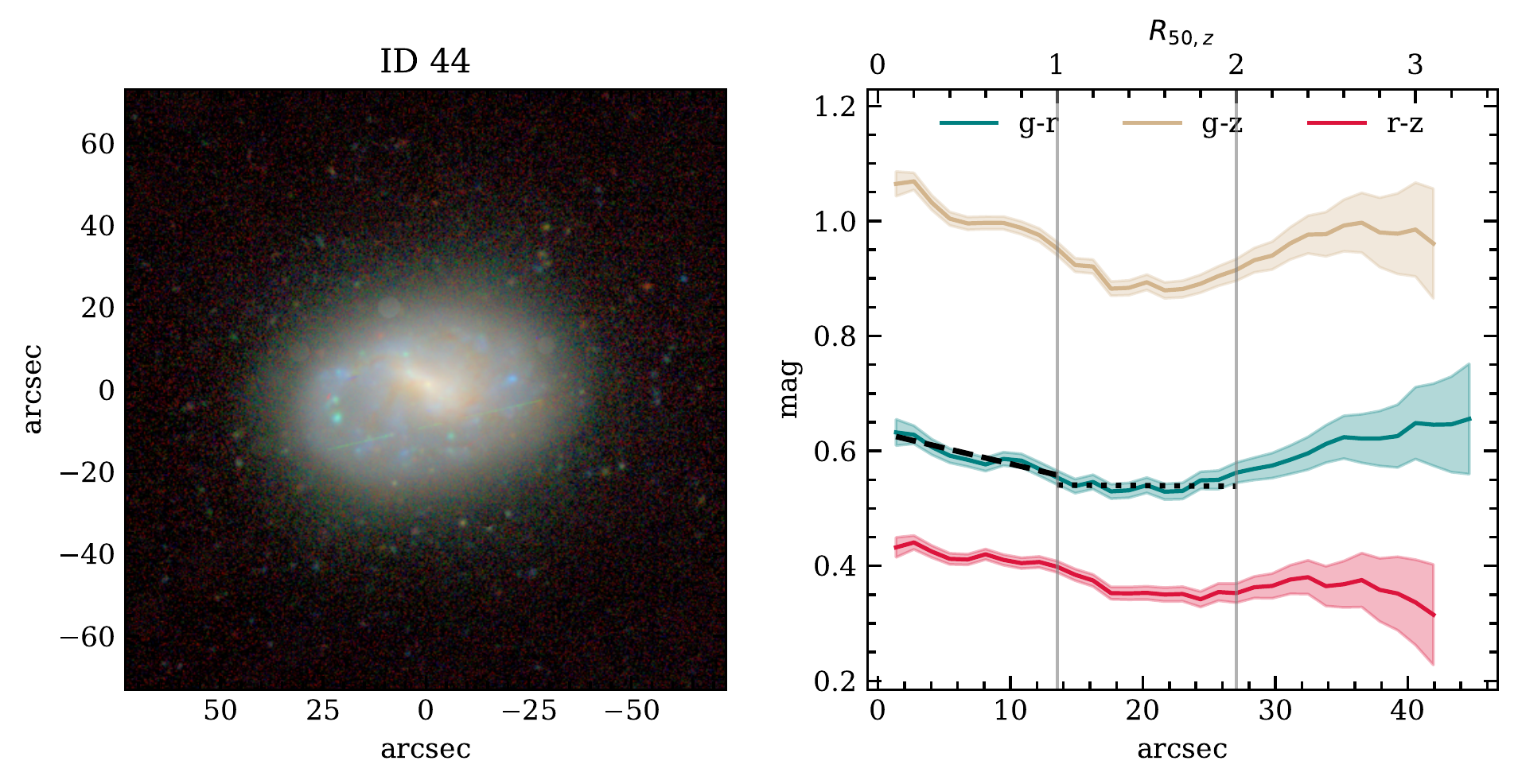}
\includegraphics[width=8.5cm]{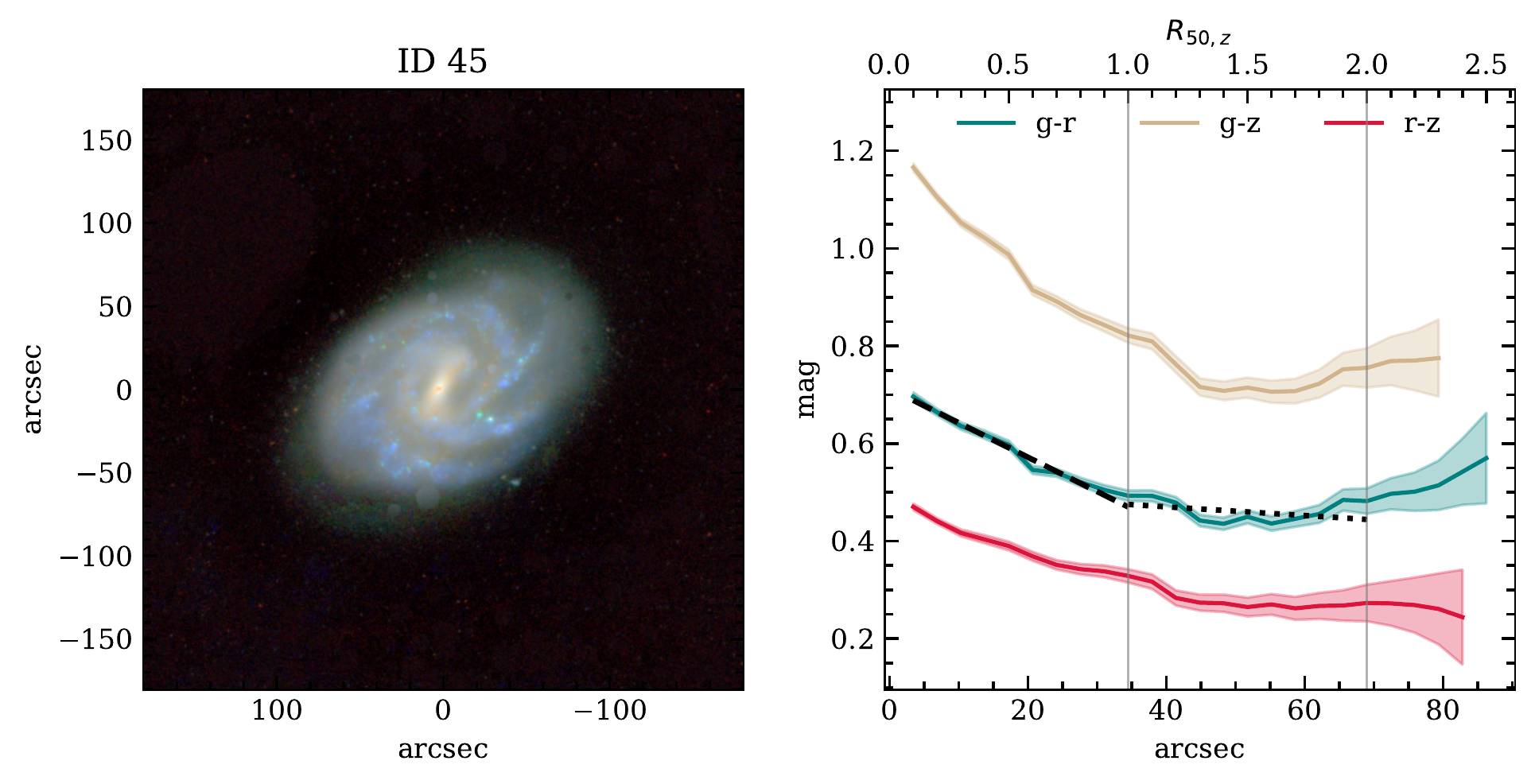}
\includegraphics[width=8.5cm]{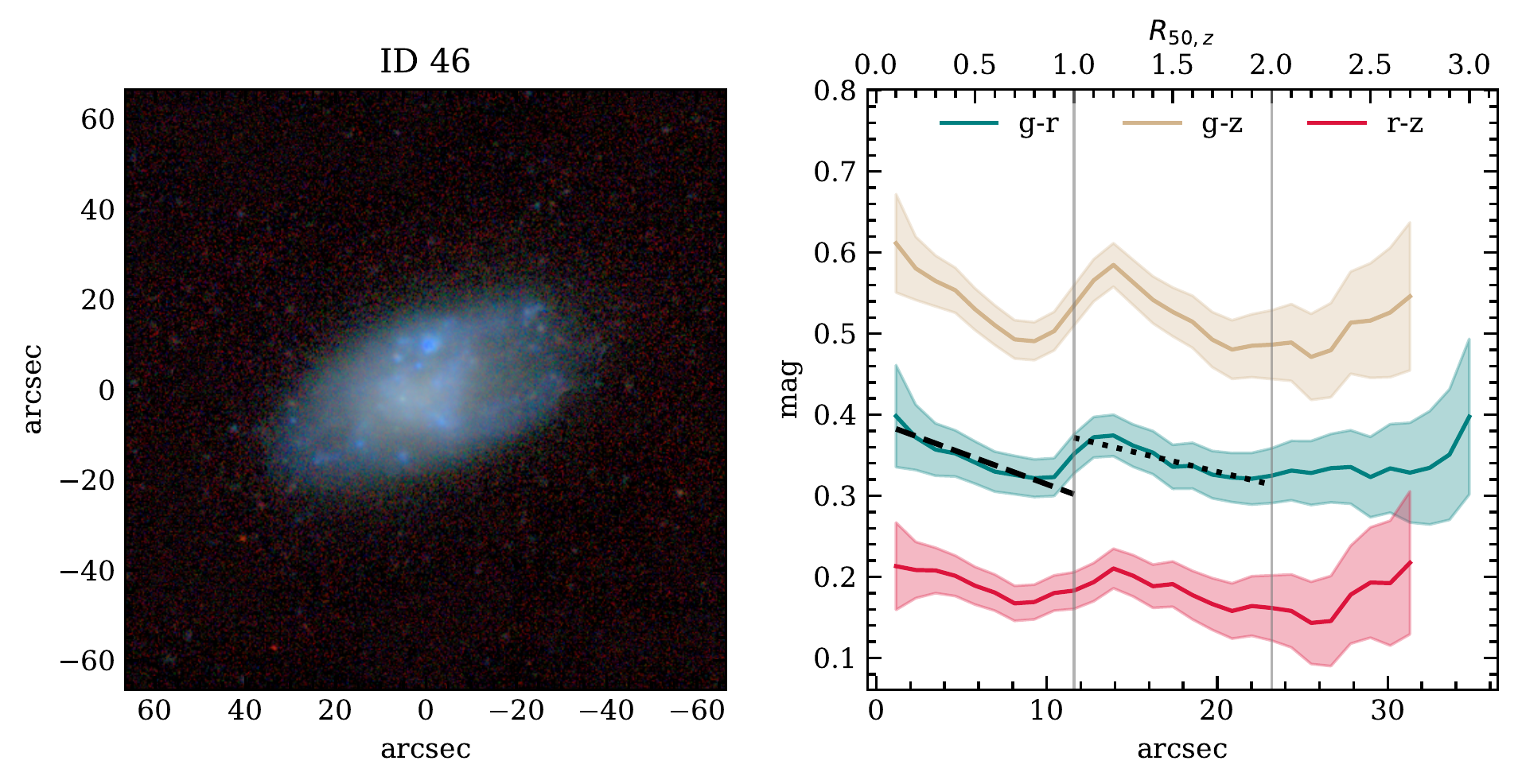}
\includegraphics[width=8.5cm]{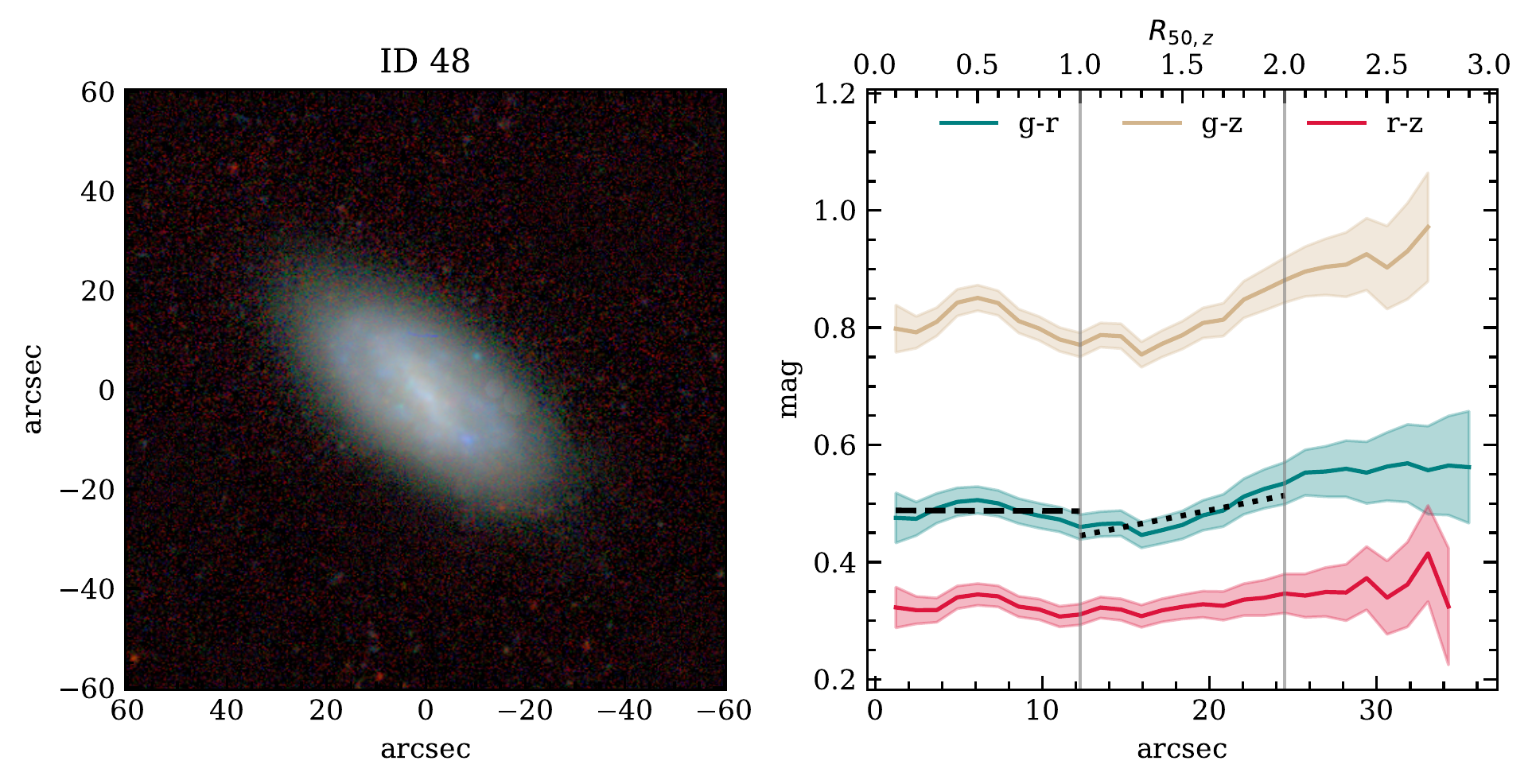}
\includegraphics[width=8.5cm]{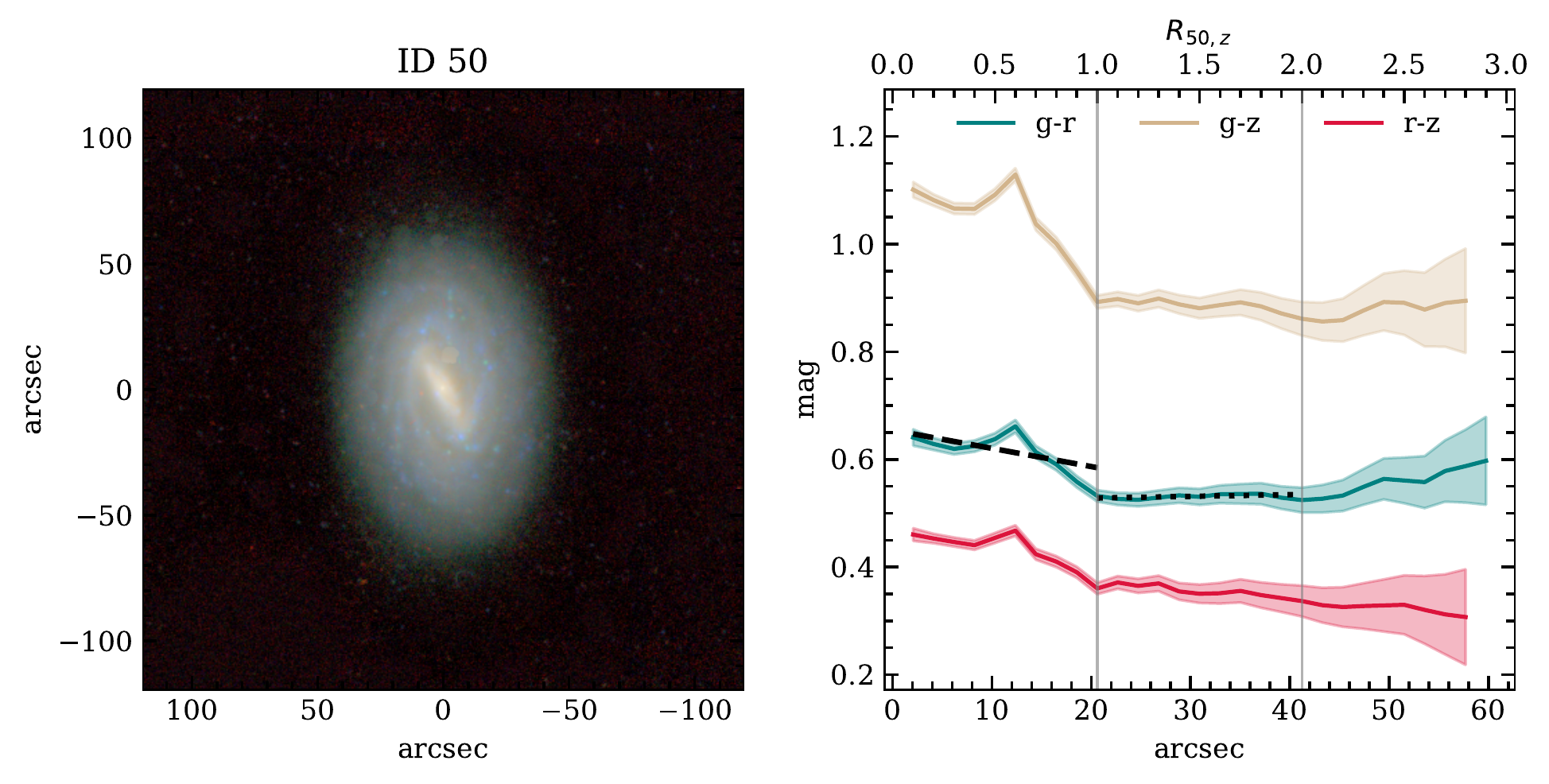}
\includegraphics[width=8.5cm]{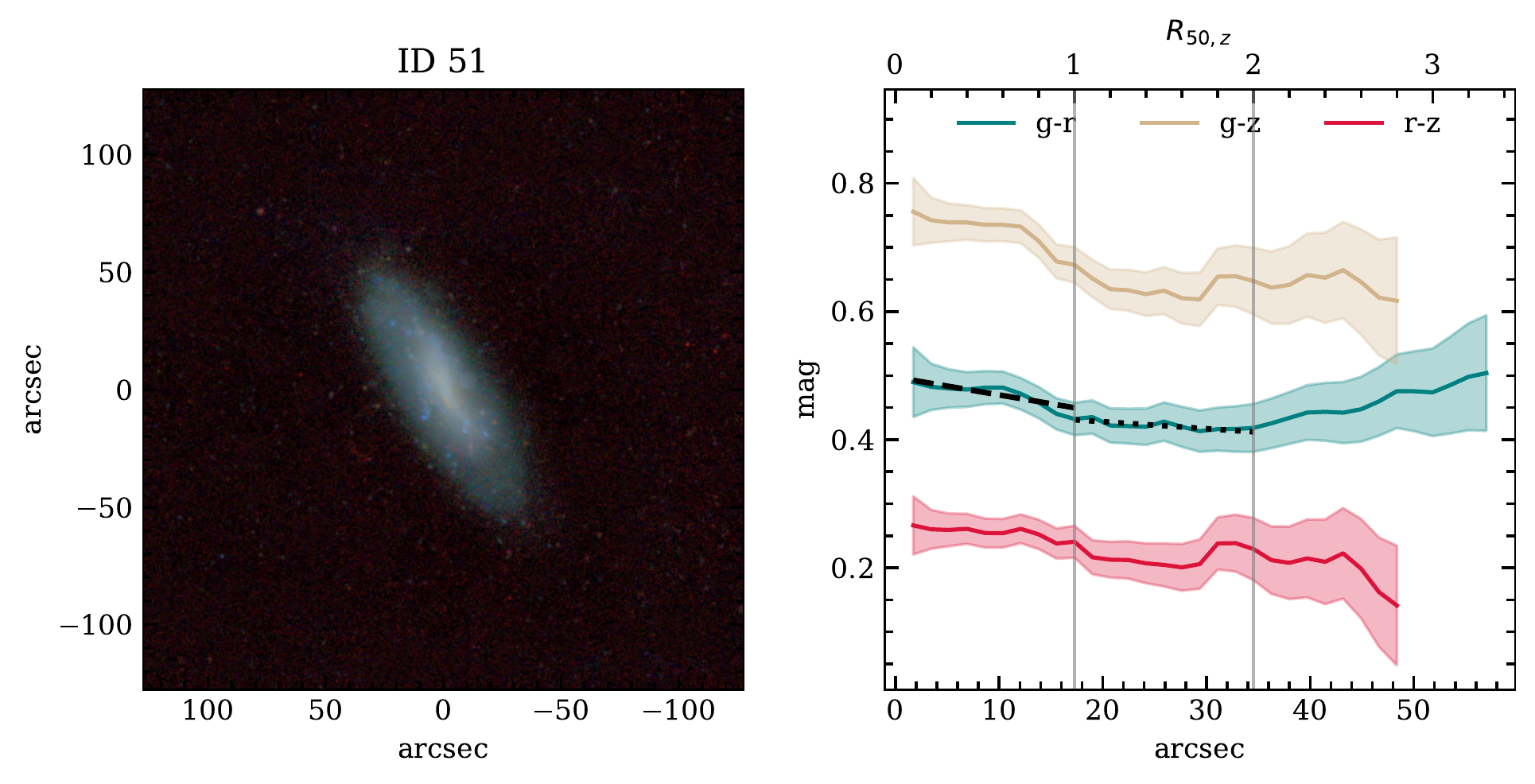}
    \caption{Continued.}
    \label{fig:atlas_2}
\end{figure}

\begin{figure}
    \centering
\includegraphics[width=8.5cm]{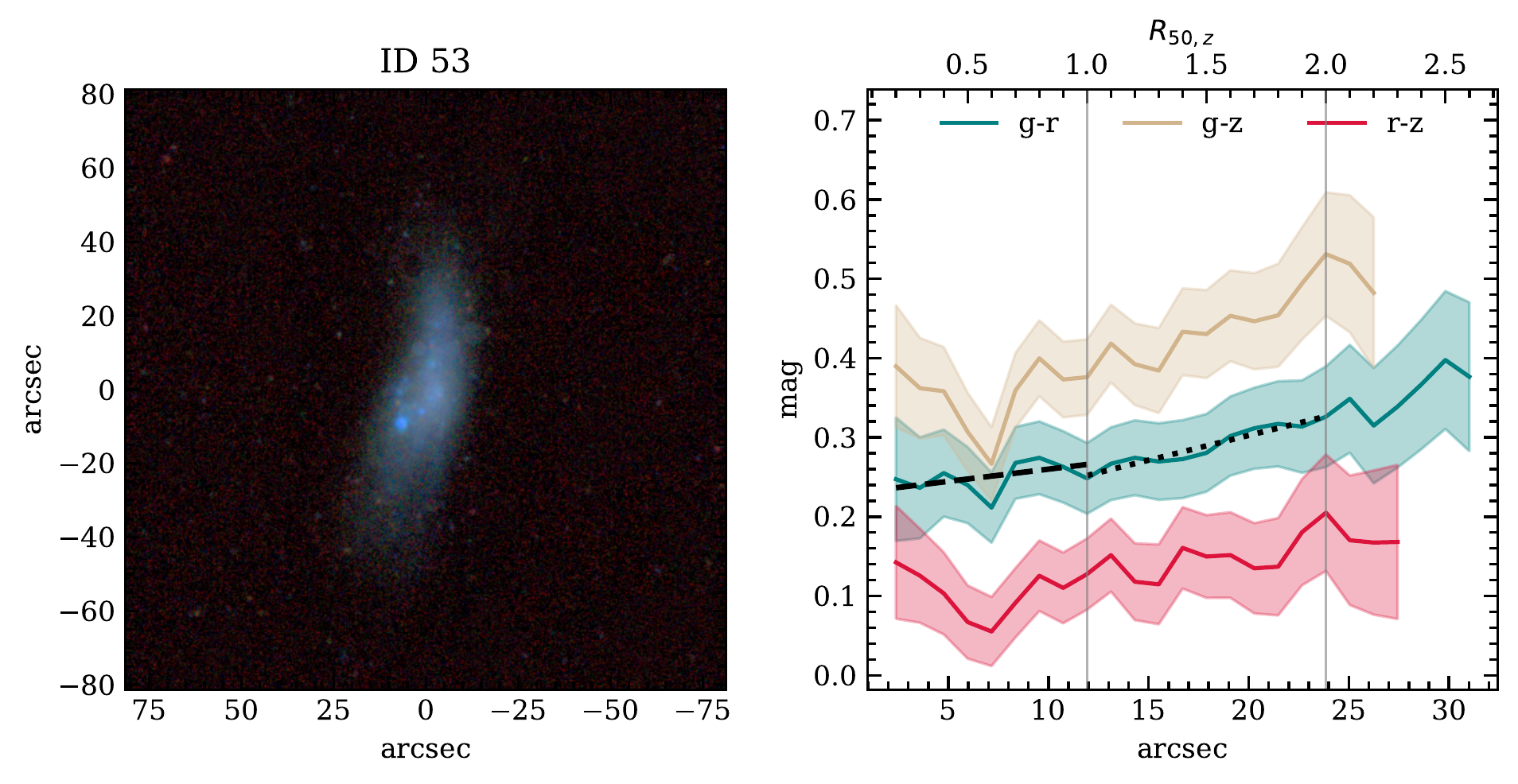}
\includegraphics[width=8.5cm]{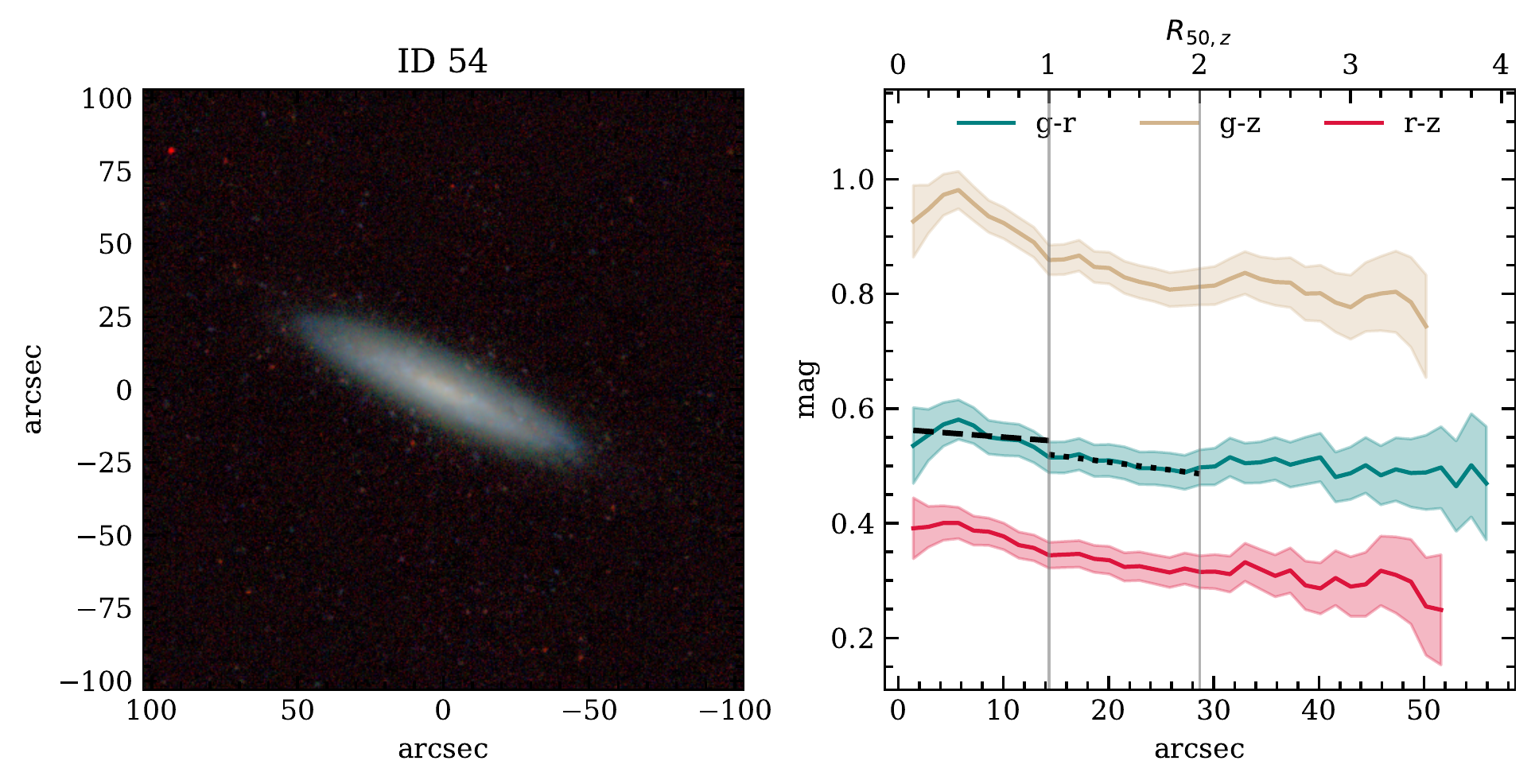}
\includegraphics[width=8.5cm]{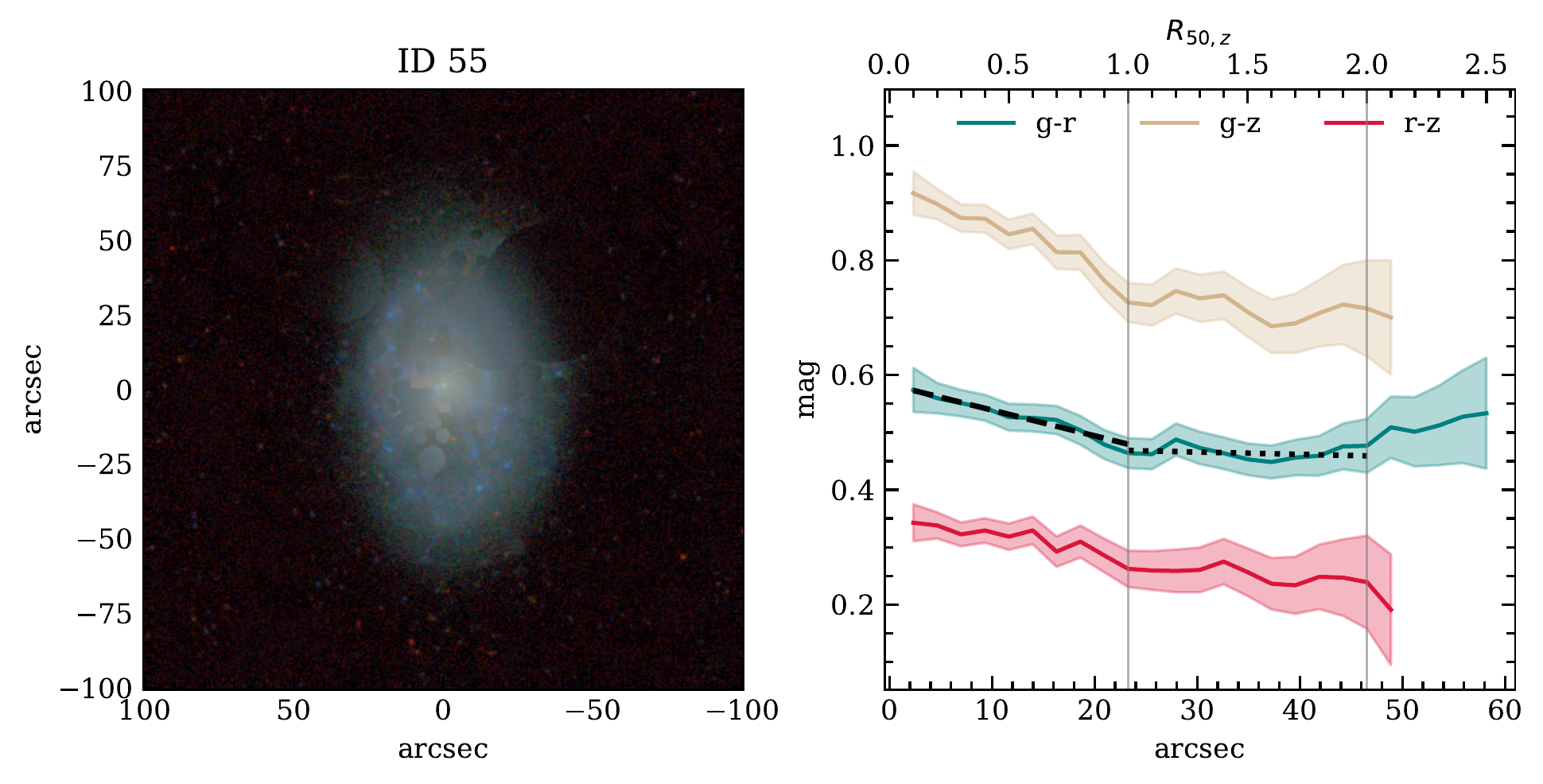}
\includegraphics[width=8.5cm]{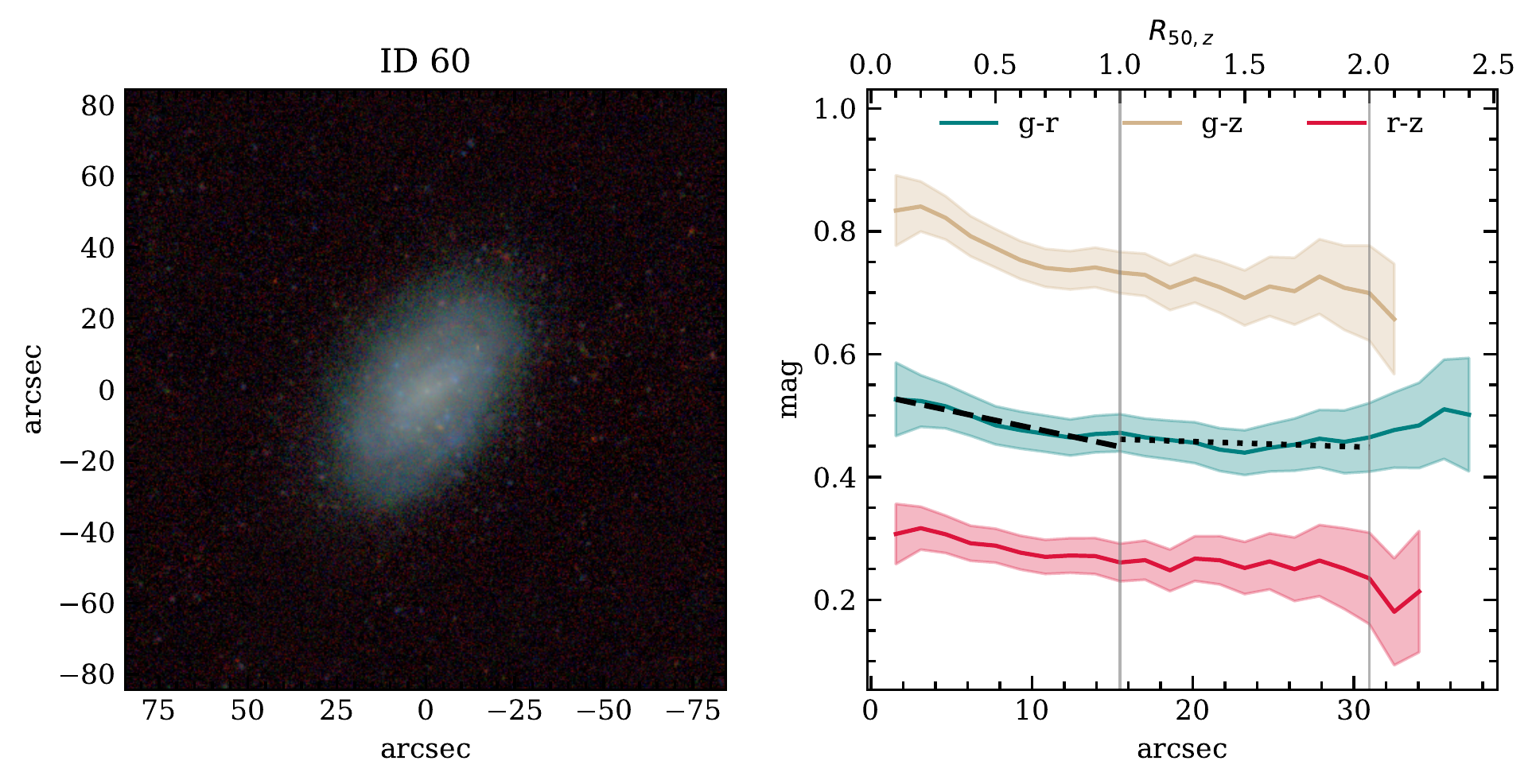}
\includegraphics[width=8.5cm]{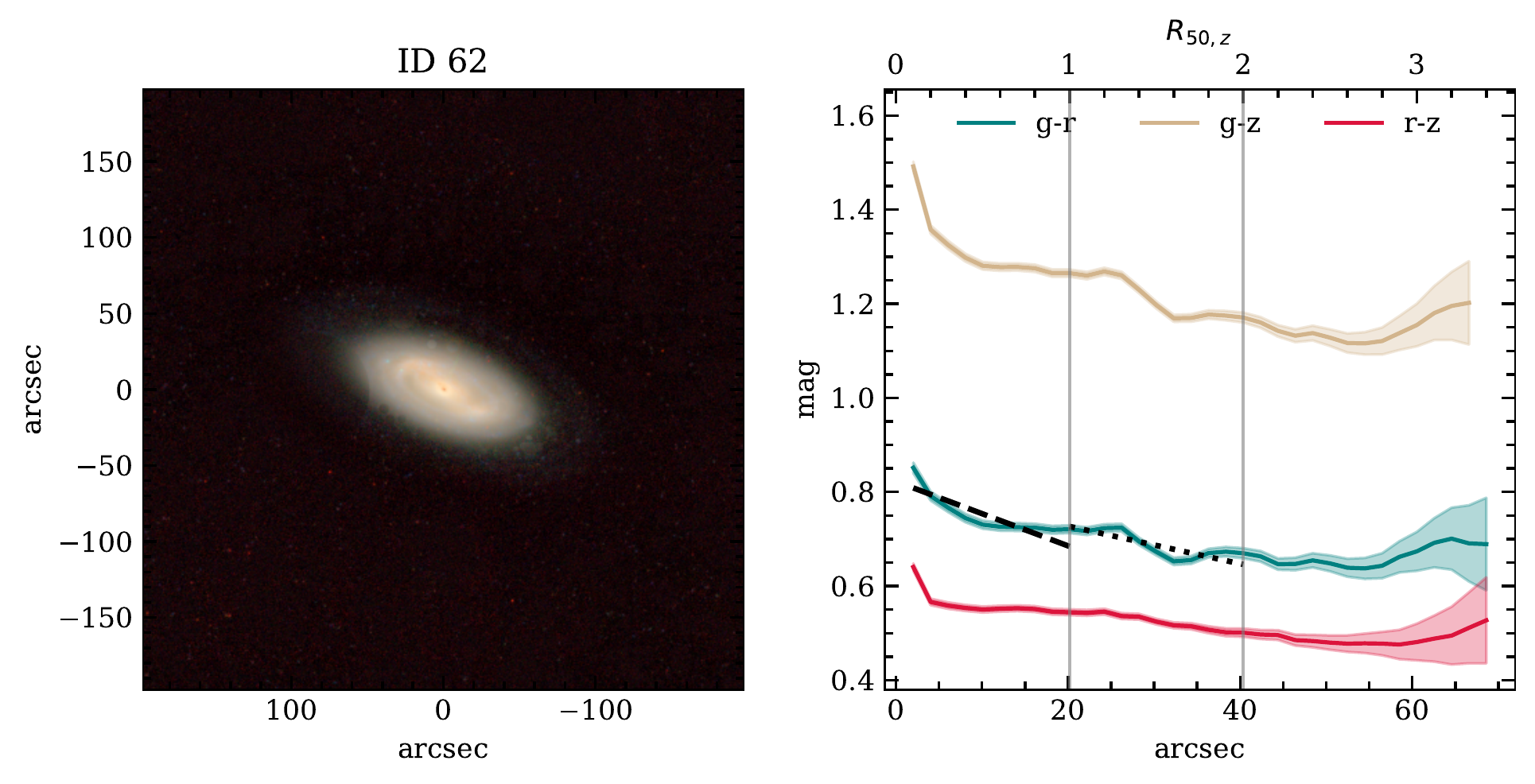}
\includegraphics[width=8.5cm]{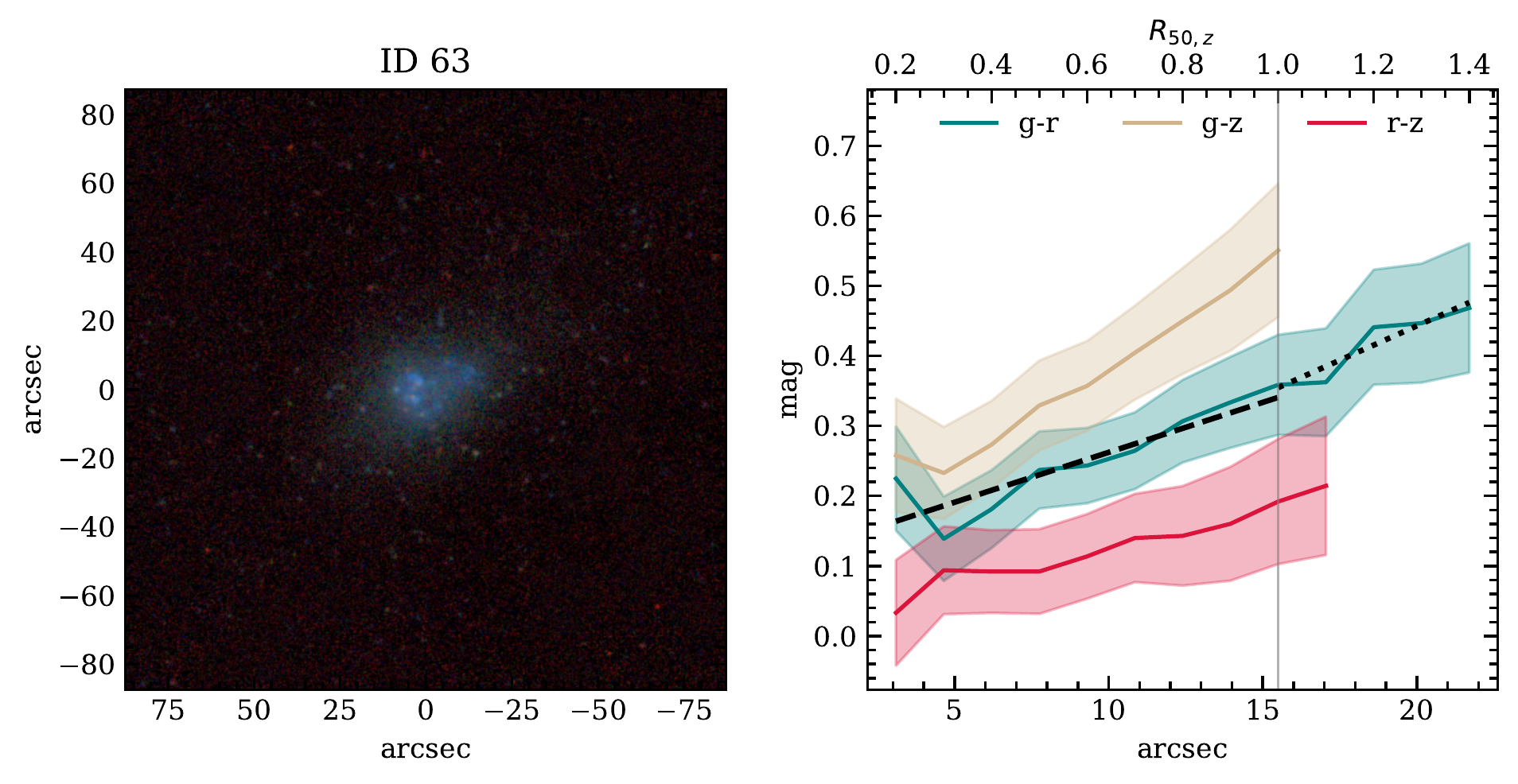}
\includegraphics[width=8.5cm]{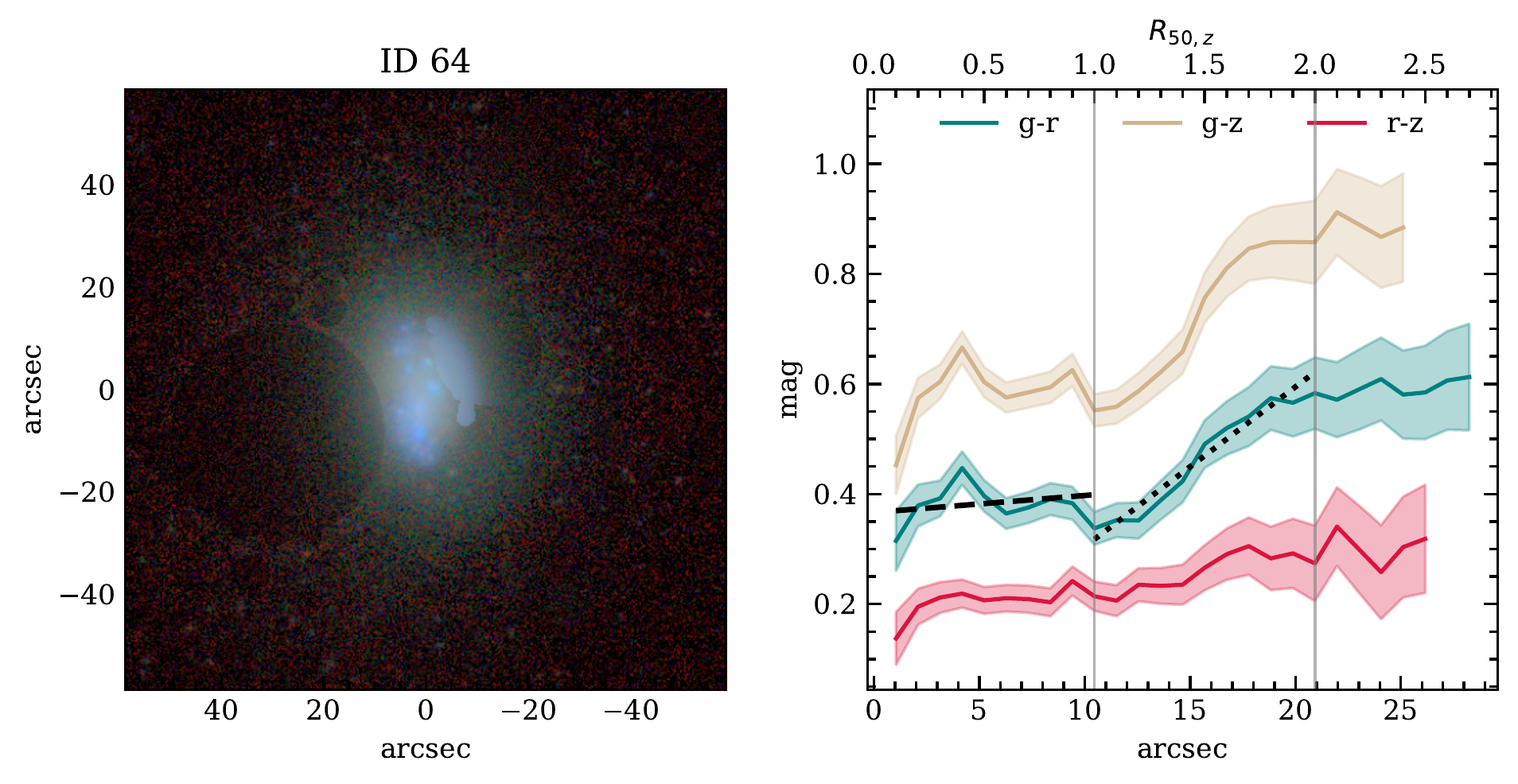}
\includegraphics[width=8.5cm]{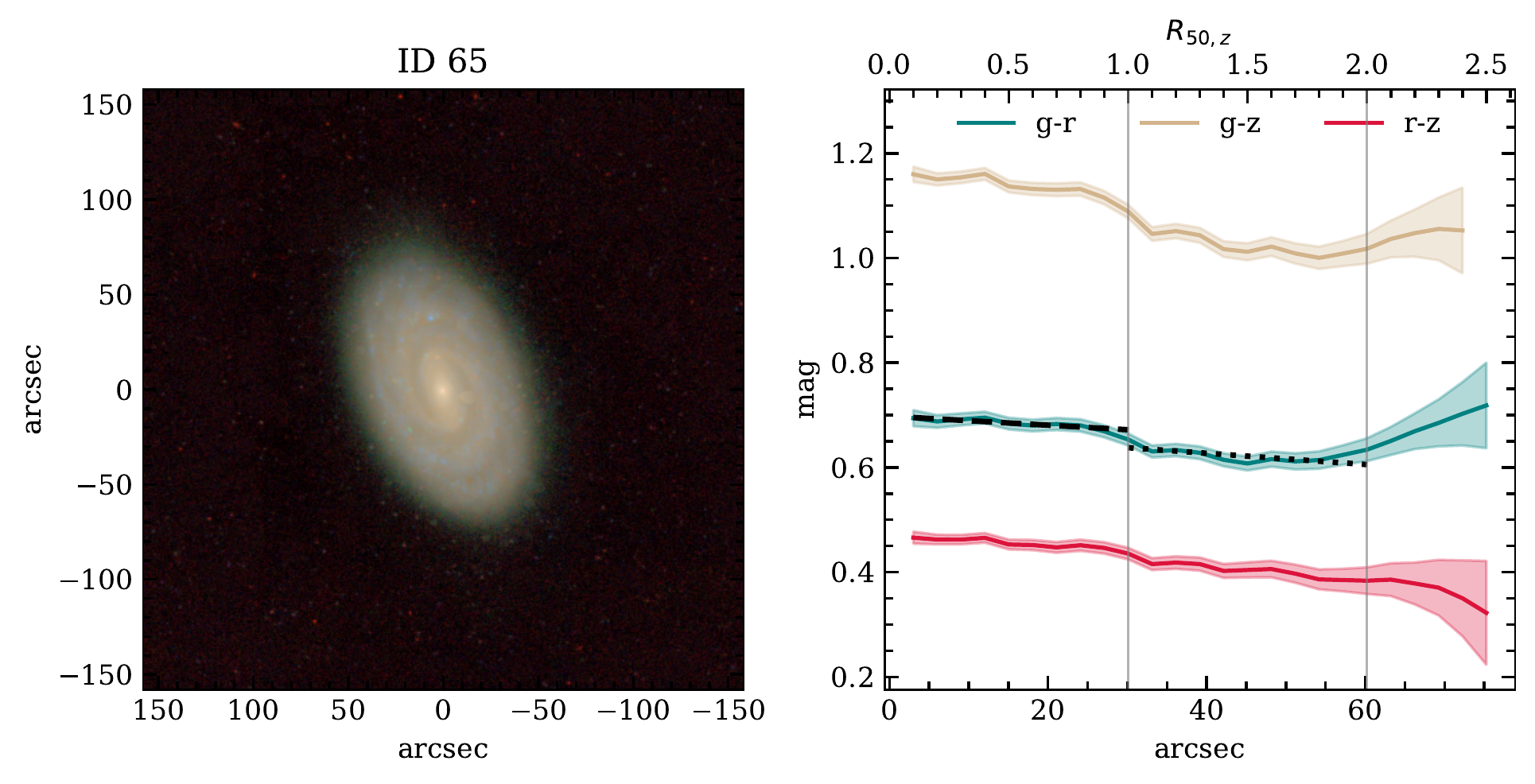}
\includegraphics[width=8.5cm]{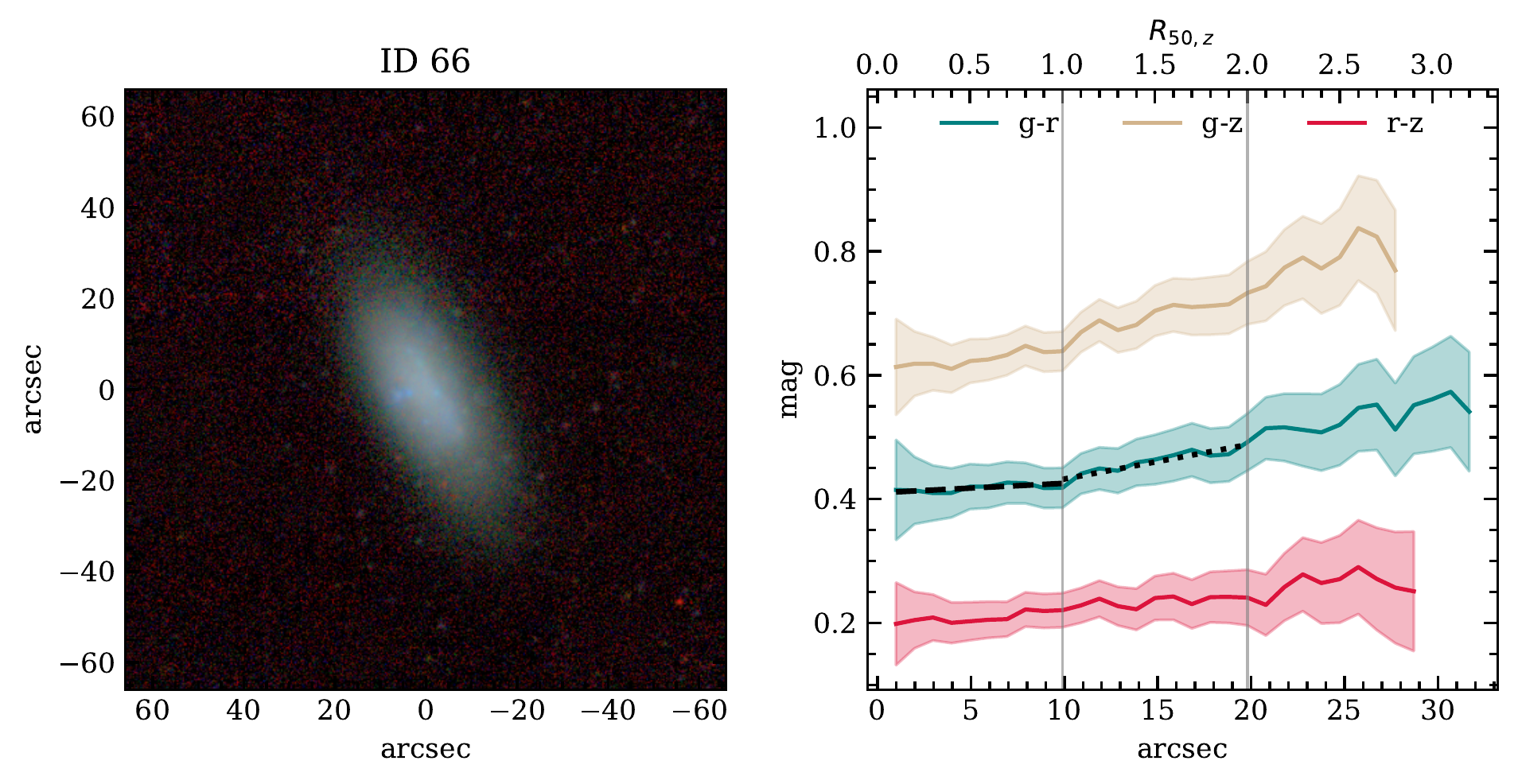}
\includegraphics[width=8.5cm]{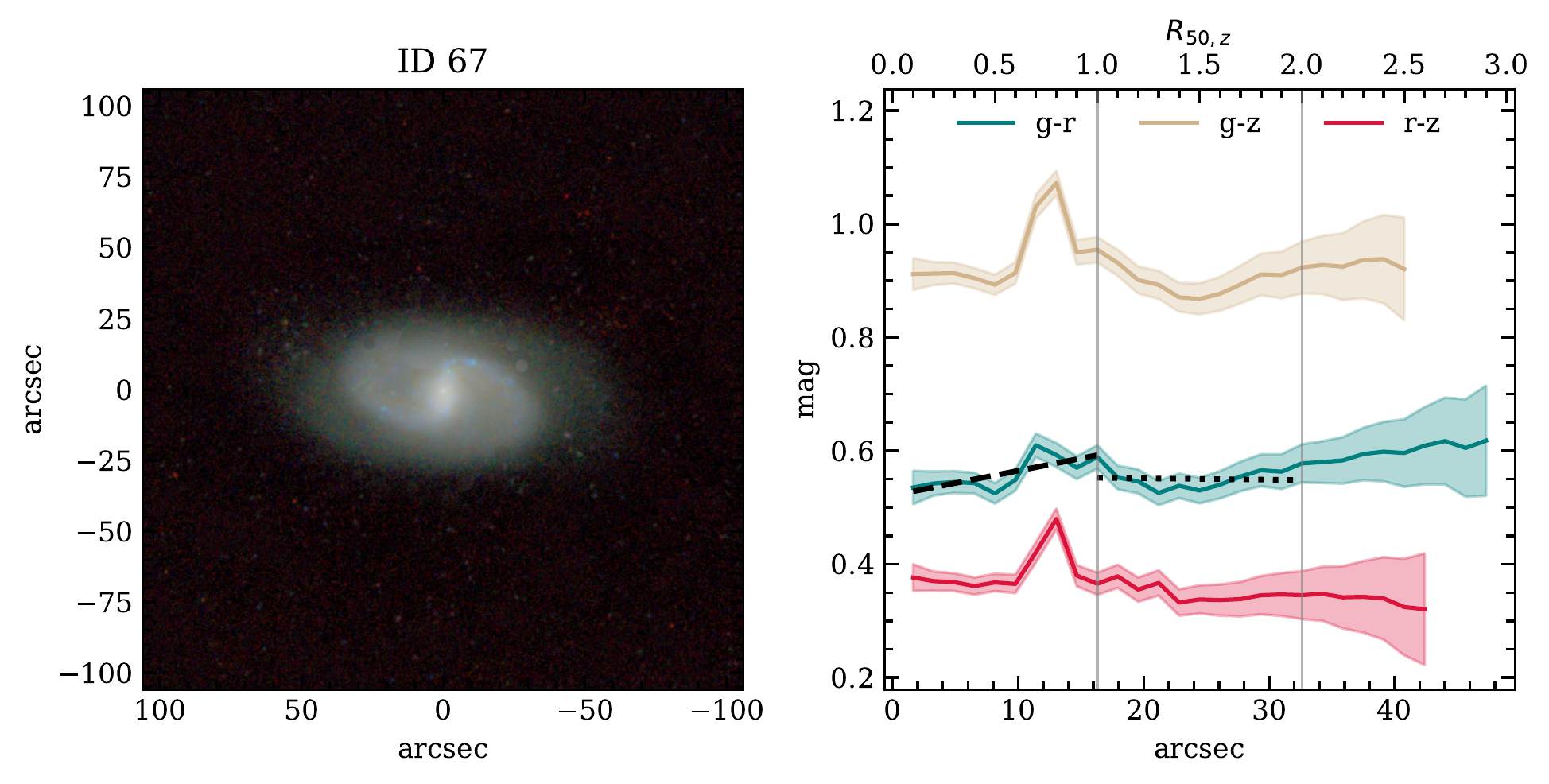}
    \caption{Continued.}
    \label{fig:atlas_3}
\end{figure}

\begin{figure}
    \centering
\includegraphics[width=8.5cm]{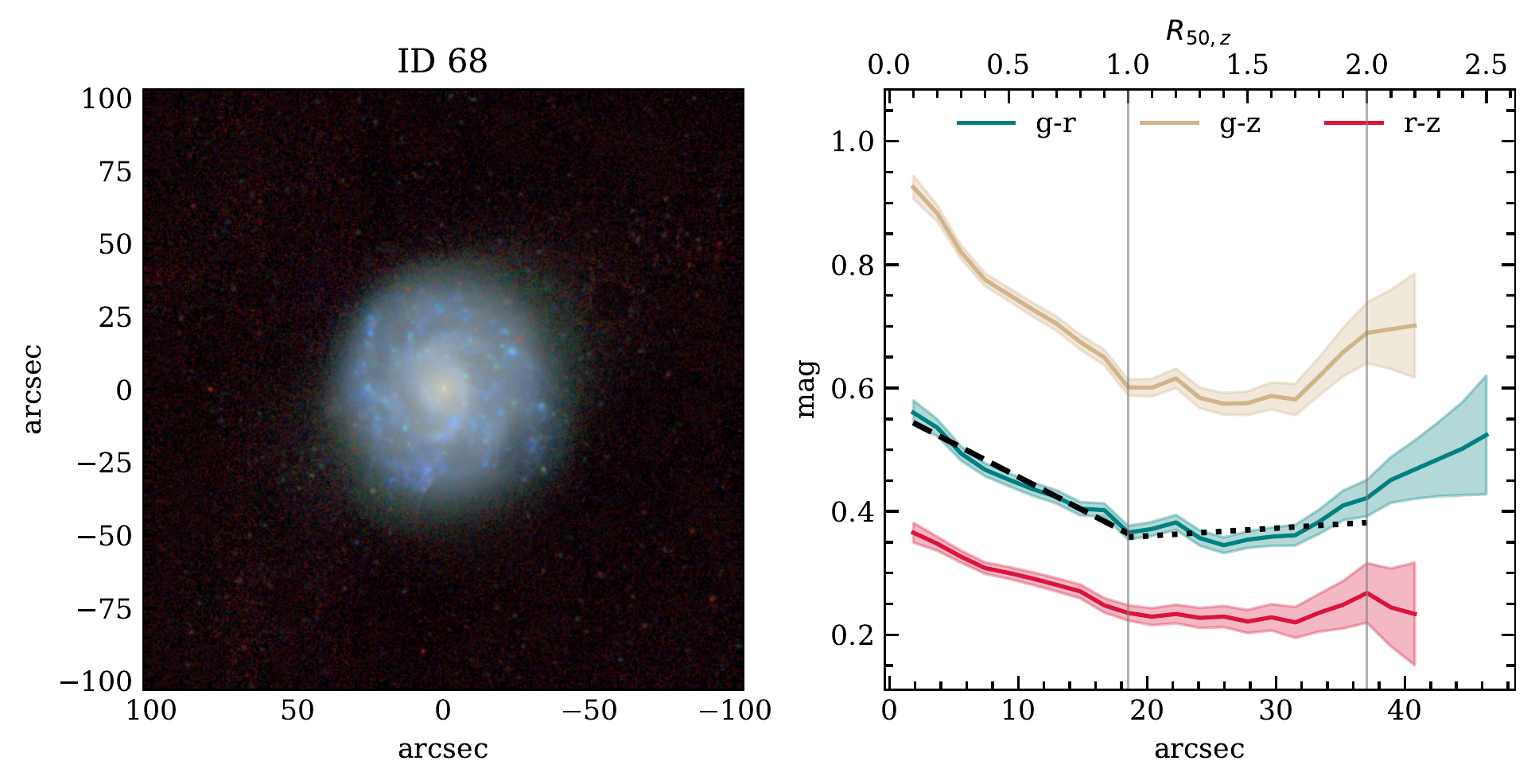}
\includegraphics[width=8.5cm]{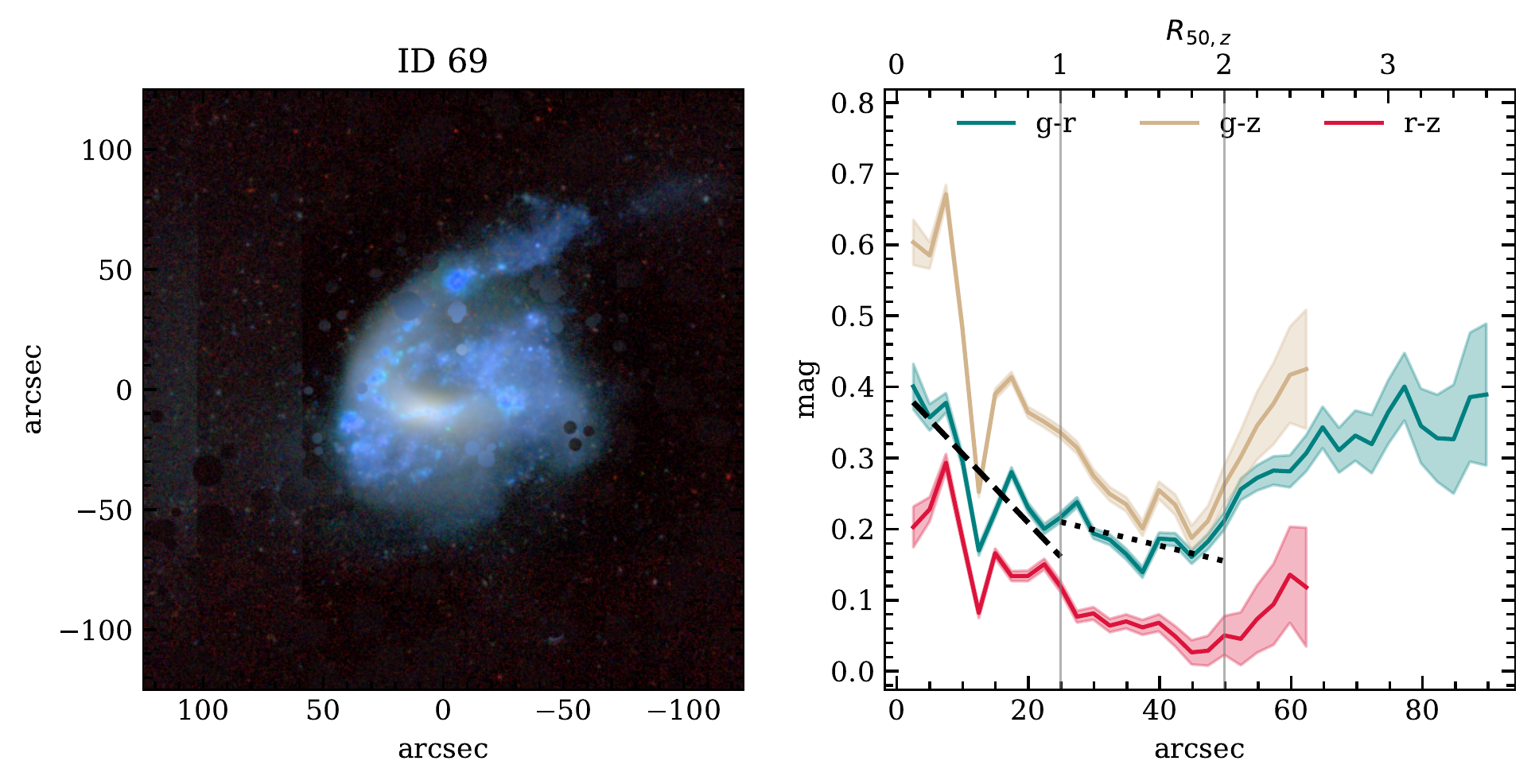}
\includegraphics[width=8.5cm]{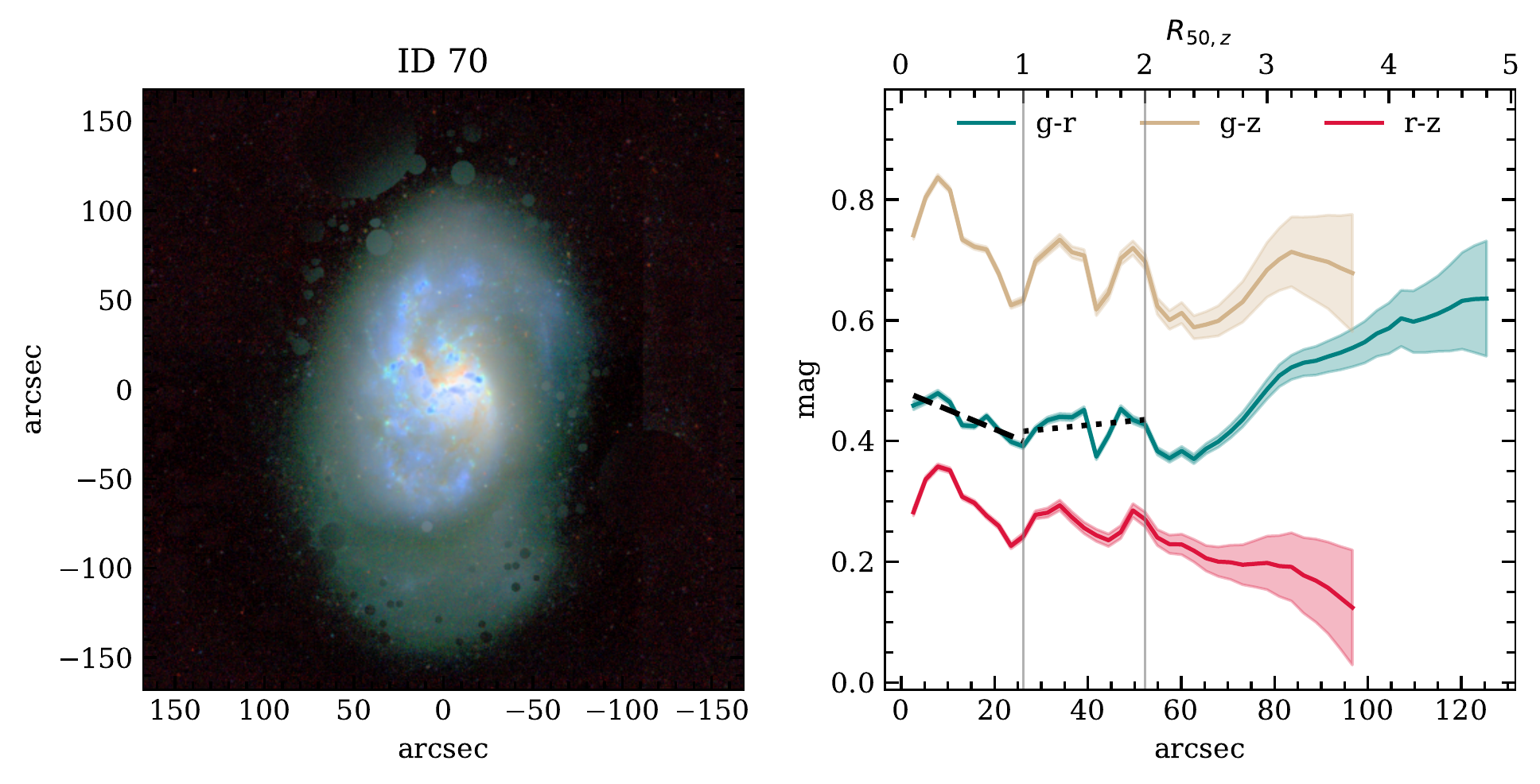}
\includegraphics[width=8.5cm]{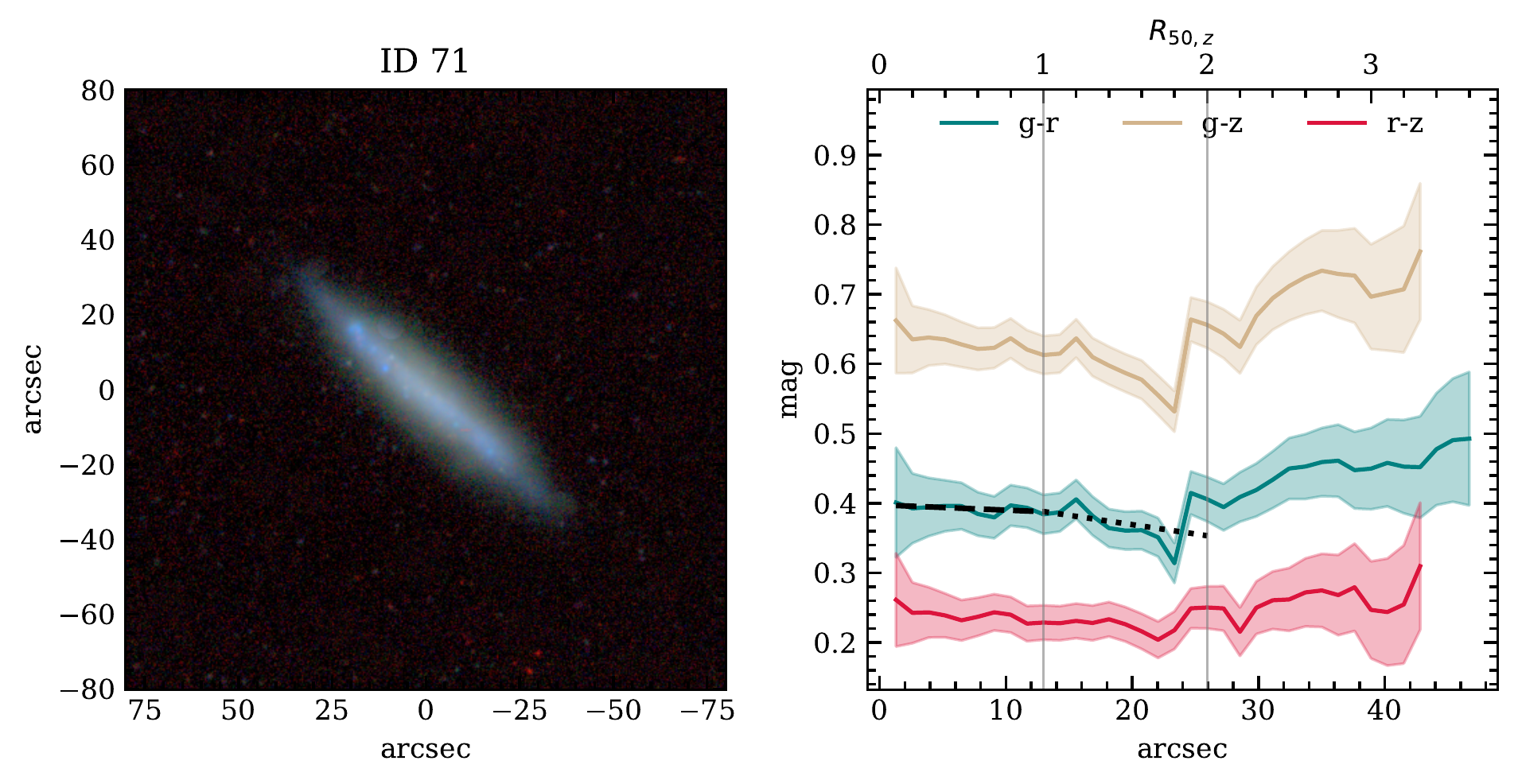}
\includegraphics[width=8.5cm]{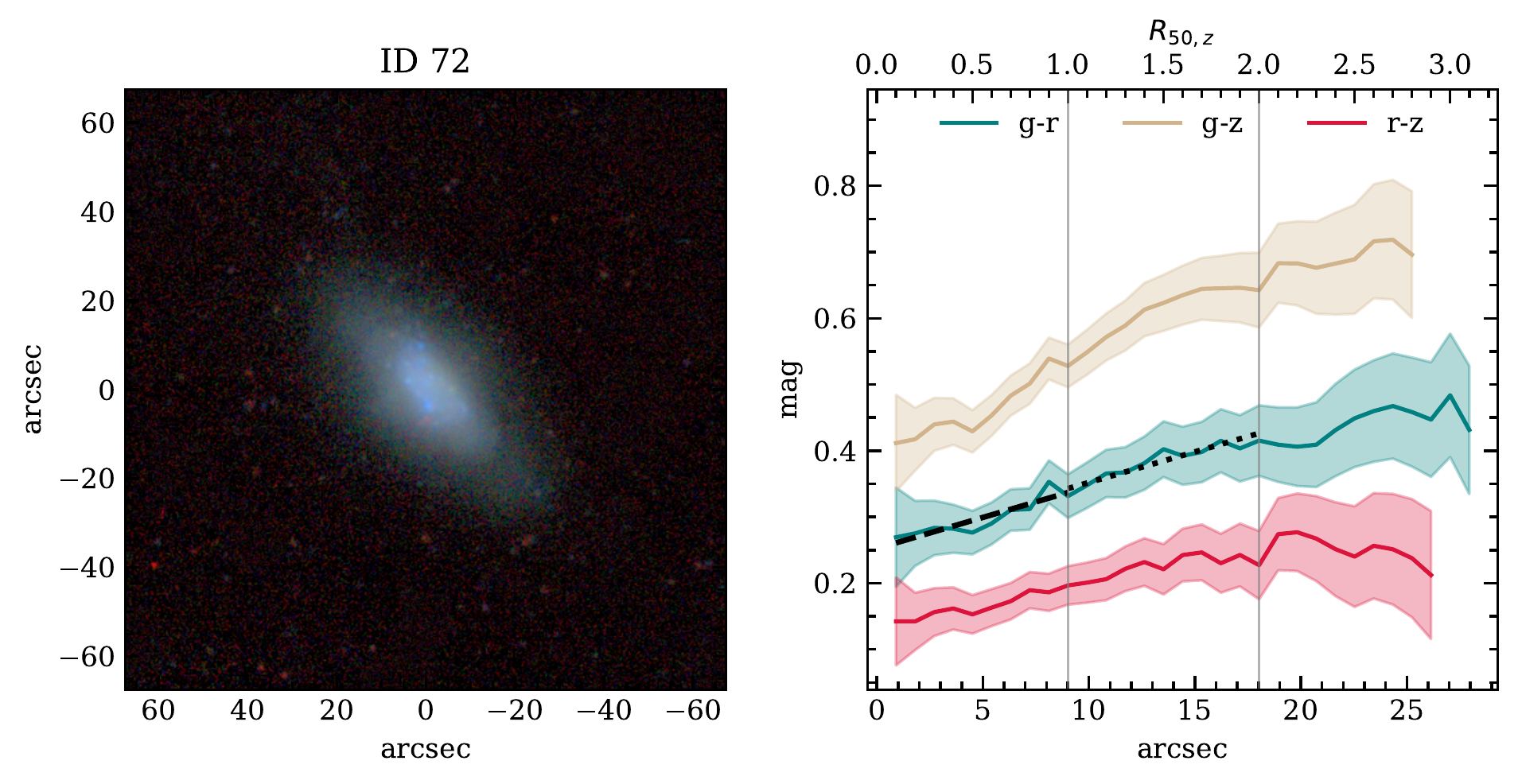}
\includegraphics[width=8.5cm]{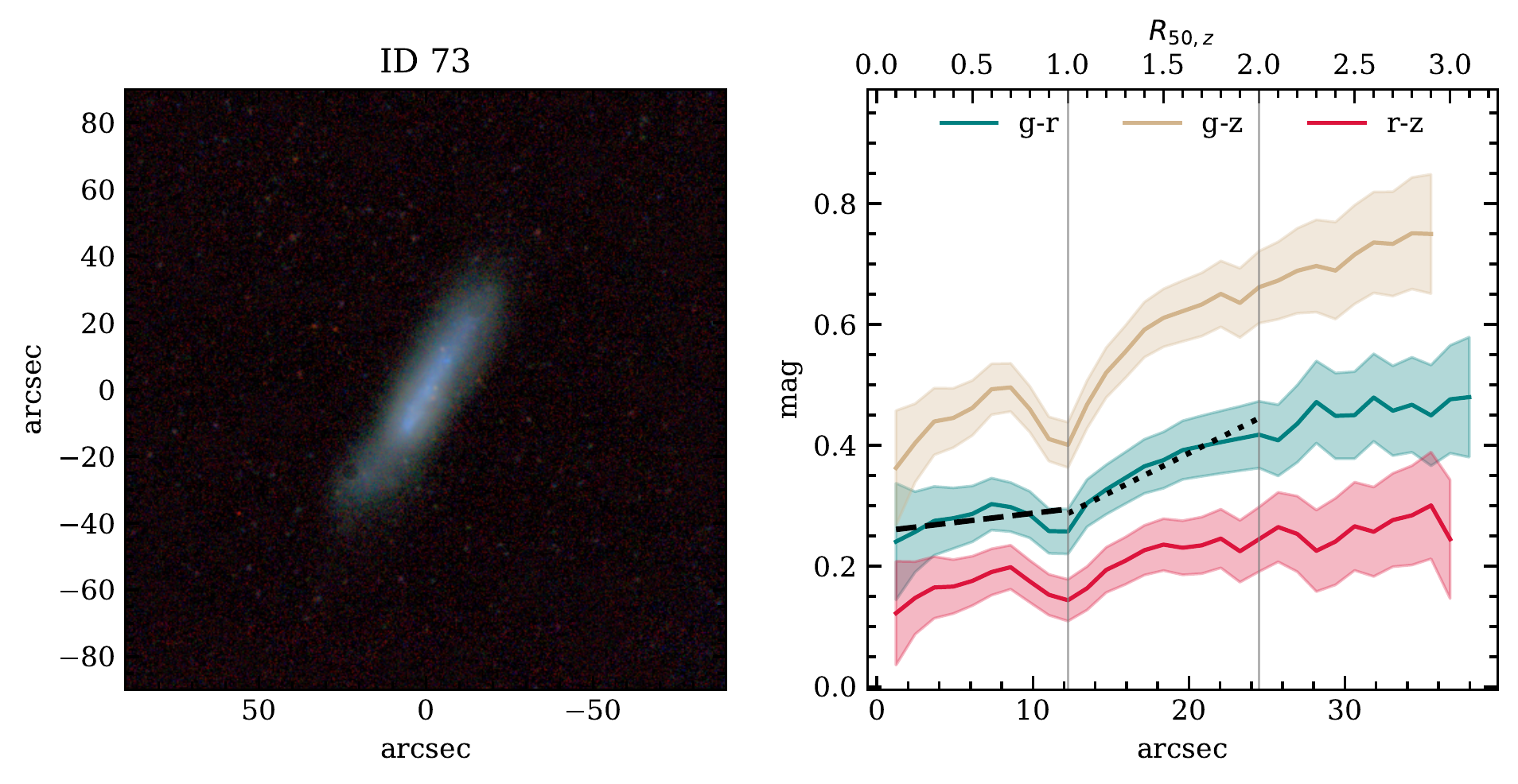}
\includegraphics[width=8.5cm]{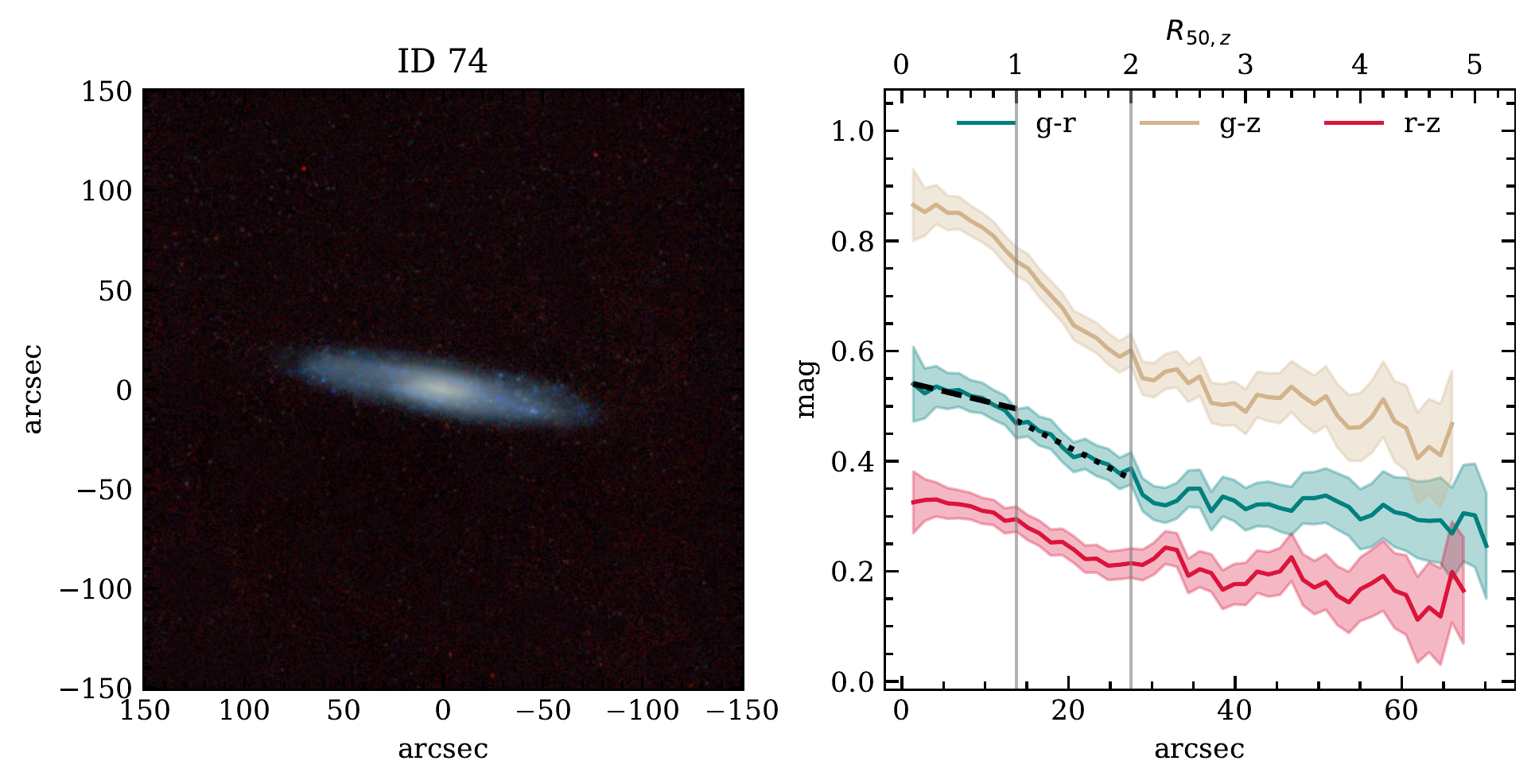}
\includegraphics[width=8.5cm]{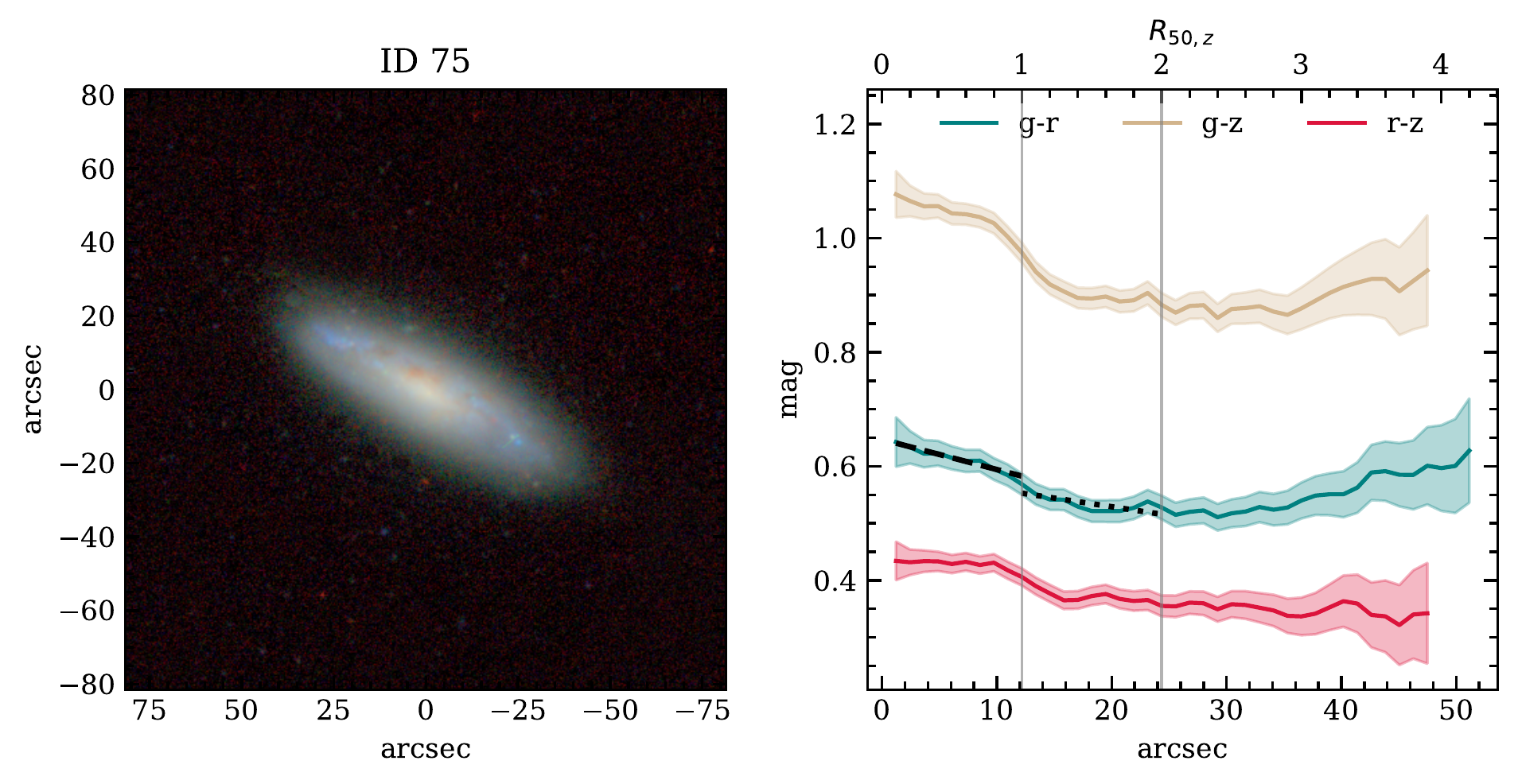}
    \caption{Continued.}
    \label{fig:atlas_4}
\end{figure}

\bibliography{sample63}{}
\bibliographystyle{aasjournal}

\end{document}